\documentclass[12pt]{article}
\usepackage[T1]{fontenc}
\usepackage[utf8]{inputenc}
\usepackage[margin=1in]{geometry}
\IfFileExists{newtxtext.sty}{\usepackage{newtxtext}}{\usepackage{mathptmx}}
\usepackage{graphicx}
\usepackage[authoryear,round]{natbib}
\usepackage{amsmath,amssymb}
\usepackage{booktabs,tabularx,longtable,array,threeparttable}

\usepackage{xcolor}
\usepackage{setspace}
\usepackage{endnotes}

\usepackage[font={small,stretch=2},labelfont=bf]{caption}
\usepackage{float,placeins,etoolbox,rotating}
\usepackage{microtype}
\usepackage{xurl}
\usepackage[hidelinks]{hyperref}
\renewcommand{\textsuperscript}[1]{\mbox{\raisebox{.6ex}{\scriptsize #1}}}
\graphicspath{{Figure/}}

\newcommand{\source}[1]{\par\smallskip{\footnotesize\doublespacing\emph{Reproducible source:} #1\par}}
\newcommand{\code}[1]{\nolinkurl{#1}}
\providecommand{\doi}[1]{\href{https://doi.org/#1}{doi: #1}}
\newcommand{\alttxt}[1]{\par\smallskip{\small\doublespacing\noindent\textbf{Alt text:} #1\par}}
\author{Huy Nguyen\textsuperscript{1} \and Dawoon Jeong\textsuperscript{2,3,*}}
\title{\textbf{\LARGE Anchored Exploration in Product Evolution:\\A Phylogenetic Network Framework for Semiconductor Designs}}
\date{}
\hypersetup{pdftitle={Anchored Exploration in Product Evolution: A Phylogenetic Network Framework for Semiconductor Designs},pdfauthor={Huy Nguyen and Dawoon Jeong},pdfsubject={},pdfkeywords={}}
\begin{document}
\begin{singlespace}
\begin{center}
{\LARGE\bfseries Anchored Exploration in Product Evolution:\\A Phylogenetic Network Framework for Semiconductor Designs\par}
\vspace{1em}
{\large Huy Nguyen\textsuperscript{1} \quad Dawoon Jeong\textsuperscript{2,3,*}\par}
\vspace{0.75em}
{\small
\textsuperscript{1}Augustana College, Rock Island, IL, USA\par
\textsuperscript{2}Department of Sociology, University of Chicago, Chicago, IL, USA\par
\textsuperscript{3}Knowledge Lab, University of Chicago, Chicago, IL, USA\par}
\end{center}
\vspace{0.5em}
{\small\noindent\textsuperscript{*}\textbf{Corresponding author:} Dawoon Jeong\par
\noindent Department of Sociology and Knowledge Lab, University of Chicago\par
\noindent Chicago, IL, USA\par
\noindent Email: \href{mailto:jdwoon0523@uchicago.edu}{jdwoon0523@uchicago.edu}\par}
\end{singlespace}

\thispagestyle{plain}
\section*{Abstract}
We develop a product phylogenetic network framework for examining anchored exploration. Using 102,465 semiconductor products grouped into 45,996 design families, we distinguish accumulated design continuity (Anchor) from externally precedented feature additions (horizontal transfer, HT). Fitted regressions yield non-monotonic relationships for HT occurrence and descendant carriage, alongside increasing within-family carriage over observed support. These patterns connect technological trajectories and recombination by distinguishing the continuity associated with incorporating external features from that associated with reproducing them in subsequent products.

\noindent\textbf{Keywords:} anchored exploration; product phylogeny; technological recombination; horizontal transfer; design continuity; semiconductor products.

\noindent\textbf{JEL codes:} O31, O33, L23, L63.

\clearpage
\section{Introduction}
\label{sec:introduction}

Technological evolution combines continuity with the incorporation of elements developed elsewhere. Products inherit configurations that make their histories recognisable, while new combinations extend what those configurations can do. The resulting question is not simply how much a product changes, but how its accumulated design history relates to the external functions it incorporates and carries into subsequent products. We develop a product phylogenetic network framework to examine this relationship through the concept of anchored exploration.

Evolutionary accounts of technological change emphasise the cumulative organisation of search. Routines preserve experience, and technological trajectories direct problem solving towards developments connected to earlier solutions \citep{nelsonwinter1982,dosi1982}. Exploration and exploitation, in turn, distinguish the extension of an established repertoire from its refinement \citep{march1991exploration}. Recombination connects these processes: a function that is well established in one technical setting can be new to another. Research on preadaptation and evolutionary chimeras shows why the history of an element and the history of its receiving configuration must be considered together \citep{cattani2005preadaptation,cattani2006speciation,carignani2019chimeras}.

Anchored exploration brings the receiving product's history to the centre of this account. An inherited configuration can provide a familiar setting for incorporating external features, yet strong continuity can also orient development towards existing functions and interfaces. Moreover, incorporating a feature and reproducing it are different achievements. A function may appear in one product variant, recur across many variants within the same family, or become part of several later design families. The continuity associated with its first appearance need not be associated in the same way with either form of subsequent reproduction. This distinction connects the cumulative character of technological trajectories with the recombinant character of product development.

Making this argument observable requires a representation of both overall design continuity and the histories of particular functions. Product phylogenies organise dated products into histories of descent with modification \citep{jeong2024where,lee2024mapping,park2025phylogenetic}. A single-parent tree, however, can assign a product to its closest overall antecedent while leaving an externally precedented feature unrepresented. Conversely, following only a feature's earlier carriers leaves the receiving design's accumulated history unspecified. We combine these perspectives in a two-layer network: a primary-continuity backbone for design ancestry and feature-labelled horizontal links for additions with precedents outside that ancestry.

Our research question is: \emph{How does accumulated design continuity relate to the occurrence of externally precedented feature additions and their subsequent reproduction within and across product families?} We operationalise continuity as \emph{Anchor}, a pre-focal measure combining historical similarity with accumulated generations. We operationalise \emph{horizontal transfer} (HT) as the first recorded appearance of a feature in a receiving history when an earlier carrier exists outside that history. We then follow the focal feature into subsequent stock-keeping units (SKUs) within the receiving family and into later carrier families among its backbone descendants. These outcomes distinguish the occurrence, local reproduction and lineage reproduction of a recombinant feature.

The empirical setting comprises 102,465 semiconductor product records grouped into 45,996 design families. The network preserves actual SKU configurations: it compares families using products observed at their first recorded dates, rather than synthetic combinations of features drawn from different variants or later years. A separate event history retains the dates of subsequent SKU observations. The resulting framework links a receiving design's prior history, the identity of an external addition and the later products that carry it. Three regression models use the subsets with the historical and feature evidence required for each outcome.

The fitted relationships differ across these three dimensions. HT occurrence is hump-shaped in Anchor, reaching its fitted maximum at an intermediate level of accumulated continuity. Conditional on a recorded addition, within-family carriage increases throughout its observed Anchor range. Descendant carriage is also hump-shaped, with a fitted maximum at an intermediate Anchor level. The product-level pattern is therefore more specific than a general claim that either continuity or novelty is beneficial: the fitted configuration associated with more new additions differs from that associated with more local variants carrying an addition, and local reproduction differs again from reproduction across descendant families.

The study makes three connected contributions. First, it develops a phylogenetic network framework for evaluating anchored exploration at the product level, defining Anchor and HT on a common, temporally ordered technical history. Second, it joins the study of recombination to the subsequent reproduction of the recombined feature, distinguishing later variants of a receiving family from later families in its lineage. Third, its semiconductor application characterises outcome-specific nonlinear relationships between continuity and recombinant development. These relationships motivate an evolutionary interpretation in which inherited configurations provide both a context for extending a product's repertoire and a basis for carrying selected additions forward.

Section~\ref{sec:theoretical_background} develops anchored exploration and its product-level implications. Section~\ref{sec:methods} describes the data, network construction and regression models. Section~\ref{sec:results} presents the network and the three fitted relationships. Section~\ref{sec:robustness} reports reconstruction checks. Section~\ref{sec:discussion} discusses implications for technological evolution and the study's limitations, and Section~\ref{sec:conclusion} concludes.

\section{Anchored exploration and product evolution}
\label{sec:theoretical_background}

\subsection{Cumulative trajectories and recombinant novelty}
\label{subsec:exploration_exploitation}

Products evolve along multiple dimensions. A design can improve the speed, capacity or efficiency of an inherited function, extend its functional repertoire, or reproduce an existing combination in a different configuration. The characteristics approach to technological change makes these dimensions explicit by treating products as bundles of technical attributes and services \citep{saviotti1984theoretical}. Product-growth and trajectory perspectives likewise connect observed development to the historical organisation of technical possibilities \citep{tellis1981evolutionary,dosi1982}. Continuity and change are consequently properties of a sequence of configurations, rather than opposite classifications assigned to an isolated product.

Recombination introduces a second history into that sequence. An element can have an established use elsewhere before appearing in the receiving design. Preadaptation explains why an existing development can acquire a new role, while the evolutionary-chimera perspective emphasises the combination of elements with different antecedents \citep{cattani2005preadaptation,cattani2006speciation,carignani2019chimeras}. The novelty of the resulting product is relational: it depends on what was previously present in the receiving history, not only on whether the added element is itself new to the world.

This distinction also connects product histories to exploration and exploitation. Refining a recorded functional repertoire corresponds to continuity in the possibilities embodied in a design; introducing a previously unrecorded external function extends that repertoire. A sequence of released products thus provides a technical counterpart to studies of organisational search and new-product introduction \citep{march1991exploration,katila2002something,lee2024mapping}. It permits the observation of what becomes embodied in products while keeping the product, rather than the organisation, as the unit through which technical continuity is reconstructed.

We use \emph{lineage} for the temporally ordered ancestry of a product family on the primary-continuity backbone. An addition is external to that lineage when an earlier carrier is located outside its prior ancestry. Technical and organisational boundaries can therefore differ. Another lineage within the same manufacturer can provide an external precedent, while a history of closely related designs can cross manufacturer labels. This definition locates novelty relative to the receiving design and makes it possible to distinguish the evolution of technical configurations from the ownership of the organisations producing them.

\subsection{Anchors as the receiving context of exploration}
\label{subsec:anchored_exploration}

\emph{Anchored exploration} describes the extension of a product's repertoire through external elements in the context of an inherited configuration. Conceptually, an anchor combines a persistent functional core with experience in developing the relationships around that core. Architectural knowledge concerns relationships among components, while knowledge crossing a boundary may require translation and transformation to fit a new setting \citep{henderson1990architectural,carlile2004transferring}. A receiving design is therefore more than an empty location into which an available function can be inserted. Its accumulated development shapes which combinations are technically familiar and which require reconfiguration.

The argument has two sides. At low continuity, repeated changes in the receiving configuration may leave fewer established interfaces and complementary arrangements around which an external feature can be developed. Some continuity can provide a basis for combining the new function with existing ones. At high continuity, however, development may be increasingly organised around an established repertoire. The configuration that makes a product reproducible can also make substantial additions less attractive or more demanding. These countervailing considerations motivate a possible non-monotonic relationship between anchoring and the occurrence of external additions.

The situation changes after an addition has appeared. A receiving family can carry the feature into further variants while reusing much of its existing configuration. In this setting, the accumulated continuity of the receiving history may be associated with a larger number of later products carrying the feature. Reproduction across descendant families poses a different question because the focal function is then observed in further design-family configurations. Continuity may provide a common basis for that reproduction, while increasing attachment to a particular configuration may limit the range of later families in which the addition recurs. The balance can therefore differ between reproduction within a family and reproduction across its lineage.

This perspective complements absorptive-capacity reasoning. Prior related knowledge provides a basis for recognising and using external knowledge \citep{cohenlevinthal1990}; anchored exploration specifies the technical receiving configuration and the product histories through which an addition is reproduced. Its distinctive empirical object is the combination of an inherited design history, an externally precedented feature and that same feature's subsequent carriers. The framework directs attention to the product-level manifestations of cumulative learning without equating a product's similarity score with an organisational capability.

Consider, for illustration, a family whose earlier designs record an established set of interfaces. A later variant adds another named interface with earlier carriers elsewhere. One question is whether the addition appears at all; another is how many subsequent variants of that family carry it; a third is how many later descendant families record it. These are three observations about the development of the same focal function. An abundant set of variants within the initial family need not coincide with extensive reproduction across later families. Anchored exploration is consequently evaluated through these separate relationships rather than through a single count of novel products.

\subsection{A phylogenetic network for three dimensions of recombinant development}
\label{subsec:phylogenetic_networks}
\label{sec:phylogeny_background}

Product phylogenies use dated characteristics to reconstruct technical ancestry and branching \citep{jeong2024where,lee2024mapping}. Related network approaches connect trajectories, product references and diversity dynamics \citep{jeong2025mapping,park2025phylogenetic}. The biological analogy offers a way to distinguish continuity from recombination while keeping technological evolution grounded in its own empirical record \citep{wagner2014spaces}. Here a primary-continuity backbone and feature-labelled HT links perform different analytical tasks.

Suppose a focal product resembles configuration A overall but includes a function previously recorded in B. The backbone locates A as its primary antecedent; the feature layer records B as a candidate external precedent. Selecting A does not erase the feature's earlier history, and recording B does not replace the receiving design's principal continuity. Anchor is calculated on the prior backbone, while HT is identified by comparing the focal repertoire with prior ancestry and outside-path carriers. This separation allows a product to be highly continuous overall while incorporating a function with a different history.

The temporal sequence is
\begin{equation}
\text{prior Anchor}_{t^-}\ \longrightarrow\
\text{HT occurrence}_{t}\ \longrightarrow\
\text{feature reproduction}_{t:t+W}.
\label{eq:anchored_sequence}
\end{equation}
The sequence organises the measurement of the receiving history, the addition and its follow-up. We operationalise \emph{HT occurrence} as first recorded appearance under an external-precedent rule. We operationalise \emph{within-family reproduction} as the number of later SKUs in the receiving family that record the focal feature, and \emph{lineage reproduction} as the number of later carrier families among backbone descendants. We use \emph{carriage} for these recorded instances of feature reproduction.

The distinction extends novelty-profile approaches beyond the initial configuration. Conventional and unconventional combinations can characterise the novelty of an invention \citep{uzzi2013atypical,kim2016technological}; following the focal feature shows how a combination becomes represented in subsequent products. The count outcomes measure the extent of recorded reproduction, combining the number of later products with the occurrence of the focal feature in those products. They are quantities of carriers, rather than retention fractions or sales-based diffusion measures.

The three dimensions motivate separate nonlinear regression models. Quadratic Anchor terms allow a relationship to turn within the observed range or to remain increasing or decreasing over that range. The central comparison is therefore between the fitted relationships for occurrence, within-family reproduction and lineage reproduction. A common inverse U is not imposed: the receiving history can relate differently to entering a new combination, reproducing it locally and carrying its elements into further families.

\section{Data and methods}\label{sec:methods}

\subsection{The semiconductor catalogue and design families}\label{sec:data_catalogue}
The catalogue combines marketed semiconductor part numbers from distributor records and manufacturer specifications. Each record contains a manufacturer label, product subcategory, recorded year, and published characteristics. The inventory comprises 102,465 stock-keeping units (SKUs), grouped into 45,996 design families, with 135 original manufacturer labels and 19 subcategories. Finite recorded years span 1976--2024; 45,464 families are dated and 532 are undated. Recorded dates supply the temporal ordering used in the network and feature-event histories.

A family is the analytical node, while its constituent SKUs supply the evidence about product configurations. Part numbers are normalised within manufacturer, and adjacent names are joined under a declared common-prefix and option-suffix rule. Accepted joins are transitive. This retains stable family identities while avoiding a separate genealogical node for every orderable variant. The resulting families are retrospective, name-based analytical groupings. There are 23,147 single-listing and 22,849 multiple-listing families; 10,618 span recorded years.

The distinction between a family and its members is consequential. A union of all features across all members can represent a configuration that no SKU has, while later members can introduce characteristics absent at the family's earliest date. We therefore use two clocks. The family birth profile contains only SKUs dated to its earliest recorded year. All later SKU records retain their actual dates in a separate capability-event history. The birth graph uses 78,139 launch-year SKUs; 23,550 later and 776 undated listings do not enter birth profiles. Later records can establish subsequent carriage of a capability without retrospectively changing the earlier family node.

\begin{figure}[tbp]\centering
\includegraphics[width=\linewidth,height=.54\textheight,keepaspectratio]{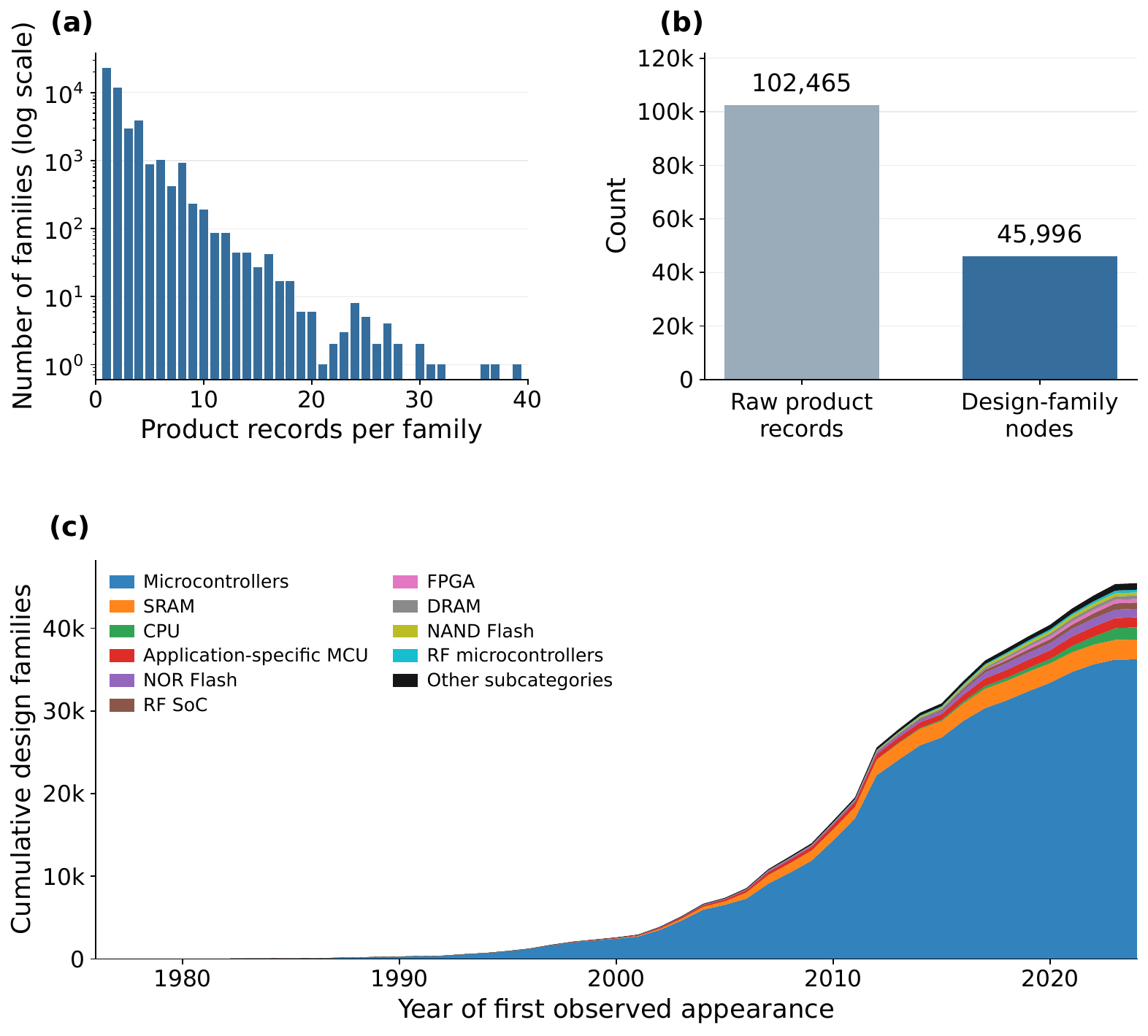}
\caption{\textbf{Coverage and composition of the semiconductor catalogue.} (a) SKU membership per design family. (b) Counts of catalogue records and design families. (c) Cumulative design families by earliest recorded year, separating the ten largest subcategories and pooling the remainder as Other. This panel uses the 45,464 dated families; 532 undated families are excluded. Years refer to catalogue observations.}\label{fig:coverage}
\alttxt{Three panels describe catalogue coverage. Most design families contain few product records. The inventory contains 102,465 records and 45,996 families. Cumulative dated families increase over 1976--2024 and are dominated by the broad Microcontrollers category.}
\end{figure}

Figure~\ref{fig:coverage} describes the family inventory. Table~\ref{tab:chip_subcategories} reports subcategory coverage and the availability of named launch capabilities. Microcontrollers account for approximately 79.6\% of families, but the prepared fields for that broad category provide no named launch-interface profile. Missing capability evidence in this and other categories is not coded as absence of HT.

Supplementary Appendix A provides the complete retained specification-field dictionary and traceable part-number, SKU and family examples. Supplementary Appendix B displays the corresponding networks for every subcategory.

\begin{table}[tbp]\centering
\begin{threeparttable}
\caption{Semiconductor subcategories and observed launch capabilities}\label{tab:chip_subcategories}
\small\doublespacing\setlength{\tabcolsep}{5pt}
\begin{tabularx}{\linewidth}{@{}Xrr@{}}\toprule
Subcategory & Families & Named launch profile\\\midrule
Microcontrollers & 36,631 & 0\\
SRAM & 2,389 & 4\\
CPU & 1,502 & 0\\
Application-specific microcontrollers & 1,291 & 874\\
NOR Flash & 1,030 & 763\\
RF SoC & 773 & 520\\
FPGA & 423 & 0\\
DRAM & 420 & 0\\
NAND Flash & 399 & 43\\
RF microcontrollers & 332 & 195\\
ARM microcontrollers & 233 & 114\\
16-bit microcontrollers & 137 & 104\\
8-bit microcontrollers & 136 & 38\\
Microprocessors & 84 & 37\\
EEPROM & 76 & 20\\
32-bit microcontrollers & 74 & 28\\
DSP / DSC & 47 & 21\\
Application-specialized processors & 17 & 13\\
CPLD & 2 & 0\\\midrule
Total families & 45,996 & \\\bottomrule
\end{tabularx}
\begin{tablenotes}\footnotesize\doublespacing
\item A named profile requires positive parsed launch-SKU evidence for a named capability. Zero indicates no such recorded profile, not a verified absence of communication functions. Categories follow the prepared catalogue assignments. These counts describe birth-profile observation and are not the three regression samples.
\end{tablenotes}
\end{threeparttable}\end{table}

\subsection{Specification features and capability values}\label{sec:data_functions}
The similarity calculation retains 25 numerical and 11 categorical fields from the prepared vocabulary. Numerical fields include capacities, frequencies and current ratings; categorical fields include recorded technology, interface and other specification attributes. Values remain attached to individual SKUs. Categorical multi-value fields become sets after declared lexical normalisation, while missing numerical values and empty sets remain unobserved. Similarity can use a broader specification vocabulary than the capability-event definition.

For the latter, the lexical parser identifies 24 broad named capability types, including I2C, SPI, USB, CAN and UART/USART, with variant evidence retained separately. Named evidence occurs in 5,550 SKUs, approximately 5.42\% of the catalogue. These labels identify recorded functional capabilities. Generic serial/parallel descriptions, package attributes and unresolved text are excluded from capability introductions. Missing entries and zero-filled flags remain unobserved rather than being coded as capability absence.

Keeping member configurations matters even within named families. In an illustrative ATSAMR30E18A-MU grouping, one SKU records I2S and USB, while another records I2C, SPI, UART/USART and USB. The family union contains five broad types, but neither observed member records all five. A union is appropriate for asking whether a capability has been recorded anywhere in prior ancestry; it is inappropriate as a synthetic SKU for measuring design similarity.

\subsection{The continuity backbone}\label{sec:methods_network}
For dated family $i$, possible predecessors satisfy $t_j<t_i$ and $c_j=c_i$, where $t$ is recorded birth year and $c$ is subcategory. Candidate selection is within subcategory and allows different manufacturer labels. We select the earlier family whose launch-SKU distribution is most similar to the recipient's distribution.

Feature weights depend only on families in the recipient's subcategory dated strictly before $t_i$. Each observed family contributes equal historical mass for a feature, divided among its observed launch SKUs; categorical mass is further divided among recorded tokens. Numerical states use five equal-width bins over the strictly historical range. With $p_{f\ell}(c_i,t_i)$ denoting the resulting state probabilities, the feature weight is Shannon entropy \citep{shannon1948}:
\begin{equation}\label{eq:entropy}
H_f(c_i,t_i)=-\sum_{\ell:p_{f\ell}>0}p_{f\ell}\log_2p_{f\ell}.
\end{equation}
A field requires at least two prior observed families. A historically concentrated field receives less weight, while a zero-entropy field receives none. Entropy thus weights a field by its ability to distinguish historically observed configurations.

For parent SKU $a$ and child SKU $b$, a numerical field contributes
\begin{equation}
s_f(a,b;t_i)=1-\min\left\{1,\frac{|x_{af}-x_{bf}|}{R_f(c_i,t_i)}\right\},
\end{equation}
where $R_f$ is the strictly historical range. Categorical sets contribute their Jaccard similarity. The mixed-data similarity follows the Gower comparison principle \citep{gower1971}, with historical entropy and set-valued extensions:
\begin{equation}\label{eq:sku-gower}
s_{ab}(t_i)=\frac{\sum_f I_{abf}H_f(c_i,t_i)s_f(a,b;t_i)}{\sum_f I_{abf}H_f(c_i,t_i)}.
\end{equation}
Here $I_{abf}$ requires observations at both endpoints and adequate historical support. At least two comparable positive-entropy fields are required. Inadequate information yields an unsupported comparison, not a zero distance.

Family similarity preserves the actual joint SKU configurations through balanced optimal transport \citep{peyre2019transport}. Assigning each family's launch SKUs equal mass, we calculate
\begin{align}\label{eq:ot}
S_{ji}(t_i)&=\max_{\Pi}\sum_{a\in\mathcal S_j^0}\sum_{b\in\mathcal S_i^0}\Pi_{ab}s_{ab}(t_i),\\
\sum_b\Pi_{ab}&=\alpha_a,\qquad \sum_a\Pi_{ab}=\beta_b,\qquad \Pi_{ab}\geq0.\nonumber
\end{align}
Unsupported SKU pairs carry no mass. All mass in both families must be matched; an infeasible full coupling leaves the family comparison undefined. This prevents poorly observed members from being dropped merely to improve the match. The solution uses no entropic regularization.

The selected parent maximises positive finite $S_{ji}$. Ties are resolved by feature-weight coverage, comparable-field support, more recent parent year and a stable identifier. All earlier same-subcategory candidates are considered, using valid transport upper bounds to avoid unnecessary optimisation. One predecessor and strictly increasing dates produce a directed forest; undated or unsupported families remain roots. The resulting backbone represents inferred technical ancestry from dated product characteristics.

\subsection{Horizontal transfer and dated feature events}\label{sec:methods_donor}
The backbone represents primary continuity, while horizontal transfer (HT) identifies an externally precedented addition to the receiving repertoire. HT is an operational, feature-level link rather than documentary evidence of an actual technology transaction. For named capabilities, direct-parent carriage is recorded inheritance; an addition relative to the parent that already appears in an older ancestor is recorded reintroduction. Let $M_i^0$ denote the union of named capabilities in family $i$'s launch SKUs and $\mathcal A_i$ its complete prior backbone path. A launch capability with no recorded carrier in $\mathcal A_i$ is eligible for an external precedent if a strictly earlier launch carrier exists outside that path and has a supported positive similarity to the recipient. The best-supported candidate is selected under the declared similarity ranking. Its actual supporting SKU and year are retained.

Externality in this birth graph is genealogical: a different lineage within the same manufacturer is eligible. The regression risk set below imposes the additional requirement that an earlier outside-path carrier belong to another recorded manufacturer. Features without adequate ancestral evidence or a prior carrier remain separately classified. Numerical changes on same-manufacturer backbone links are descriptive vertical-change attributes; a higher numerical value is not automatically better performance.

For later developments, let $\tau_{im}$ be the first dated SKU record of capability $m$ in family $i$. An event with $\tau_{im}>t_i$ remains a later SKU event rather than an attribute of family birth. There are 6,783 first family--capability records, including 127 after-launch first records in 56 families. Of these later records, 113 occur in families without named launch evidence. The event date is the first positive record of the focal feature under the applicable observation rules.

\subsection{Historical paths and their interpretation}\label{sec:validation}
Main path analysis identifies structurally prominent histories in the reconstructed network. For an edge $u\to v$ in the acyclic graph, the search-path-count weight is $L(u)R(v)$, where $L(u)$ counts source-to-$u$ routes and $R(v)$ counts $v$-to-sink routes \citep{batagelj2003,liulu2012}. The global main path maximises the sum of edge weights. Equally optimal routes are retained, together with every complete root-to-leaf path. An isolate contributes one zero-edge path.

We distinguish paths in the continuity backbone from paths in the combined backbone--HT graph. Only the unique backbone ancestry defines the prior anchor and descendant sets used below. Adding donor routes expands the set of network paths. The regressions retain the original family--capability events as their observations, independently of how often a family occurs on overlapping paths. Documentary comparisons and alternative reconstructions assess the plausibility and stability of the inferred histories in Section~\ref{sec:robustness}.

\subsection{Anchor and three regression outcomes}\label{sec:methods_stats}\label{sec:stage_methods}
\subsubsection{Prior continuity and accumulated anchoring}\label{sec:line_anchor}
Let $v_0,\ldots,v_d=i$ be the backbone path to family $i$. Its strictly prior edges end at $v_{d-1}$ and exclude the focal incoming edge $v_{d-1}\to i$. With $G_i^-$ complete prior edges, define
\begin{equation}\label{eq:updated_preanchor}
A_i^-=\frac{1}{G_i^-}\sum_{r=1}^{d-1}S_{v_{r-1},v_r}(t_{v_r}),
\qquad C_i^-=A_i^-\log(1+G_i^-).
\end{equation}
The mean $A_i^-$ measures prior continuity; $C_i^-$ combines that continuity with a diminishing contribution from accumulated generations. The resulting Anchor can exceed one. Each edge retains its original historical feature weights. The analysis requires at least three complete prior edges and freezes Anchor at family birth, including for later SKU events. Excluding the focal incoming link keeps the receiving history temporally prior to the focal configuration. The measure operationalises the accumulated continuity component of anchored exploration.

\begin{table}[tbp]\centering\begin{threeparttable}
\caption{Variables for Anchor, HT occurrence and feature reproduction}\label{tab:staged_variables}
\small\doublespacing\begin{tabularx}{\linewidth}{@{}>{\raggedright\arraybackslash}p{.23\linewidth}X@{}}\toprule
Variable & Definition and timing\\\midrule
Prior continuity, $A_i^-$ & Mean launch-SKU transport similarity over backbone links before the focal incoming link.\\
Accumulated Anchor\newline $C_i^-$ & $A_i^-\log(1+G_i^-)$; at least three complete prior generations; frozen at family birth.\\
Recorded HT, $T_{im,t}$ & First focal-capability mention in an eligible family--capability--release-year opportunity with a strictly earlier cross-manufacturer outside-path carrier.\\
Within-family carriage\newline $Y^W(5)$ & Unique later focal-capability-positive SKUs in the receiving family within five years of the first record.\\
Descendant carriage\newline $Y^D(5)$ & Later recorded focal-capability-positive backbone descendant families within five years; not the count of all descendants.\\
Common controls & Prior generations; prior outside-path carrier-family count; release/event year; capability, subcategory and manufacturer fixed effects.\\
Detection-only control & Log current-release SKU count.\\\bottomrule
\end{tabularx}
\begin{tablenotes}\footnotesize\doublespacing\item Observation and horizon restrictions differ across stages. A recorded carrier is not necessarily a newly adopting product. Missing profiles are not negative outcomes. Continuous predictors are standardised as described in the text; the dependent variables retain binary or count units.\end{tablenotes}
\end{threeparttable}\end{table}

\subsubsection{Three conditional regression models}\label{sec:main_regression_controls}
The analysis separates first recorded appearance under the external-precedent rule from subsequent recorded carriage. The first sample comprises family--capability--actual-release-year opportunities with strictly prior outside-path availability, including a carrier from another manufacturer, and no earlier own-family or ancestral record of that capability. It compares first focal mentions with nonmentions in otherwise clean recorded profiles under symmetric current and ancestral evidence requirements. Completely missing evidence does not enter as nonadoption. This binary outcome is termed recorded HT occurrence.

The second and third samples condition on a recorded externally precedented first-introduction candidate, a five-year horizon and the required later named-profile observation. They count subsequent carrier SKUs within the family and subsequent carrier families among backbone descendants, respectively (Table~\ref{tab:staged_variables}). Already established carriers can contribute later carriage in the selected descendant outcome. A zero denotes no recorded carriers in the admitted window. The follow-up models use the respective samples meeting their event and subsequent-observation requirements.

For stage $s\in\{T,W,D\}$, $z_s(C)=(C-\overline C_s)/\sigma_s$ standardises the accumulated anchor using that stage's eligible sample before separation. The quadratic term is the square of this standardised variable. The three models are
\begin{align}
\operatorname{logit}p(C,X_T)&=\alpha_T+\beta_Tz_T(C)+\kappa_Tz_T(C)^2+X_T'\gamma_T,\label{eq:staged-detection}\\
\log\mu_W(C,X_W)&=\alpha_W+\beta_Wz_W(C)+\kappa_Wz_W(C)^2+X_W'\gamma_W,\label{eq:staged-within}\\
\log\mu_D(C,X_D)&=\alpha_D+\beta_Dz_D(C)+\kappa_Dz_D(C)^2+X_D'\gamma_D.\label{eq:staged-descendant}
\end{align}
Here $p$ is the detection probability, while $\mu_W$ and $\mu_D$ are conditional expected counts among the respective observed follow-up populations. The count models use Poisson pseudo-maximum likelihood \citep{santos2006gravity}. $X_s$ includes standardised $\log(1+G_i^-)$, standardised $\log(1+\text{prior outside-path carriers})$, standardised linear year, and capability, subcategory and manufacturer fixed effects. Detection additionally controls for log current-release SKU count. The count models use no exposure offset because the outcomes are numbers of recorded carriers.

These controls distinguish prior lineage depth, external capability availability, temporal trends, group differences and, in the detection model, current recording opportunity. Because $C_i^-=A_i^-\log(1+G_i^-)$, controlling for generations makes the Anchor association conditional on lineage length: varying $C_i^-$ at fixed $G_i^-$ varies prior similarity. Table~\ref{tab:stage_regressions} displays the numerical-control coefficients as well as the Anchor terms; fixed effects are indicated separately.

For each outcome, the sign of the slope is determined by $\beta_s+2\kappa_s z_s(C)$ because the inverse-logit and exponential links are increasing. We describe fitted turning points on the raw Anchor scale and locate them relative to the observed support. A negative quadratic coefficient alone does not establish a turning point within that support. The directional comparisons evaluate the HT slope at the fitted 95th Anchor percentile and the follow-up slopes at their fitted 5th percentiles. These one-sided endpoint comparisons are distinct from the two-sided coefficient tests in Table~\ref{tab:stage_regressions}.

Coefficient tables report root-cluster standard errors and unadjusted tests. Figure~\ref{fig:staged-curves} displays family-cluster pointwise intervals. Supplementary Appendix D reports the alternative dependence units, the specification-selection record and the full inferential diagnostics; their implications are summarised in Section~\ref{sec:limitations}.

\subsubsection{Linking occurrence and reproduction}\label{sec:joint-scenarios}
Supplementary Appendix C combines fitted occurrence probabilities with conditional carriage counts on fixed target profiles. The calculation separates the expected number of carriers following an observed addition from the probability-weighted number across introduction opportunities. It multiplies predictions within each profile and then averages over a fixed target population. Because the component models use different conditional samples, these calculations are presented as model-based scenarios with explicit support restrictions.

\FloatBarrier

\section{Results}\label{sec:results}
\subsection{The semiconductor phylogenetic network}\label{sec:pivot_results}
The reconstructed backbone contains 40,713 edges and 5,283 roots among 45,996 families. Candidate launch HT comprises 1,462 capability events on 959 distinct donor--recipient arcs, including 298 cross-subcategory arcs; 643 capability events connect families with the same manufacturer label. Several capabilities can share an arc, and several arcs can reach a recipient. Combining the two layers yields 41,672 distinct directed edges. The network thus represents a product family's overall continuity together with external precedents for particular functions.

\begin{figure}[tbp]\centering
\includegraphics[width=.98\linewidth,height=.54\textheight,keepaspectratio]{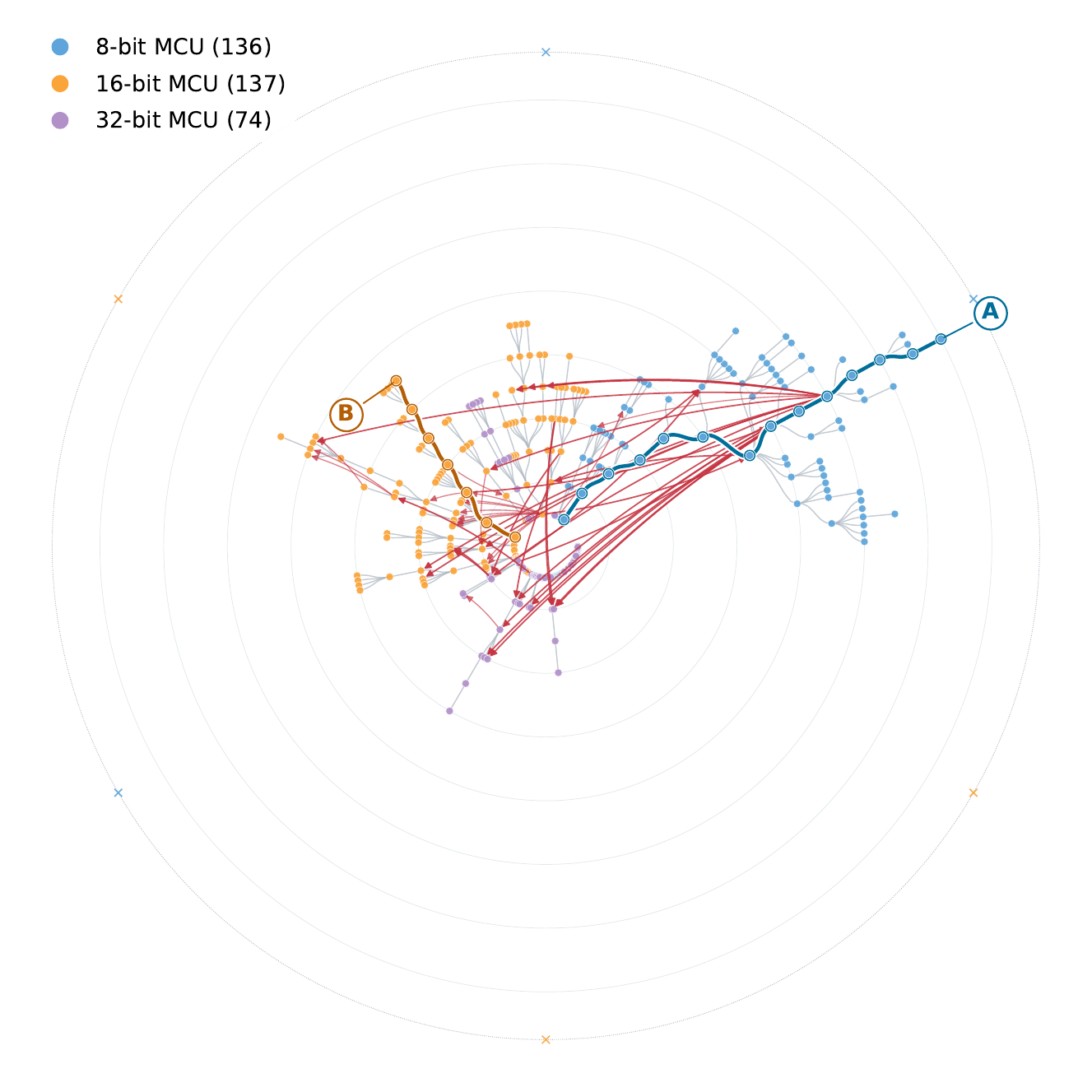}
\caption{\textbf{8-bit, 16-bit and 32-bit MCU families in a shared network.} The upper-left legend identifies node colours. Grey curves represent primary continuity; red arrows represent candidate horizontal precedents, including 29 cross-subcategory links. Emphasized paths A (8-bit MCU) and B (16-bit MCU) are expanded in Figure~\ref{fig:combined-path}. Radius indicates backbone generation, not elapsed time; crosses on the outer dotted ring mark six undated families. The two link types distinguish overall design continuity from feature-specific external precedents.}\label{fig:network_category}
\alttxt{A radial network displays 347 families in three MCU categories. Curved backbone links organise inferred continuity, while directed external-precedent links cross some branches and category groups. Highlighted 8-bit and 16-bit paths are expanded in the next figure. Distance from the centre indicates generation, not calendar time.}
\end{figure}

\begin{figure}[tbp]\centering
\includegraphics[width=\linewidth,height=.54\textheight,keepaspectratio]{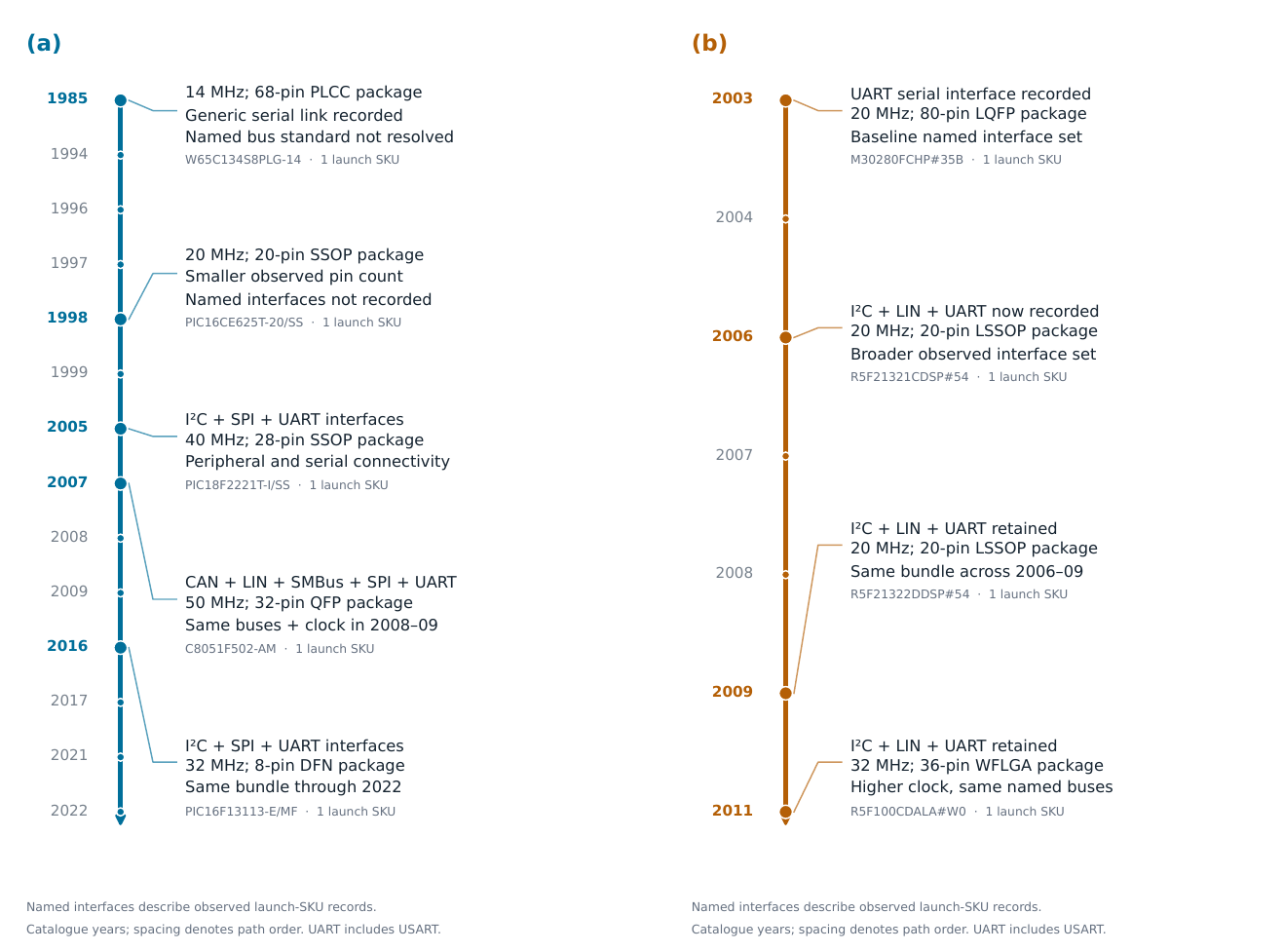}
\caption{\textbf{Interpreting the highlighted MCU histories.} The two panels expand paths A (8-bit MCU) and B (16-bit MCU) from Figure~\ref{fig:network_category}. All path nodes and links are retained; selected checkpoints explain recorded launch-SKU clock ratings and communication interfaces. Vertical position follows path order rather than proportional elapsed time. Years are catalogue observations. The annotations distinguish continuing interface bundles from changes in clock ratings and packages.}\label{fig:combined-path}
\alttxt{Two vertical paths annotate recorded MCU histories. The 8-bit path changes interface bundles and includes a later clock decrease from 50 to 32 MHz. The 16-bit path retains I2C, LIN and UART at later checkpoints while the recorded clock increases from 20 to 32 MHz. Spacing indicates path order.}
\end{figure}

Figure~\ref{fig:network_category} places 8-bit, 16-bit and 32-bit microcontrollers in one coordinate space, with node colours distinguishing the three catalogue subcategories. The view includes all 347 families in those categories (136, 137 and 74, respectively), 227 backbone links and 47 candidate HT links. Of the latter, 29 connect different displayed subcategories and 18 connect families within one subcategory. The highlighted 8-bit and 16-bit paths are expanded and annotated in Figure~\ref{fig:combined-path}. The radial layout shows backbone generation rather than calendar time and introduces no synthetic common ancestor. Word-width categories follow the catalogue assignments. Their separation in the backbone and connections through HT show how category-specific continuity and cross-category feature precedents can be represented together.

Mean selected parent--recipient similarity is 0.983. Supplementary Appendix D reports similarity distributions, comparison support and alternative parent assignments.

The complete backbone contains 37,401 root-to-leaf paths; including candidate horizontal links yields 85,845 complete routes. There are 17 tied backbone main paths and 11 tied combined-network main paths at the whole-network level. Figure~\ref{fig:combined-path} follows two saved subcategory-specific backbone main paths highlighted in Figure~\ref{fig:network_category}: 14 families in 8-bit MCU (1985--2022) and seven in 16-bit MCU (2003--2011). Their launch-SKU annotations describe recorded clock and communication capabilities, making the inferred continuity interpretable beyond part identifiers. These routes connect the network representation to the recorded functions and specifications of individual products.

The 8-bit case includes a CAN, LIN, SMBus, SPI and UART interface bundle with a 50\,MHz clock, followed later on the path by I2C, SPI and UART with a 32\,MHz clock. The 16-bit case records an I2C, LIN and UART bundle from 2006 through the 2011 checkpoint, while the observed clock changes from 20 to 32\,MHz and the package changes. Read at the product level, the first case combines changes in the functional repertoire with changes in a performance characteristic; the second retains a named interface bundle across changing specifications. These histories illustrate why functional continuity, performance improvement and repertoire extension should be followed separately. Their annotations describe the product configurations; HT classification additionally uses the ancestry and external-precedent rules.

Candidate HT recipients often appear late along completed histories, but the result depends on the counting unit. The recipient-adjusted mean position is 92.05\% in application-specific MCUs and 94.03\% in RF SoC. In the former category, weighting every repeated path occurrence instead gives 63.17\%. Catalogue endpoints and shared paths can generate such differences. Likewise, length--mean-similarity correlations are negative in Microcontrollers and RF SoC but positive in NOR Flash, EEPROM and SRAM. The coexistence of positive and negative associations across categories motivates measuring historical similarity and lineage depth separately before combining them in Anchor.
\FloatBarrier

\subsection{Anchor and three dimensions of recombinant development}\label{sec:anchoring_consequences}\label{sec:staged-primary}
\begin{figure}[tbp]\centering
\includegraphics[width=\linewidth]{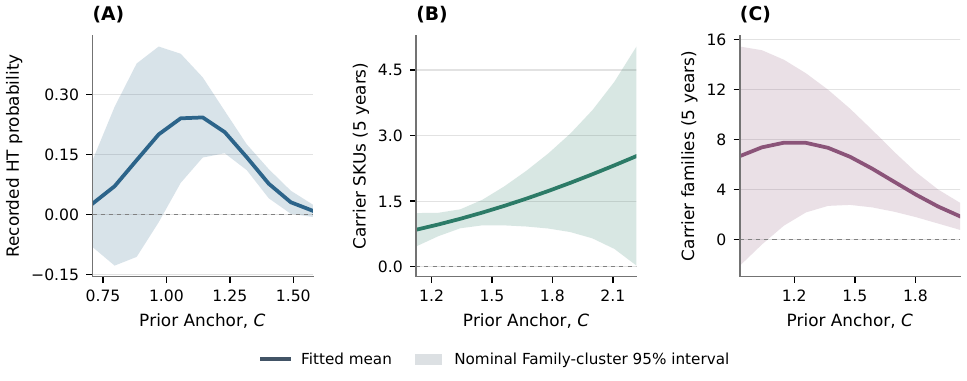}
\caption{\textbf{Anchor, HT occurrence and subsequent feature reproduction.} From left to right: recorded HT probability, later within-family carrier SKUs and later descendant carrier families. Lines show fitted means; shading shows unadjusted family-cluster pointwise 95\% delta intervals. Each panel uses its outcome-specific sample and standardisation. The untransformed delta intervals are local approximations and can extend outside the feasible outcome range.}\label{fig:staged-curves}
\alttxt{The HT-occurrence curve rises and then falls, with a fitted peak near Anchor 1.10. Within-family carriage increases across its displayed range. Descendant carriage rises at the low end and then falls, peaking near Anchor 1.20. Shaded intervals show uncertainty around each fitted curve.}
\end{figure}

Table~\ref{tab:stage_regressions} reports the three regression models. The HT-occurrence model retains 389 opportunities, including 52 positive first mentions, after separation removes 500 of 889 eligible observations. The within-family and descendant models retain 152 and 179 feature events, respectively. Each sample supplies the historical, capability and follow-up evidence required for its outcome. The fitted relationships in Figure~\ref{fig:staged-curves} distinguish the occurrence of an addition from its subsequent carriage at two product scales.

\begingroup\small\doublespacing\setlength{\tabcolsep}{4pt}
\begin{longtable}{@{}>{\raggedright\arraybackslash}p{.36\linewidth}*{3}{>{\centering\arraybackslash}p{.19\linewidth}}@{}}
\caption{Anchor, HT occurrence and subsequent feature carriage}\label{tab:stage_regressions}\\
\toprule & (1) & (2) & (3)\\
Outcome & Recorded HT & Within-family SKUs & Descendant families\\
Estimator & Logit & PPML & PPML\\\midrule
\endfirsthead
\caption[]{Anchor, HT occurrence and subsequent feature carriage (continued)}\\
\toprule & (1) & (2) & (3)\\
Outcome & Recorded HT & Within-family SKUs & Descendant families\\
Estimator & Logit & PPML & PPML\\\midrule
\endhead
\midrule\multicolumn{4}{r}{\emph{Continued on next page}}\\\endfoot
\bottomrule\endlastfoot
Standardized Anchor & $-1.371$ & $0.336$ & $-0.634$\\*
Root-cluster SE & $(0.211)$ & $(0.017)$ & $(0.104)$\\*
Two-sided $p$ & $<0.001$ & $<0.001$ & $<0.001$\\\addlinespace
Anchor squared & $-1.244$ & $-0.022$ & $-0.330$\\*
Root-cluster SE & $(0.284)$ & $(0.011)$ & $(0.047)$\\*
Two-sided $p$ & $0.003$ & $0.126$ & $<0.001$\\\addlinespace
Prior generations, $\log(1+G^-)$ & $1.299$ & $-0.284$ & $0.858$\\*
Root-cluster SE & $(0.625)$ & $(0.037)$ & $(0.118)$\\*
Two-sided $p$ & $0.076$ & $0.001$ & $<0.001$\\\addlinespace
Prior outside-path carriers, $\log(1+N)$ & $-3.796$ & $-0.155$ & $-0.579$\\*
Root-cluster SE & $(2.048)$ & $(0.046)$ & $(0.097)$\\*
Two-sided $p$ & $0.106$ & $0.027$ & $<0.001$\\\addlinespace
Release / event year & $0.218$ & $0.109$ & $0.193$\\*
Root-cluster SE & $(0.193)$ & $(0.019)$ & $(0.096)$\\*
Two-sided $p$ & $0.296$ & $0.004$ & $0.077$\\\addlinespace
Current-release SKU count, log & $-0.098$ & --- & ---\\*
Root-cluster SE & $(0.160)$ & --- & ---\\*
Two-sided $p$ & $0.562$ & --- & ---\\\midrule
Eligible observations & 889 & 153 & 179\\
Removed by separation & 500 & 1 & 0\\
Fitted observations & 389 & 152 & 179\\
Positive outcomes & 52 & 144 & 173\\
Root clusters & 8 & 5 & 10\\
Largest-root share & 57.326\% & 93.421\% & 81.006\%\\
Capability fixed effects & Yes & Yes & Yes\\
Subcategory fixed effects & Yes & Yes & Yes\\
Manufacturer fixed effects & Yes & Yes & Yes\\
\end{longtable}
\noindent\footnotesize\doublespacing Coefficients and standard errors are on the link scale; tests are unadjusted, two-sided and root-cluster based. Anchor is $C=A^-\log(1+G^-)$, standardised within each eligible stage sample. Follow-up is five years. Dashes indicate an excluded predictor. Fixed-effect contrasts are omitted and nonidentified contrasts are aliased. Specification-selection and covariance diagnostics are reported in Supplementary Appendix D and discussed in Section~\ref{sec:limitations}.\par
\endgroup

\subsubsection{HT occurrence: an intermediate-Anchor fitted peak}\label{sec:stage_reach}
The HT model has negative linear and quadratic Anchor coefficients on the standardised logit scale, $-1.371$ and $-1.244$, respectively. In raw Anchor units, the fitted probability increases below $C=1.104$ and decreases above that point. The curve therefore describes a non-monotonic relationship: recorded external additions are most frequent around an intermediate level of accumulated design continuity in the fitted model.

This pattern gives a product-level interpretation to the receiving side of recombination. At low Anchor, the receiving history contains less accumulated continuity around which an additional function might be incorporated. At the high end, development is closely connected to an established configuration, and extensions of the recorded repertoire become less frequent in the fitted relationship. The intermediate peak is consistent with a receiving context that combines continuity with scope for additions. In this interpretation, an external feature extends an inherited design rather than replacing the importance of inheritance.

\subsubsection{Within-family reproduction: increasing carriage over observed support}\label{sec:stage_retention}
The within-family model has a positive linear Anchor coefficient of $0.336$ and a small negative quadratic coefficient of $-0.022$. Its fitted count increases throughout the observed Anchor range. The quadratic's implied turning point, $C=4.089$, lies beyond that range; the observed relationship is therefore increasing rather than an interior inverse U.

The outcome counts later SKUs that record the focal feature within the receiving family. A positive fitted relationship means that, among the observed recipients, a more continuous receiving history is associated with more subsequent variants carrying the addition. From an anchored-exploration perspective, this is the local reproduction side of recombinant development: once an addition appears, the inherited configuration can provide a setting in which it is carried into further variants. The product question has changed from extending the repertoire to reproducing the extended repertoire within a family.

\subsubsection{Lineage reproduction: a hump-shaped descendant relationship}\label{sec:stage_growth}
The descendant model has linear and quadratic Anchor coefficients of $-0.634$ and $-0.330$ on its standardised log scale. The fitted slope is positive at the low-Anchor end and becomes negative after a turning point near raw $C=1.200$. Thus, the fitted number of descendant carrier families is highest at an intermediate Anchor level.

This outcome extends the observation beyond variants of the initial recipient. It follows the focal feature into later families on the primary-continuity backbone. The hump-shaped curve is consistent with a balance between a recognisable technical basis for reproduction and the scope for the function to recur in further configurations. The increase in within-family carriage and the hump-shaped descendant curve distinguish local product variation from reproduction across a lineage. The same amount of prior continuity need not be associated with the greatest extent of both.

\FloatBarrier

\subsection{Reading the three relationships at the product level}\label{sec:stage_comparison}
Table~\ref{tab:product_interpretation} summarises the substantive distinctions. HT occurrence concerns whether the receiving repertoire is extended by an externally precedented feature. Within-family carriage concerns how extensively that addition is represented in later variants. Descendant carriage concerns how extensively it is represented in further families connected through the technical backbone. The fitted non-monotonicity of occurrence and descendant carriage, alongside increasing local carriage, gives these three dimensions different fitted patterns.

\begin{table}[tbp]\centering\begin{threeparttable}
\caption{Three product-level dimensions of anchored exploration}\label{tab:product_interpretation}
\small\doublespacing\setlength{\tabcolsep}{4pt}
\begin{tabularx}{\linewidth}{@{}p{.20\linewidth}p{.29\linewidth}X@{}}\toprule
Dimension & Fitted Anchor relationship & Product-level interpretation\\\midrule
HT occurrence & Hump-shaped; peak near $C=1.10$ & Extension of a receiving repertoire by a feature with an external precedent.\\
Within-family reproduction & Increasing over observed support & Repetition of the focal addition in subsequent variants of its receiving family.\\
Lineage reproduction & Hump-shaped; peak near $C=1.20$ & Carriage of the focal addition in later families on the receiving lineage.\\\bottomrule
\end{tabularx}
\begin{tablenotes}\footnotesize\doublespacing\item Shapes and peak locations summarise the point estimates in Table~\ref{tab:stage_regressions}; they refer to the respective conditional samples.\end{tablenotes}
\end{threeparttable}\end{table}

The interpretation is outcome-specific. In the follow-up samples, an external addition has already been recorded; in the occurrence sample, its appearance is the outcome being modelled. Product-category composition also differs. Application-specific MCUs supply 141 within-family and 145 descendant observations, while RF SoC supplies 223 HT-occurrence observations and five within-family and 14 descendant observations. Supplementary Appendix C reports the complete category coverage and combines the fitted models on explicitly defined target profiles. The primary findings remain the three separate relationships, which distinguish the entry of an addition from its later product representation.
\FloatBarrier

\section{Network reconstruction checks}\label{sec:robustness}\label{app:stage_checks}
\label{sec:backbone_robustness}
We reconstruct parents and HT donors under six settings: the baseline, three or seven entropy bins, three required comparable fields, equal mass for distinct SKU configurations, and exclusion of the two interface fields from similarity ranking. The capability parser remains unchanged. These comparisons examine how the information used to infer continuity and external precedents affects the network.

\begin{table}[tbp]\centering\begin{threeparttable}
\caption{Sensitivity of the reconstructed backbone and horizontal links}\label{tab:robustness}
\small\doublespacing\setlength{\tabcolsep}{3pt}
\begin{tabularx}{\linewidth}{@{}Xrrrr@{}}\toprule
Setting & Links & All-node & Both-linked & HT Jaccard\\\midrule
Baseline & 40,713 & 100.00\% & 100.00\% & 1.000\\
Three entropy bins & 40,713 & 98.84\% & 98.69\% & 0.901\\
Seven entropy bins & 40,713 & 99.28\% & 99.19\% & 0.962\\
Three comparable fields & 40,191 & 62.36\% & 58.22\% & 0.535\\
Distinct-configuration mass & 40,713 & 98.56\% & 98.38\% & 0.943\\
Interface fields excluded & 40,644 & 98.17\% & 98.10\% & 0.282\\\bottomrule
\end{tabularx}
\begin{tablenotes}\footnotesize\doublespacing\item All-node parent agreement includes unchanged roots. Both-linked agreement conditions on a parent being present in both reconstructions. HT Jaccard compares distinct donor--recipient links, not feature-event counts.\end{tablenotes}
\end{threeparttable}\end{table}

Changing the entropy bin count preserves 98.84--99.28\% of parent assignments, and distinct-configuration weighting preserves 98.56\%. Requiring a third comparable field yields 62.36\% parent agreement. Excluding interface fields preserves 98.17\% of parents and yields an HT-link Jaccard of 0.282. Thus, the primary-continuity backbone and the assignment of feature-specific donors respond differently to the available specification information. Table~\ref{tab:robustness} reports both, rather than using aggregate parent agreement as a substitute for feature-link stability.

Supplementary Appendix D provides the full graph and main-path comparisons, including the alternative histories generated by each reconstruction. It also reports the documentary comparisons and date audit. These checks distinguish the broad continuity structure from particular donor assignments and prominent paths, making explicit the measurement choices on which a product-level interpretation depends.
\FloatBarrier

\section{Discussion}\label{sec:discussion}

\subsection{Continuity as a context for recombination}
The framework links two central features of technological evolution: development follows inherited trajectories, yet new functions can arrive from outside those trajectories. In the network, the backbone records the receiving design's history and HT records the feature's external precedent. Anchor makes the continuity of that receiving history measurable before the focal addition. Anchored exploration therefore becomes a product-level question about how accumulated continuity relates to repertoire extension and subsequent reproduction, rather than a contrast between products that are simply continuous or novel.

The fitted HT-occurrence curve gives this question a non-monotonic form. Intermediate anchoring is associated with the highest fitted probability of an externally precedented addition. An evolutionary reading is that inheritance can supply a context for recombination while also concentrating development around an established repertoire. This connects the cumulative search emphasised by trajectory theory with the new uses of existing elements emphasised by preadaptation and evolutionary chimeras \citep{dosi1982,cattani2006speciation,carignani2019chimeras}. The receiving configuration is an active part of the explanatory problem: the availability of a feature elsewhere is only one component of its appearance in a focal product.

This interpretation also refines the relationship between exploration and exploitation. At the product level, exploration can extend a configuration whose earlier development embodies substantial continuity. The empirical task is then to follow how that extension becomes part of further development. Continuity need not imply a complete absence of external additions, and a novel addition need not imply the abandonment of the receiving trajectory. A phylogenetic network accommodates their coexistence by allowing the overall design and its individual features to have different antecedents.

\subsection{From a recombinant product to its later carriers}\label{sec:discussion_stage_results}
The follow-up models add a temporal dimension to the study of technological combinations. The first appearance of an external feature identifies a recombinant event; later carrier products reveal the extent to which the feature is represented in continued development. Studies of novelty profiles focus attention on the configuration at the point of combination \citep{uzzi2013atypical,kim2016technological}. Following carriers complements that perspective by distinguishing a combination's occurrence from its subsequent reproduction.

Within-family and descendant carriage then separate two scales of reproduction. The increasing within-family fitted relationship associates stronger receiving-history continuity with a larger number of later variants carrying the feature. The hump-shaped descendant relationship associates intermediate continuity with the largest fitted number of carrier families. In product terms, reproducing an addition within an established family and carrying it into further family configurations pose different developmental questions. A configuration can offer a basis for local repetition without being associated with the broadest later representation across its lineage.

Anchored exploration provides a way to interpret this distinction. An inherited configuration may help organise the repeated development of a newly extended repertoire, while the scope for further recombination and reconfiguration may matter more as the analysis moves to later families. Architectural knowledge and prior related knowledge offer complementary accounts of why the receiving context could matter \citep{henderson1990architectural,cohenlevinthal1990}. The present product-level measures specify where those accounts can be connected to observable technical histories: before the addition, at its first record and in its later carriers.

\subsection{Implications for studying product and industry evolution}
A product-level view changes how innovation is described. The semiconductor histories show that an interface bundle can persist while clock ratings and packages change, and that a later product on an inferred path can have a lower value of a performance characteristic. Technical development is consequently better represented by following several dimensions together: continuity in the existing repertoire, additions to that repertoire, changes in performance and the reproduction of specific functions. A single measure of product novelty or descendant abundance compresses these distinct developments.

The network also separates technical boundaries from organisational ones. The reconstructed launch-time HT layer includes 643 feature events between families carrying the same manufacturer label. These events are external relative to a receiving lineage even though they remain within a manufacturer label. Conversely, the backbone allows similar designs from different manufacturers to be linked. For industrial analysis, this distinction opens a route to studying how product histories intersect with firms' portfolios, organisational boundaries and opportunities for reuse, while preserving the technical configuration as an explicit unit of observation.

The substantive implication is to distinguish the generation of combinations from the development through which their elements become reproduced. The three fitted relationships describe different aspects of this process; they do not reduce to a single scale of innovative success. Research on technological trajectories can use the framework to follow external additions within otherwise cumulative histories. Research on recombination can use it to follow the receiving configuration and the later carriers of a focal element. Linking these records to engineering documentation, organisational histories and product-use evidence would connect the observed technical sequence to the processes that sustain it.

\subsection{Limitations and future research}\label{sec:limitations}\label{sec:composition_checks}
The measures depend on catalogue coverage and inferred ancestry. Named-feature records are sparse, dates reflect observed records, and retrospective family grouping and unverified units can affect comparisons. Anchor measures accumulated similarity rather than an independently observed invariant core or integration knowledge; HT records an external precedent rather than an authenticated transfer. Alternative histories can change ancestry and descendant sets, and this graph uncertainty is not propagated through every regression. Documentary validation and consistent feature-presence and absence records would strengthen the product-level interpretation.

The regressions estimate conditional associations rather than causal effects. Their samples differ in composition and are concentrated in few, unequal roots; within-family carriage places 142 of 152 events in one root and has rank-one joint Anchor covariance. Precision varies with the clustering unit \citep{cameron2015guide,mackinnon2022cluster}. The reported models were selected from 128 combinations involving 288 fits; the joint directional comparison has Holm-adjusted $p=1.000$. These inferential constraints apply to the fitted patterns, especially the follow-up relationships. Supplementary Appendix D retains the complete selection and covariance diagnostics. Evaluation on additional independent histories under a pre-specified model would provide a stronger test of their generality.

\section{Conclusion}\label{sec:conclusion}
We develop a product phylogenetic network framework for examining anchored exploration. The framework reconstructs primary design continuity, identifies feature-specific external precedents and defines a prior Anchor on the receiving history. It then links HT occurrence to the subsequent carriage of the focal feature within its receiving family and among descendant families. This makes inheritance and recombination observable within the same product history while preserving their distinct roles.

The semiconductor application yields outcome-specific fitted relationships. HT occurrence and descendant carriage are non-monotonic in Anchor, whereas within-family carriage increases over the observed range. Read together, these patterns suggest that the continuity associated with incorporating an external feature can differ from the continuity associated with reproducing it locally or across later families. Anchored exploration thus directs attention both to the inherited context of a new combination and to the product development through which its elements are carried forward.

The broader contribution is to connect technological trajectories with the continuing histories of recombinant features. Product evolution involves not only the appearance of new combinations, but also the reproduction of selected elements in subsequent configurations. A feature-labelled phylogenetic network provides a common empirical framework for studying those developments and their relationship to accumulated design continuity.

\section*{Supplementary material}
Supplementary Appendices A--D provide the specification dictionary and product examples, networks for all 19 subcategories, combined-model scenarios and their support register, and detailed reconstruction and statistical diagnostics.

\section*{Data availability}
The data supporting this study are available from the corresponding author.

\section*{Code availability}
The replication code is available from the corresponding author and will be publicly shared via a Git repository.

\section*{Competing interests}
The authors declare that they have no competing interests.

\clearpage
\appendix
\urlstyle{same}
\setcounter{figure}{0}
\setcounter{table}{0}
\setcounter{equation}{0}
\renewcommand{\thefigure}{S\arabic{figure}}
\renewcommand{\thetable}{S\arabic{table}}
\renewcommand{\theequation}{S\arabic{equation}}
\renewcommand{\theHfigure}{supplement.\arabic{figure}}
\renewcommand{\theHtable}{supplement.\arabic{table}}
\renewcommand{\theHequation}{supplement.\arabic{equation}}
\section*{Appendices}
\noindent These appendices document the product phylogenetic network framework, provide data definitions and product examples, display the complete network gallery, and report probability-weighted carriage scenarios and reconstruction and statistical diagnostics. Supplementary figures and tables are numbered in separate S-prefixed sequences. References to figures and tables in the main article are identified explicitly.\par

\section{Data variables, part numbers and design-family examples}\label{app:technical_network}\label{app:data_examples}
\subsection{Catalogue fields and analysis variables}\label{app:a_fields}

The analysed catalogue contains 102,465 product records grouped into 45,996 design families. A part number identifies a recorded SKU variant; the design identifier groups related variants and is the family node identifier used in the network. The group label need not be the earliest part number. Manufacturer labels are retained separately from abbreviated prefixes in design identifiers. The observed first catalogue year defines family birth and supplies the network's temporal ordering.

The raw KPI inventory contains 80 fields: 29 numerical, 11 categorical, and 40 flag fields. The active network retains 36 fields (25 numerical and 11 categorical). The 40 flags and four other numerical fields are outside its active KPI vocabulary. The following dictionary describes exactly those retained fields. Missing values mean unobserved information, not absence of a feature.

Field names below omit only the common \textnormal{kpi\_} prefix. A unit is stated only when explicit in the field label, or when the variable is an explicit count. `Unverified' means unverified recorded units; no physical-unit conversion is inferred from the magnitude. Similarly named frequency fields remain separate when their observation masks differ. A source label alone does not establish semantic equivalence between fields or subcategories.

\begingroup\footnotesize\doublespacing\setlength{\tabcolsep}{3pt}\doublespacing
\begin{longtable}{@{}>{\raggedright\arraybackslash}p{0.340\linewidth}>{\raggedright\arraybackslash}p{0.410\linewidth}>{\raggedright\arraybackslash}p{0.170\linewidth}@{}}
\caption{Numerical KPI fields retained in the network.}\label{app:a_numeric_dictionary}\\
\toprule Stored field & Recorded meaning & Unit / type \\ \midrule\endfirsthead
\toprule Stored field & Recorded meaning & Unit / type \\ \midrule\endhead
\textnormal{access\_\allowbreak{}time} & Recorded access latency & Unverified \\
\textnormal{active\_\allowbreak{}current\_\allowbreak{}a} & Active-state supply current & A \\
\textnormal{active\_\allowbreak{}read\_\allowbreak{}current\_\allowbreak{}max} & Maximum active-read current & Unverified \\
\textnormal{cache\_\allowbreak{}memory} & Recorded cache capacity & Unverified \\
\textnormal{core\_\allowbreak{}count} & Recorded core-related value; not a verified physical core count & Unverified \\
\textnormal{data\_\allowbreak{}bus\_\allowbreak{}width} & Recorded data-bus width & Unverified \\
\textnormal{data\_\allowbreak{}ram\_\allowbreak{}size} & Recorded data-RAM capacity & Unverified \\
\textnormal{frequency\_\allowbreak{}max\_\allowbreak{}hz\_\allowbreak{}num} & Maximum frequency, normalised field & Hz \\
\textnormal{max\_\allowbreak{}clock\_\allowbreak{}frequency} & Maximum clock frequency, source-specific field & Unverified \\
\textnormal{max\_\allowbreak{}clock\_\allowbreak{}frequency\_\allowbreak{}hz} & Maximum clock frequency, normalised field & Hz \\
\textnormal{maximum\_\allowbreak{}clock\_\allowbreak{}frequency} & Maximum clock frequency, separate observation mask & Unverified \\
\textnormal{maximum\_\allowbreak{}data\_\allowbreak{}rate} & Maximum data-transfer rate & Unverified \\
\textnormal{memory\_\allowbreak{}size} & Recorded memory capacity; units may differ by source & Unverified \\
\textnormal{number\_\allowbreak{}of\_\allowbreak{}cores} & Reported number of processor cores & count \\
\textnormal{number\_\allowbreak{}of\_\allowbreak{}io} & Reported number of input/output connections & count \\
\textnormal{number\_\allowbreak{}of\_\allowbreak{}logic\_\allowbreak{}elements} & Reported number of logic elements & count \\
\textnormal{operating\_\allowbreak{}frequency\_\allowbreak{}hz\_\allowbreak{}num} & Operating frequency, normalised field & Hz \\
\textnormal{operating\_\allowbreak{}supply\_\allowbreak{}current\_\allowbreak{}a} & Operating supply current & A \\
\textnormal{organization\_\allowbreak{}width} & Recorded memory-organization value; not a verified word width & Unverified \\
\textnormal{program\_\allowbreak{}memory\_\allowbreak{}size\_\allowbreak{}num} & Recorded program-memory capacity & Unverified \\
\textnormal{sram\_\allowbreak{}kb} & SRAM capacity in the field's declared kB scale & kB (as labelled) \\
\textnormal{supply\_\allowbreak{}current\_\allowbreak{}max} & Maximum supply current & Unverified \\
\textnormal{supply\_\allowbreak{}current\_\allowbreak{}receiving} & Supply current while receiving & Unverified \\
\textnormal{tdp\_\allowbreak{}max} & Maximum thermal-design-power value & Unverified \\
\textnormal{total\_\allowbreak{}memory} & Recorded total memory capacity & Unverified \\
\bottomrule\end{longtable}\endgroup

\begingroup\footnotesize\doublespacing\setlength{\tabcolsep}{3pt}\doublespacing
\begin{longtable}{@{}>{\raggedright\arraybackslash}p{0.340\linewidth}>{\raggedright\arraybackslash}p{0.410\linewidth}>{\raggedright\arraybackslash}p{0.170\linewidth}@{}}
\caption{Categorical KPI fields retained in the network.}\label{app:a_categorical_dictionary}\\
\toprule Stored field & Recorded meaning & Unit / type \\ \midrule\endfirsthead
\toprule Stored field & Recorded meaning & Unit / type \\ \midrule\endhead
\textnormal{cat\_\allowbreak{}applications} & Application categories or use cases & categorical tokens \\
\textnormal{cat\_\allowbreak{}architecture} & Recorded architecture label & categorical tokens \\
\textnormal{cat\_\allowbreak{}development\_\allowbreak{}kit} & Development-kit identifier or availability label & categorical tokens \\
\textnormal{cat\_\allowbreak{}interface} & Interface labels, tokenized and canonicalized where resolved & categorical tokens \\
\textnormal{cat\_\allowbreak{}interface\_\allowbreak{}type} & Interface-type labels, including named protocols and unresolved tokens & categorical tokens \\
\textnormal{cat\_\allowbreak{}moisture\_\allowbreak{}sensitive} & Recorded moisture-sensitivity category & categorical tokens \\
\textnormal{cat\_\allowbreak{}mounting\_\allowbreak{}type} & Mounting method, such as surface mount & categorical tokens \\
\textnormal{cat\_\allowbreak{}package\_\allowbreak{}case} & Recorded package/case category & categorical tokens \\
\textnormal{cat\_\allowbreak{}supplier\_\allowbreak{}device\_\allowbreak{}package} & Supplier-specific package designation & categorical tokens \\
\textnormal{cat\_\allowbreak{}technology} & Recorded technology or material label & categorical tokens \\
\textnormal{cat\_\allowbreak{}type} & Source-specific product-type category & categorical tokens \\
\bottomrule\end{longtable}\endgroup

\subsection{Part numbers within design families}\label{app:a_families}

The four examples below are descriptive illustrations of membership and within-family variation, not a random sample. Each displayed part number is verified against its exact source file, row, URL identifier, and family assignment. E1 spans two source subcategories in the inventory while retaining its prepared 8-bit MCU node category; this is an observed grouping property, not a new cross-category merge. The full memberships and launch-profile provenance are exported alongside this appendix.

\begingroup\footnotesize\doublespacing\setlength{\tabcolsep}{3pt}\doublespacing
\begin{longtable}{@{}>{\raggedright\arraybackslash}p{0.210\linewidth}>{\raggedright\arraybackslash}p{0.250\linewidth}>{\raggedright\arraybackslash}p{0.130\linewidth}>{\raggedright\arraybackslash}p{0.330\linewidth}@{}}
\caption{Actual catalogue memberships and birth-year profiles for four design families.}\label{app:a_family_examples}\\
\toprule Example / category & Design identifier & Records / launch profiles; years & Example member part numbers (observed year) \\ \midrule\endfirsthead
\toprule Example / category & Design identifier & Records / launch profiles; years & Example member part numbers (observed year) \\ \midrule\endhead
\textbf{E1}: 8-bit Microcontrollers - MCU\newline Silicon Labs & Silicon Labs::\allowbreak{}C8051F530-\allowbreak{}C-\allowbreak{}AM\newline Node 40427 & 16 / 2\newline 2008--2011 & C8051F530-\allowbreak{}IM (2008)\newline C8051F530A-\allowbreak{}AMR (2008)\newline C8051F530A-\allowbreak{}IM (2011) \\
\textbf{E2}: 16-bit Microcontrollers - MCU\newline NXP & NXP::\allowbreak{}S912ZVML31AMKF\newline Node 23910 & 8 / 8\newline 2016--2016 & S912ZVML31AMKF (2016)\newline S912ZVML31AMKFR (2016)\newline S912ZVML31AWKHR (2016) \\
\textbf{E3}: RF System on a Chip - SoC\newline Espressif & Espressif::\allowbreak{}ESP32-\allowbreak{}S2\newline Node 1580 & 5 / 1\newline 2019--2021 & ESP32-\allowbreak{}S2 (2019)\newline ESP32-\allowbreak{}S2FH4 (2020)\newline ESP32-\allowbreak{}S2FH2 (2021) \\
\textbf{E4}: DRAM\newline AP Memory Technology & AP Memory Tech::\allowbreak{}APS6404L-\allowbreak{}SQH-\allowbreak{}RA\newline Node 187 & 9 / 5\newline 2018--2019 & APS6404L-\allowbreak{}SQH-\allowbreak{}SN (2018)\newline APS6404L-\allowbreak{}SQH-\allowbreak{}ZR (2018)\newline APS6404L-\allowbreak{}SQRHX-\allowbreak{}SN (2019) \\
\bottomrule\end{longtable}\endgroup

\subsection{Observed KPI values on individual launch SKUs}\label{app:a_skus}

Every row below is one actual SKU observed in its family's first catalogue year. No row combines one variant's interface with another variant's clock or package. Frequency values with explicit Hz fields are rescaled to MHz or GHz for readability. E4's memory-size value is left in recorded units. E1 illustrates complementary source coverage: the missing named-interface entry on its second SKU is not evidence that SPI or UART is absent. E2 and E4 demonstrate package variation within a Family, and E4 also shows different observed maximum clocks at birth.

\begingroup\footnotesize\doublespacing\setlength{\tabcolsep}{3pt}\doublespacing
\begin{longtable}{@{}>{\raggedright\arraybackslash}p{0.280\linewidth}>{\raggedright\arraybackslash}p{0.280\linewidth}>{\raggedright\arraybackslash}p{0.360\linewidth}@{}}
\caption{Per-SKU birth-year KPI observations in the selected Families.}\label{app:a_sku_examples}\\
\toprule Example, part number, year & Numerical observations & Categorical observations \\ \midrule\endfirsthead
\toprule Example, part number, year & Numerical observations & Categorical observations \\ \midrule\endhead
E1\newline C8051F530A-\allowbreak{}AMR\newline 2008 & Clock: 25.000 MHz & Package: QFN-20\newline Named interfaces: SPI, UART\_\allowbreak{}USART \\
E1\newline C8051F530-\allowbreak{}IM\newline 2008 & Maximum frequency: 25.000 MHz\newline I/O connections: 16.000 count & Package: 20-VFQFN Exposed Pad\newline Named interfaces: not recorded \\
E2\newline S912ZVML31AMKF\newline 2016 & Clock: 50.000 MHz & Package: LQFP-48\newline Named interfaces: CAN, LIN, SPI\newline Additional unresolved SCI label \\
E2\newline S912ZVML31AMKH\newline 2016 & Clock: 50.000 MHz & Package: LQFP-64\newline Named interfaces: CAN, LIN, SPI\newline Additional unresolved SCI label \\
E3\newline ESP32-\allowbreak{}S2\newline 2019 & Operating frequency: 2.400 GHz & Package: QFN-56\newline Named interfaces: I2C, I2S, SPI, UART\_\allowbreak{}USART \\
E4\newline APS6404L-\allowbreak{}SQH-\allowbreak{}SN\newline 2018 & Clock: 144.000 MHz\newline Memory-size field: 64.000 recorded units & Package: SOP-8\newline Named interfaces: not recorded \\
E4\newline APS6404L-\allowbreak{}SQH-\allowbreak{}ZR\newline 2018 & Clock: 144.000 MHz\newline Memory-size field: 64.000 recorded units & Package: USON-8\newline Named interfaces: not recorded \\
E4\newline APS6404L-\allowbreak{}SQRH-\allowbreak{}SN\newline 2018 & Clock: 84.000 MHz\newline Memory-size field: 64.000 recorded units & Package: SOP-8\newline Named interfaces: not recorded \\
\bottomrule\end{longtable}\endgroup

A categorical KPI field is not itself a functional module. Named tokens such as CAN, I2C, I2S, LIN, SPI, and UART\_USART identify resolved interface capabilities in the functional-module analysis. Generic or ambiguous tokens (including the additional SCI label in E2) are not promoted to named functional modules. Package and mounting categories describe physical implementation and are not treated as new named functional capabilities. E3's 2020--2021 variants illustrate later catalogue observations: they are family members but are not inserted into its 2019 birth profile. Later first appearances and recorded capability carriage are evaluated using time-stamped SKU events, separating family-birth profiles from subsequent feature observations.

\FloatBarrier\clearpage

\section{Phylogenetic networks within each subcategory}\label{app:subcategory_networks}
The following gallery contains all 19 subcategories in the family birth graph. Node colours identify manufacturers within each panel. Grey links indicate the similarity backbone and red arrows indicate candidate launch-time HT within the displayed subcategory. Cross-subcategory context is omitted from these category-specific views; it remains visible in Figure~2 of the main article. Undated families are shown separately, and radial distance represents backbone generation rather than elapsed calendar time. No synthetic ancestor, sampling or new reconstruction is introduced by the gallery.
\clearpage
\begin{figure}[p]\centering
\includegraphics[width=.98\linewidth,height=.54\textheight,keepaspectratio]{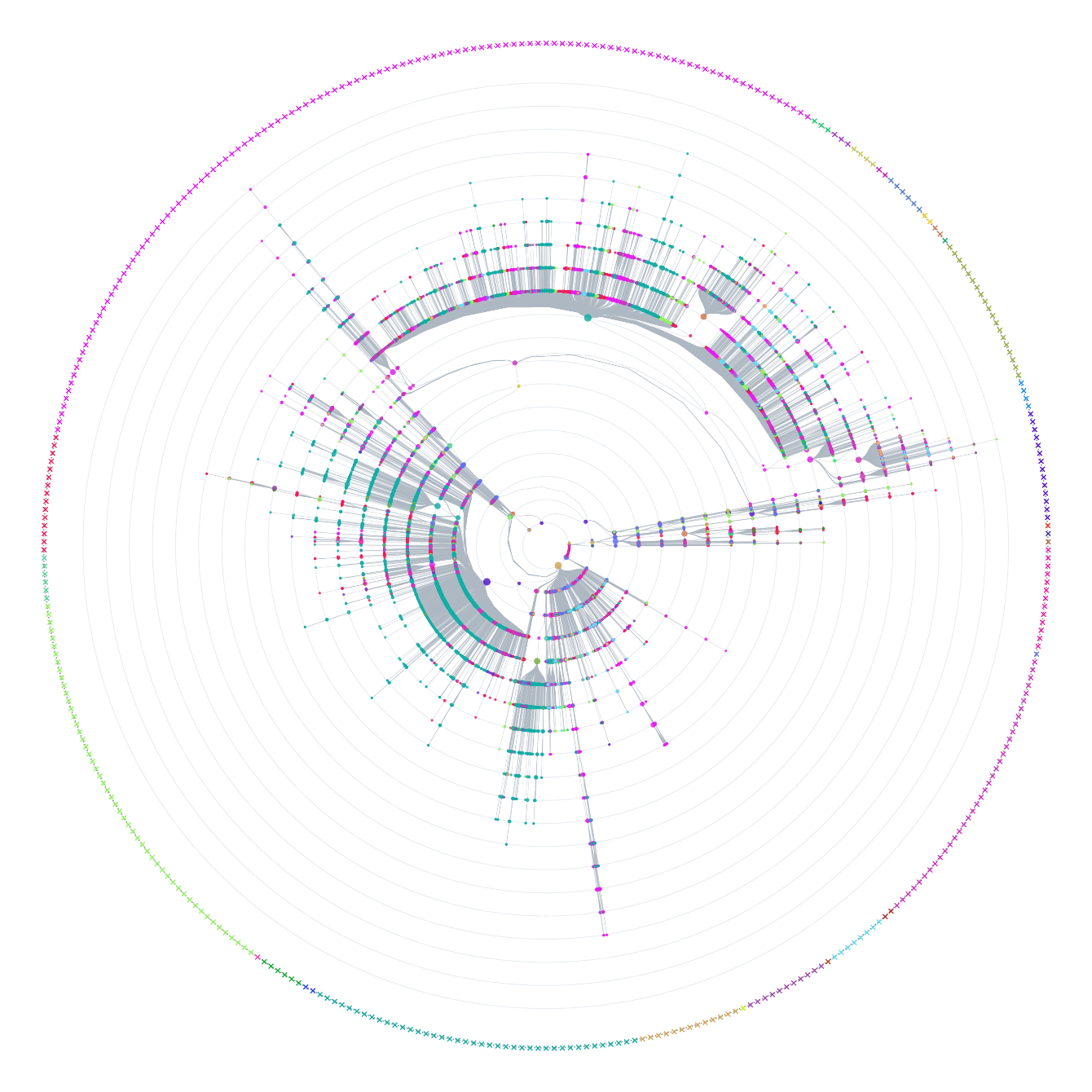}
\caption{\textbf{Microcontrollers.} All 36,631 design families, 33,801 backbone links, and 0 within-subcategory candidate-HT donor--recipient pairs are shown. Recorded first appearances span 1976--2024. Node colours identify manufacturers; grey links are inferred backbone connections and red arrows are candidate HT links. Radius measures backbone generation, not calendar time. The 390 undated families are isolated crosses on the outer dotted ring, not identified historical founders. Cross-subcategory links and the optional numerical-VI overlay are omitted.}\label{app:b_network_microcontrollers}
\alttxt{Radial network for Microcontrollers, showing 36,631 design families, 33,801 inferred backbone links and 0 within-category external-precedent pairs. Node colours distinguish recorded manufacturers. Radius indicates generation, not elapsed time.}
\end{figure}
\clearpage
\begin{figure}[p]\centering
\includegraphics[width=.98\linewidth,height=.54\textheight,keepaspectratio]{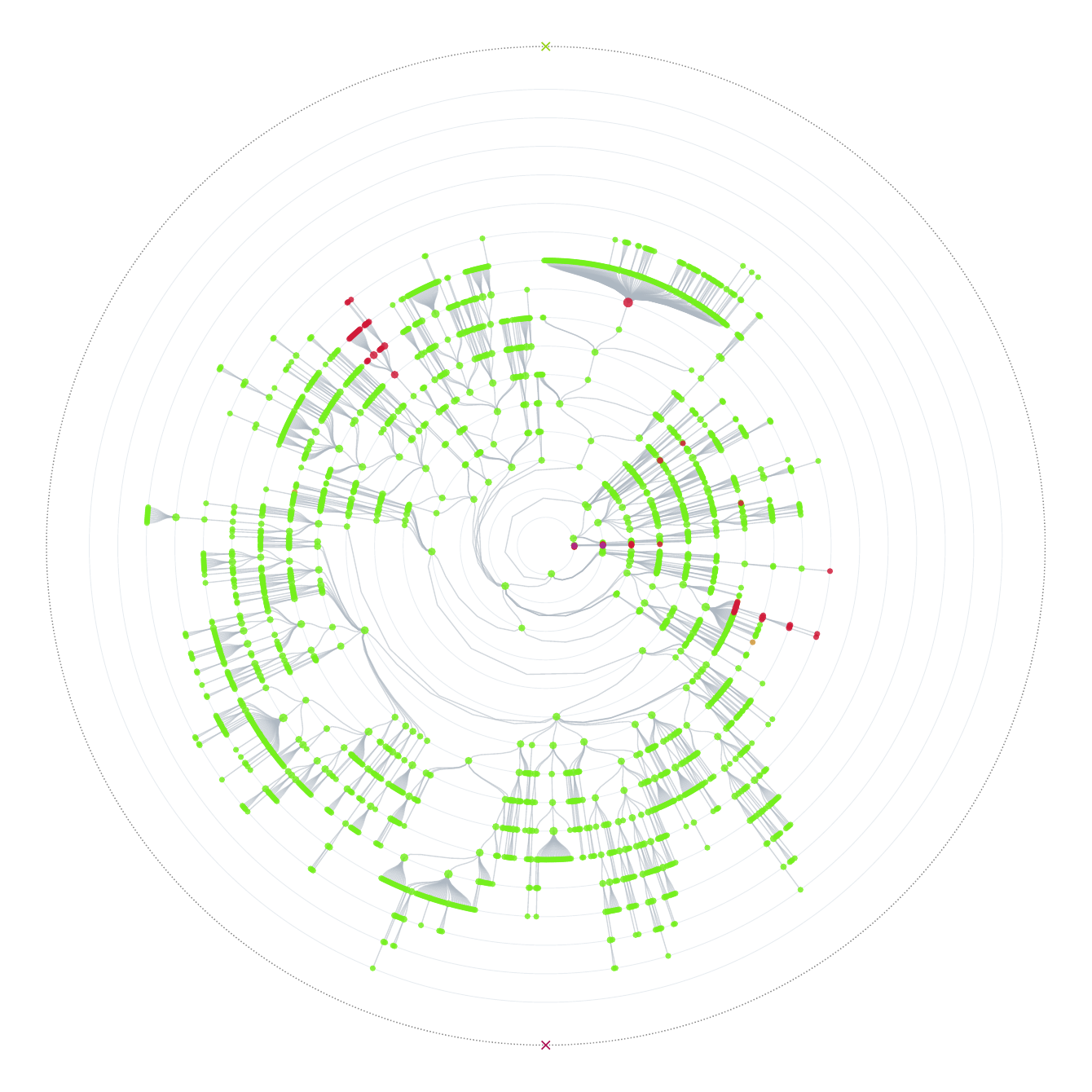}
\caption{\textbf{SRAM.} All 2,389 design families, 2,370 backbone links, and 0 within-subcategory candidate-HT donor--recipient pairs are shown. Recorded first appearances span 1993--2022. Node colours identify manufacturers; grey links are inferred backbone connections and red arrows are candidate HT links. Radius measures backbone generation, not calendar time. The 2 undated families are isolated crosses on the outer dotted ring, not identified historical founders. Cross-subcategory links and the optional numerical-VI overlay are omitted.}\label{app:b_network_sram}
\alttxt{Radial network for SRAM, showing 2,389 design families, 2,370 inferred backbone links and 0 within-category external-precedent pairs. Node colours distinguish recorded manufacturers. Radius indicates generation, not elapsed time.}
\end{figure}
\clearpage
\begin{figure}[p]\centering
\includegraphics[width=.98\linewidth,height=.54\textheight,keepaspectratio]{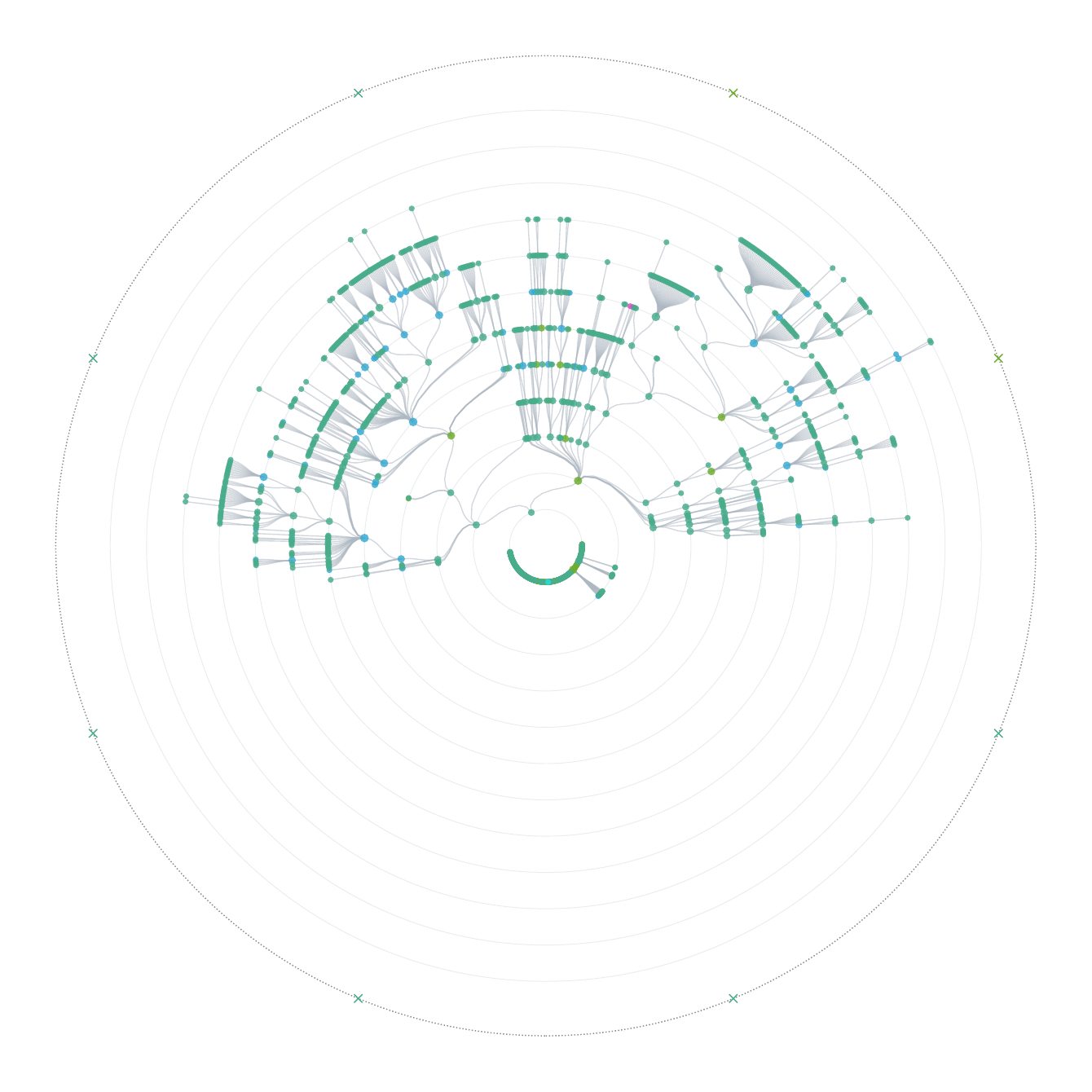}
\caption{\textbf{CPU - Central Processing Units.} All 1,502 design families, 911 backbone links, and 0 within-subcategory candidate-HT donor--recipient pairs are shown. Recorded first appearances span 2000--2024. Node colours identify manufacturers; grey links are inferred backbone connections and red arrows are candidate HT links. Radius measures backbone generation, not calendar time. The 8 undated families are isolated crosses on the outer dotted ring, not identified historical founders. Cross-subcategory links and the optional numerical-VI overlay are omitted.}\label{app:b_network_cpu_central_processing_units}
\alttxt{Radial network for CPU - Central Processing Units, showing 1,502 design families, 911 inferred backbone links and 0 within-category external-precedent pairs. Node colours distinguish recorded manufacturers. Radius indicates generation, not elapsed time.}
\end{figure}
\clearpage
\begin{figure}[p]\centering
\includegraphics[width=.98\linewidth,height=.54\textheight,keepaspectratio]{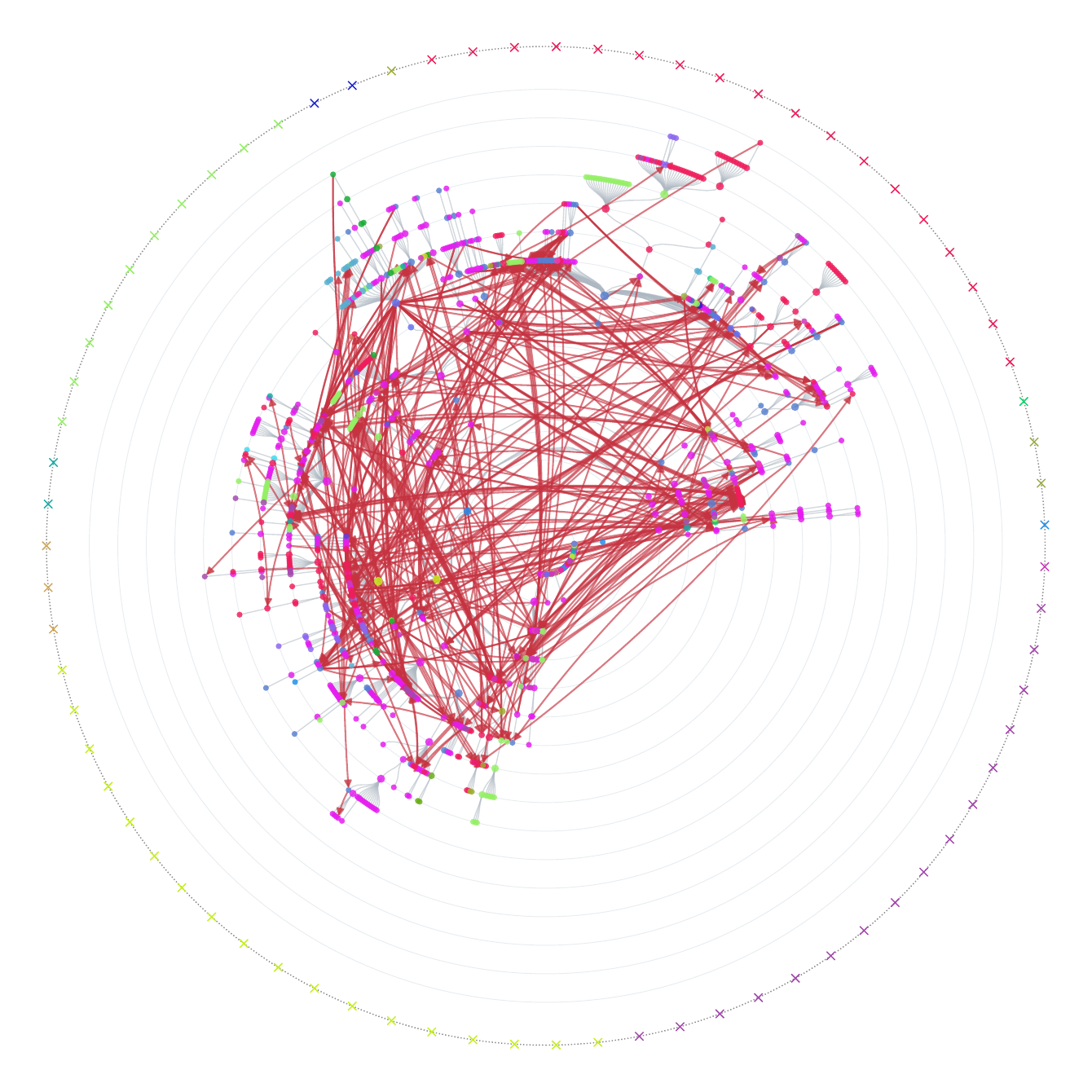}
\caption{\textbf{Application Specific Microcontrollers.} All 1,291 design families, 973 backbone links, and 446 within-subcategory candidate-HT donor--recipient pairs are shown. Recorded first appearances span 1977--2023. Node colours identify manufacturers; grey links are inferred backbone connections and red arrows are candidate HT links. Radius measures backbone generation, not calendar time. The 75 undated families are isolated crosses on the outer dotted ring, not identified historical founders. Cross-subcategory links and the optional numerical-VI overlay are omitted.}\label{app:b_network_application_specific_microcontrollers}
\alttxt{Radial network for Application Specific Microcontrollers, showing 1,291 design families, 973 inferred backbone links and 446 within-category external-precedent pairs. Node colours distinguish recorded manufacturers. Radius indicates generation, not elapsed time.}
\end{figure}
\clearpage
\begin{figure}[p]\centering
\includegraphics[width=.98\linewidth,height=.54\textheight,keepaspectratio]{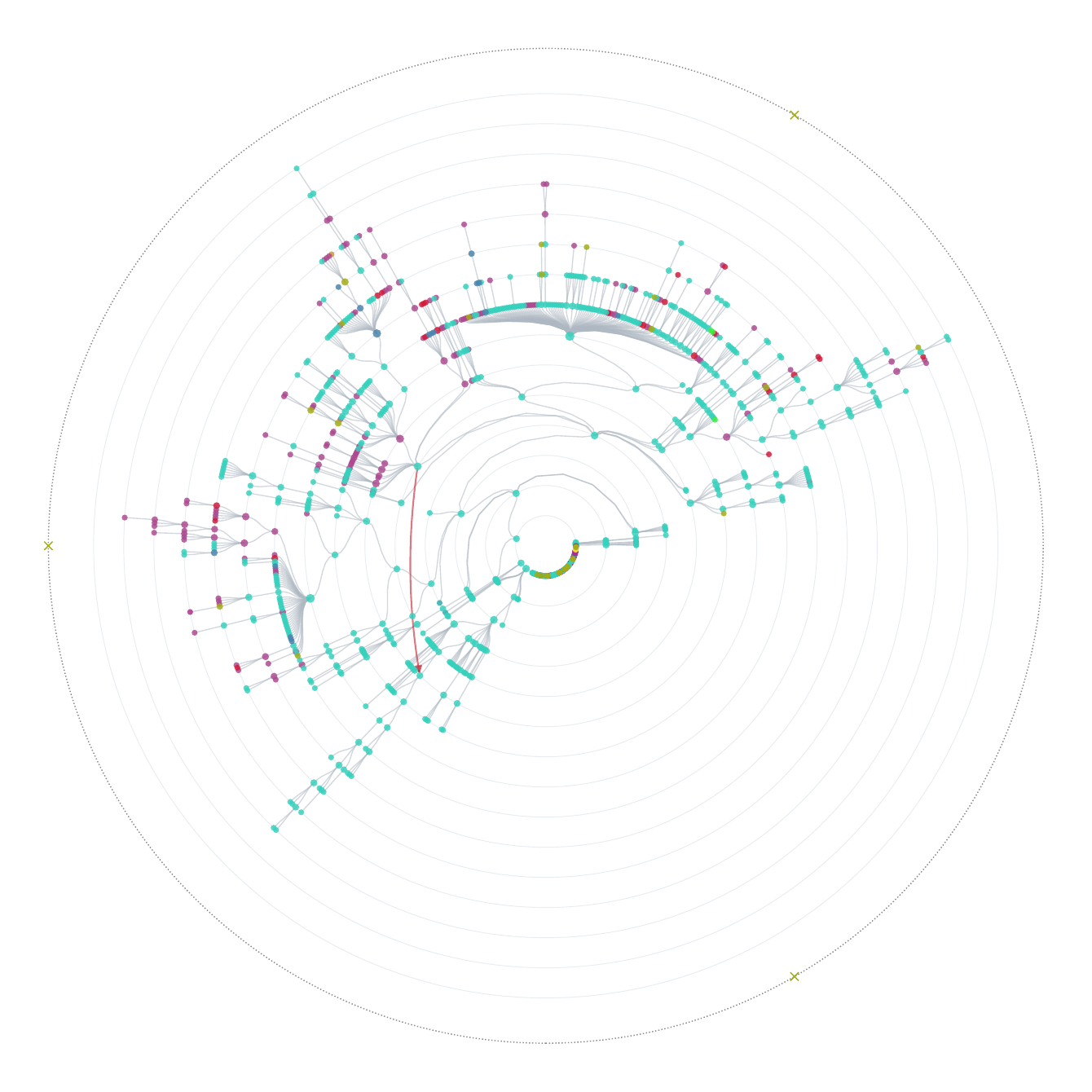}
\caption{\textbf{NOR Flash.} All 1,030 design families, 774 backbone links, and 1 within-subcategory candidate-HT donor--recipient pair are shown. Recorded first appearances span 1996--2023. Node colours identify manufacturers; grey links are inferred backbone connections and red arrows are candidate HT links. Radius measures backbone generation, not calendar time. The 3 undated families are isolated crosses on the outer dotted ring, not identified historical founders. Cross-subcategory links and the optional numerical-VI overlay are omitted.}\label{app:b_network_nor_flash}
\alttxt{Radial network for NOR Flash, showing 1,030 design families, 774 inferred backbone links and 1 within-category external-precedent pairs. Node colours distinguish recorded manufacturers. Radius indicates generation, not elapsed time.}
\end{figure}
\clearpage
\begin{figure}[p]\centering
\includegraphics[width=.98\linewidth,height=.54\textheight,keepaspectratio]{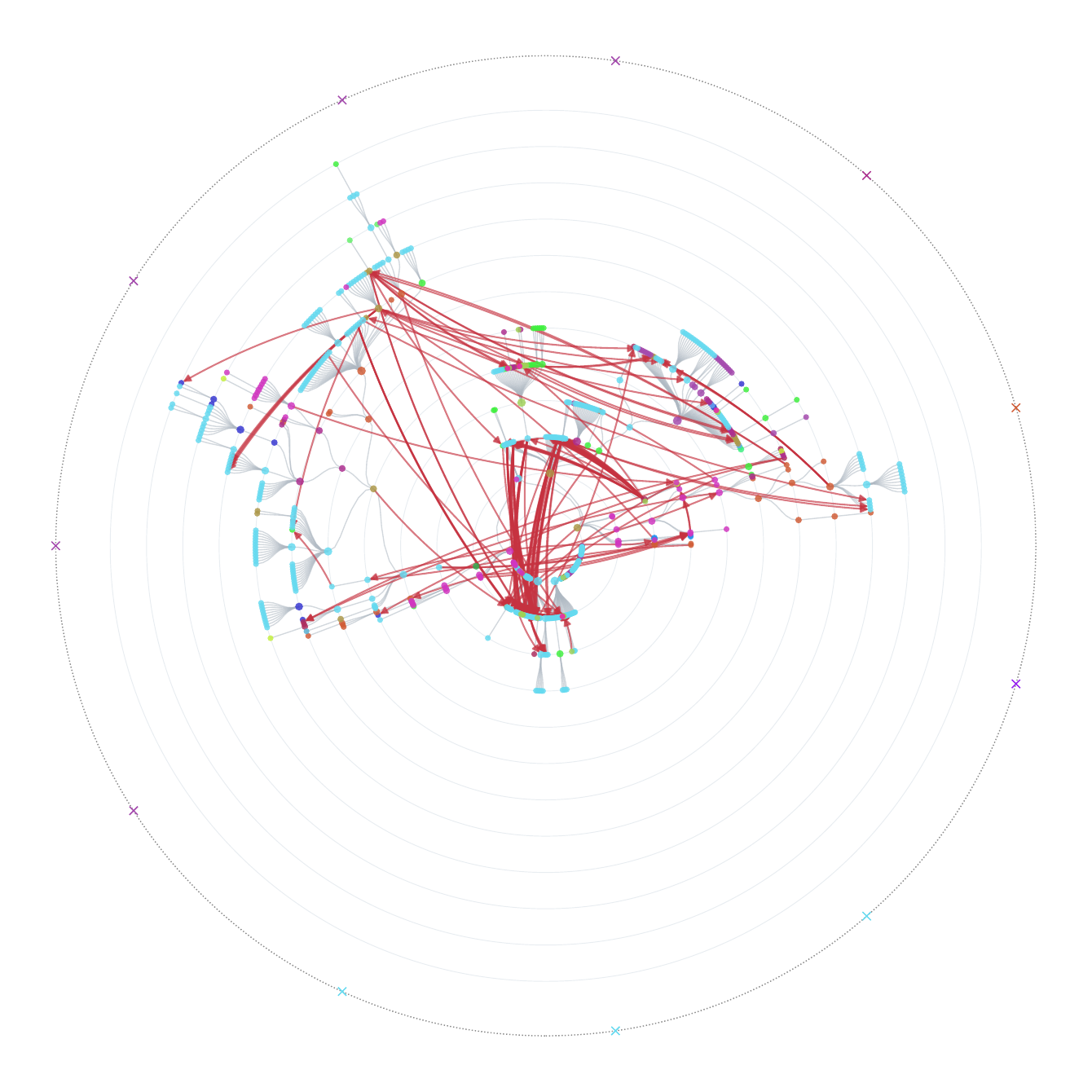}
\caption{\textbf{RF System on a Chip - SoC.} All 773 design families, 576 backbone links, and 160 within-subcategory candidate-HT donor--recipient pairs are shown. Recorded first appearances span 2004--2024. Node colours identify manufacturers; grey links are inferred backbone connections and red arrows are candidate HT links. Radius measures backbone generation, not calendar time. The 11 undated families are isolated crosses on the outer dotted ring, not identified historical founders. Cross-subcategory links and the optional numerical-VI overlay are omitted.}\label{app:b_network_rf_system_on_a_chip_soc}
\alttxt{Radial network for RF System on a Chip - SoC, showing 773 design families, 576 inferred backbone links and 160 within-category external-precedent pairs. Node colours distinguish recorded manufacturers. Radius indicates generation, not elapsed time.}
\end{figure}
\clearpage
\begin{figure}[p]\centering
\includegraphics[width=.98\linewidth,height=.54\textheight,keepaspectratio]{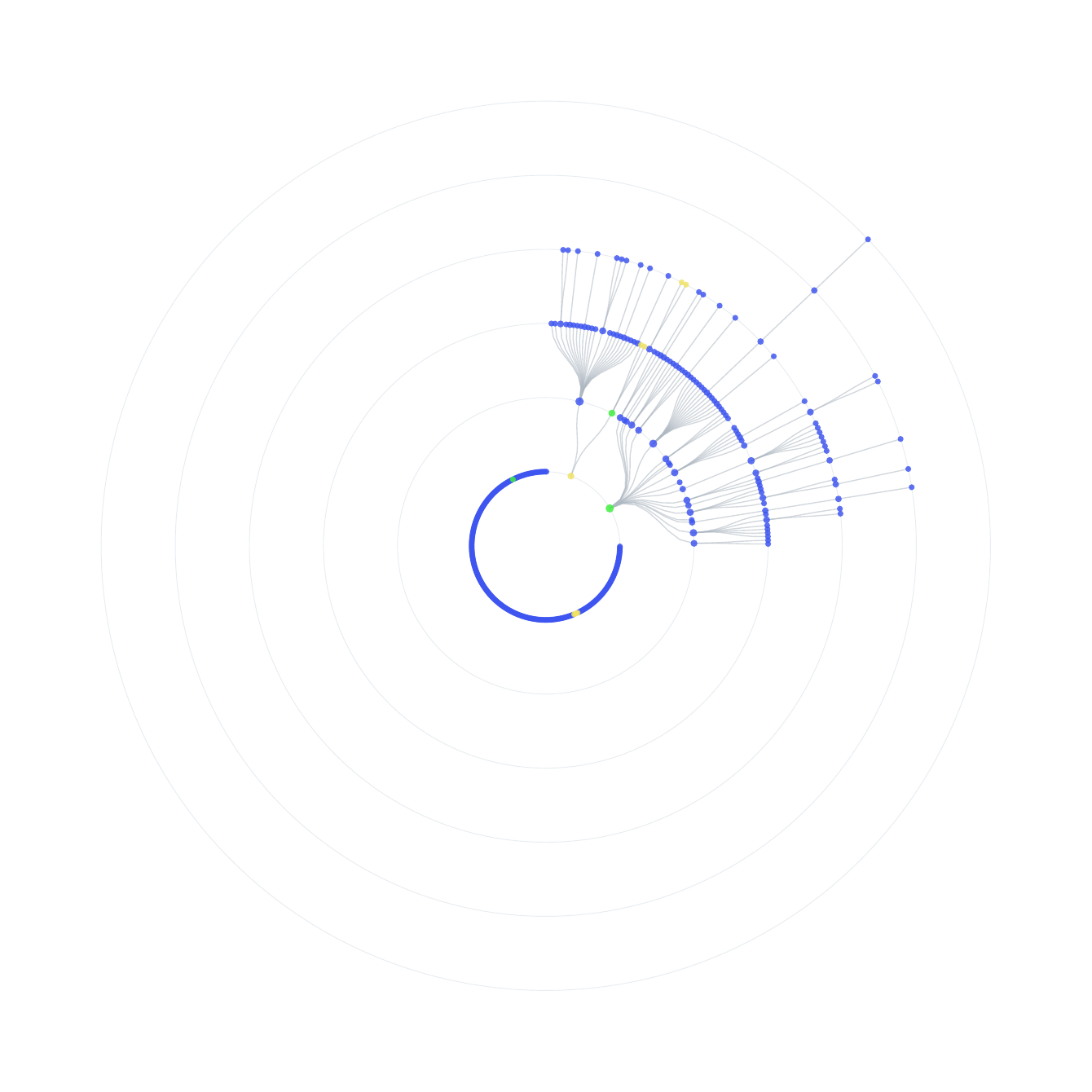}
\caption{\textbf{FPGA - Field Programmable Gate Array.} All 423 design families, 138 backbone links, and 0 within-subcategory candidate-HT donor--recipient pairs are shown. Recorded first appearances span 2007--2022. Node colours identify manufacturers; grey links are inferred backbone connections and red arrows are candidate HT links. Radius measures backbone generation, not calendar time. Cross-subcategory links and the optional numerical-VI overlay are omitted.}\label{app:b_network_fpga_field_programmable_gate_array}
\alttxt{Radial network for FPGA - Field Programmable Gate Array, showing 423 design families, 138 inferred backbone links and 0 within-category external-precedent pairs. Node colours distinguish recorded manufacturers. Radius indicates generation, not elapsed time.}
\end{figure}
\clearpage
\begin{figure}[p]\centering
\includegraphics[width=.98\linewidth,height=.54\textheight,keepaspectratio]{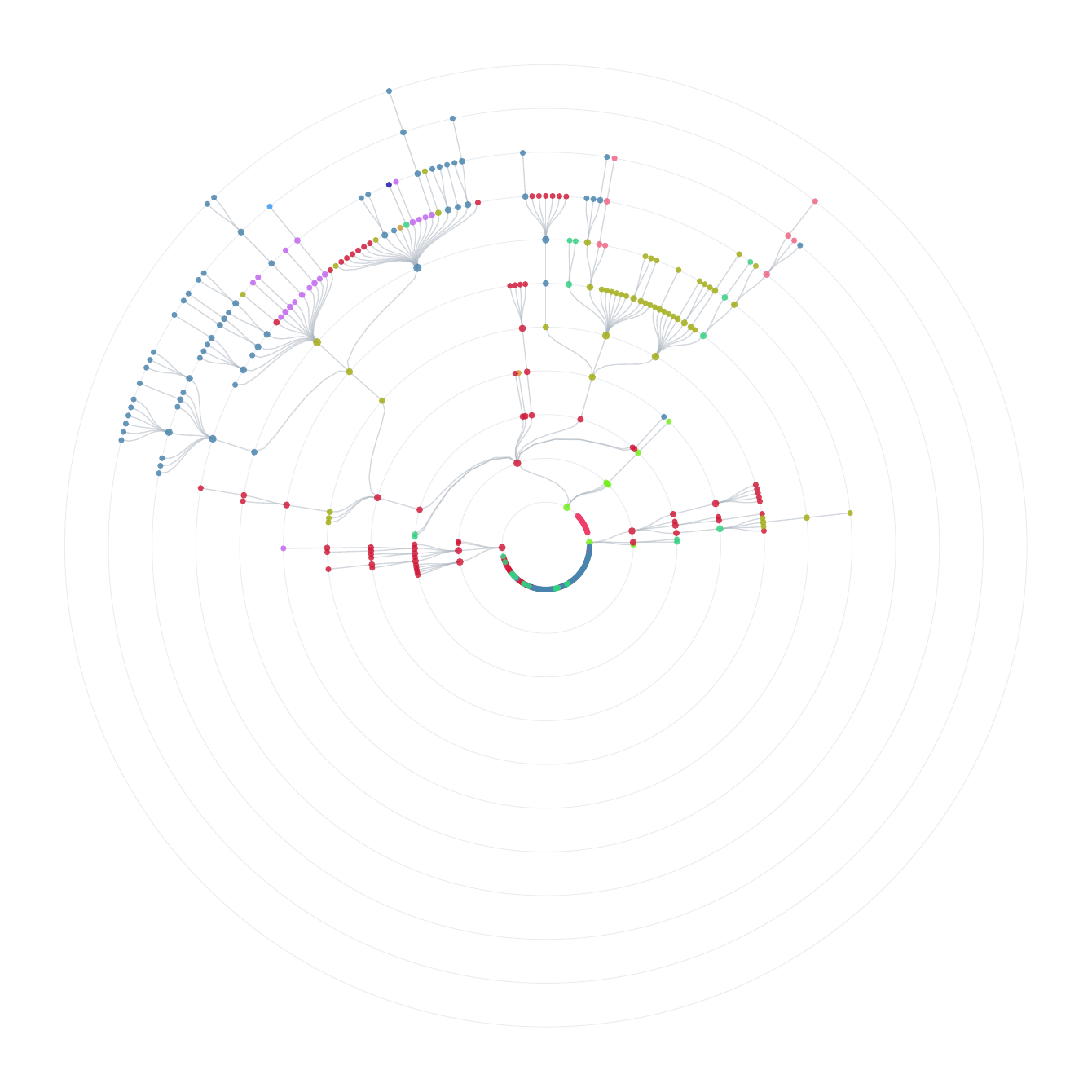}
\caption{\textbf{DRAM.} All 420 design families, 244 backbone links, and 0 within-subcategory candidate-HT donor--recipient pairs are shown. Recorded first appearances span 2003--2023. Node colours identify manufacturers; grey links are inferred backbone connections and red arrows are candidate HT links. Radius measures backbone generation, not calendar time. Cross-subcategory links and the optional numerical-VI overlay are omitted.}\label{app:b_network_dram}
\alttxt{Radial network for DRAM, showing 420 design families, 244 inferred backbone links and 0 within-category external-precedent pairs. Node colours distinguish recorded manufacturers. Radius indicates generation, not elapsed time.}
\end{figure}
\clearpage
\begin{figure}[p]\centering
\includegraphics[width=.98\linewidth,height=.54\textheight,keepaspectratio]{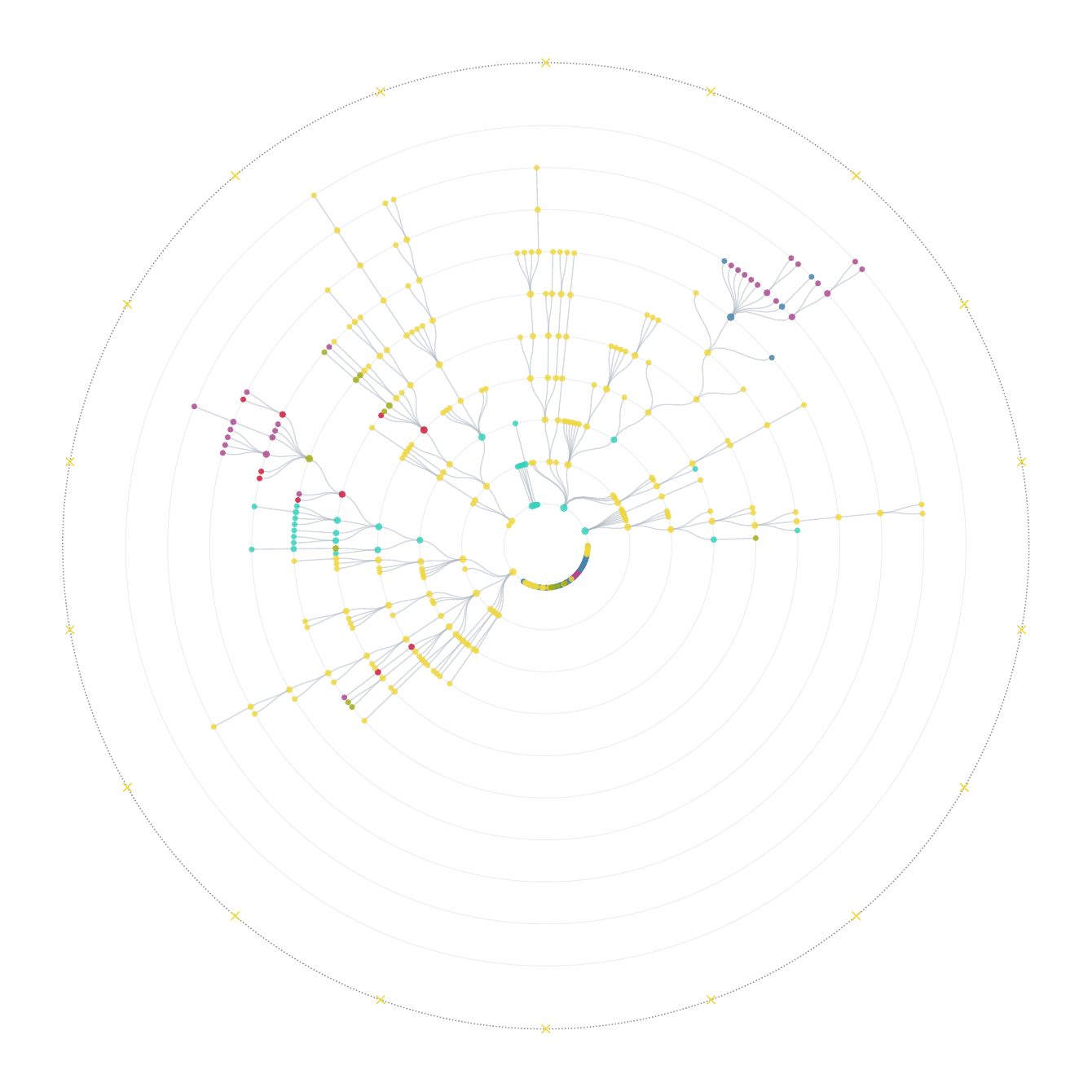}
\caption{\textbf{NAND Flash.} All 399 design families, 282 backbone links, and 0 within-subcategory candidate-HT donor--recipient pairs are shown. Recorded first appearances span 2004--2023. Node colours identify manufacturers; grey links are inferred backbone connections and red arrows are candidate HT links. Radius measures backbone generation, not calendar time. The 18 undated families are isolated crosses on the outer dotted ring, not identified historical founders. Cross-subcategory links and the optional numerical-VI overlay are omitted.}\label{app:b_network_nand_flash}
\alttxt{Radial network for NAND Flash, showing 399 design families, 282 inferred backbone links and 0 within-category external-precedent pairs. Node colours distinguish recorded manufacturers. Radius indicates generation, not elapsed time.}
\end{figure}
\clearpage
\begin{figure}[p]\centering
\includegraphics[width=.98\linewidth,height=.54\textheight,keepaspectratio]{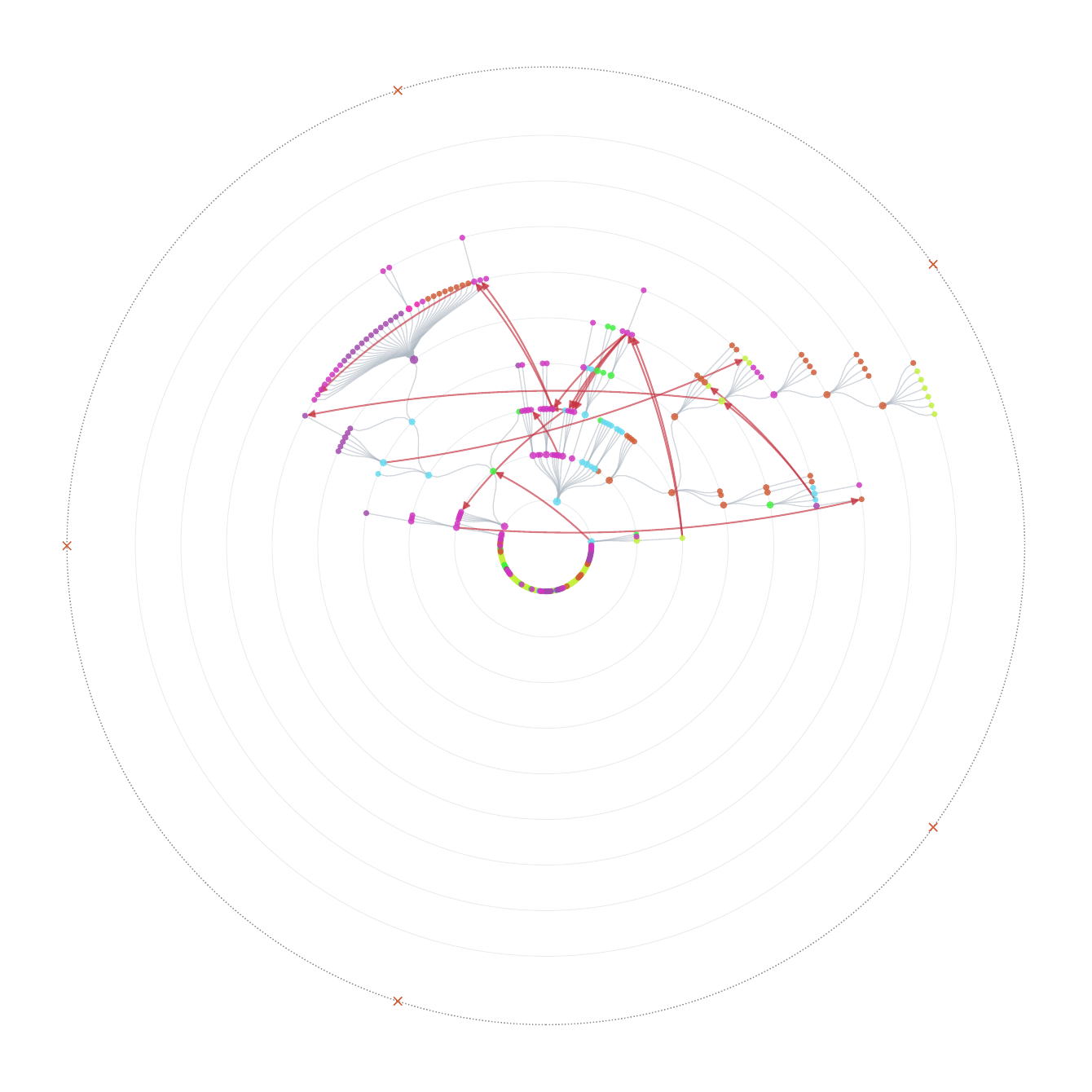}
\caption{\textbf{RF Microcontrollers - MCU.} All 332 design families, 173 backbone links, and 19 within-subcategory candidate-HT donor--recipient pairs are shown. Recorded first appearances span 2007--2023. Node colours identify manufacturers; grey links are inferred backbone connections and red arrows are candidate HT links. Radius measures backbone generation, not calendar time. The 5 undated families are isolated crosses on the outer dotted ring, not identified historical founders. Cross-subcategory links and the optional numerical-VI overlay are omitted.}\label{app:b_network_rf_microcontrollers_mcu}
\alttxt{Radial network for RF Microcontrollers - MCU, showing 332 design families, 173 inferred backbone links and 19 within-category external-precedent pairs. Node colours distinguish recorded manufacturers. Radius indicates generation, not elapsed time.}
\end{figure}
\clearpage
\begin{figure}[p]\centering
\includegraphics[width=.98\linewidth,height=.54\textheight,keepaspectratio]{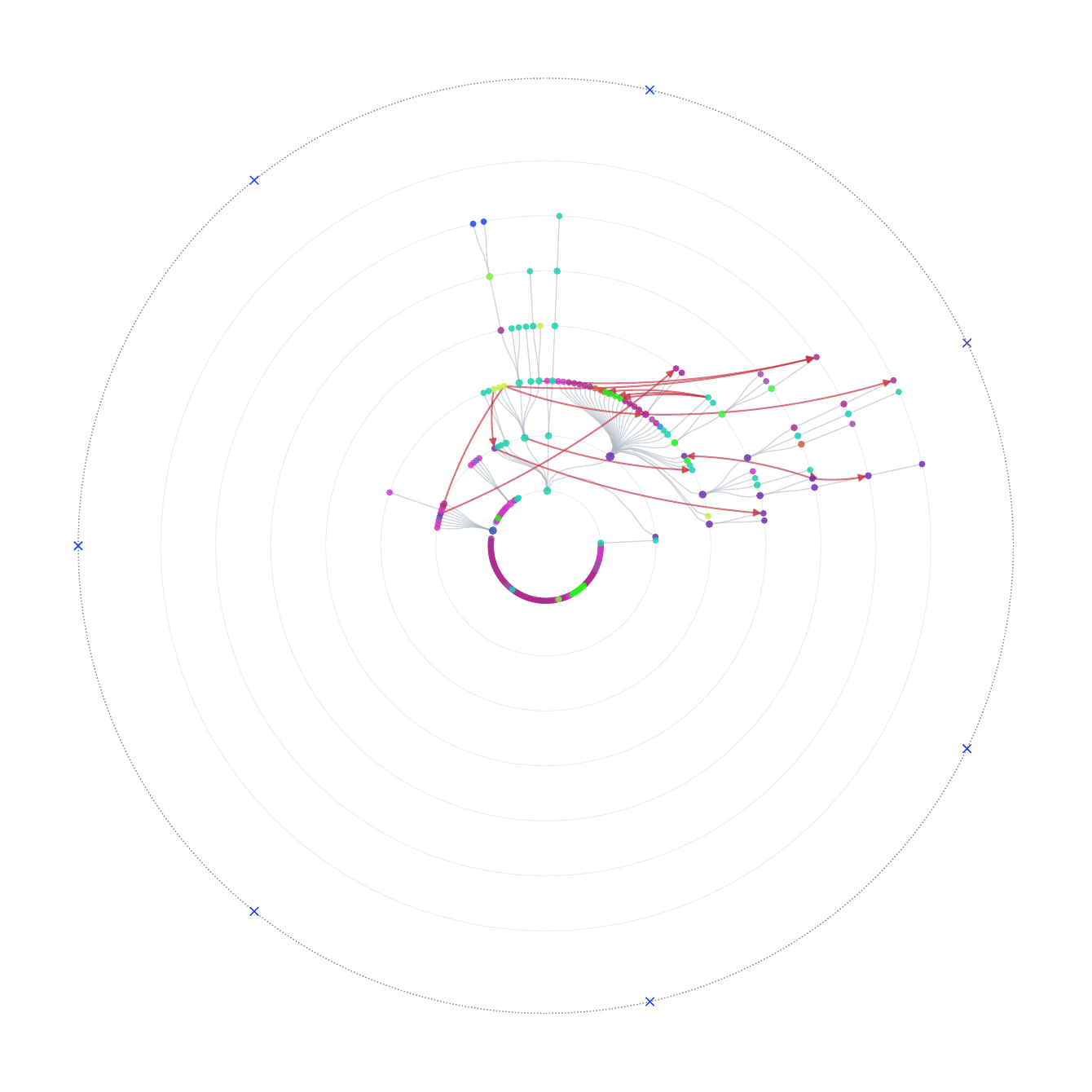}
\caption{\textbf{ARM Microcontrollers - MCU.} All 233 design families, 105 backbone links, and 17 within-subcategory candidate-HT donor--recipient pairs are shown. Recorded first appearances span 2000--2024. Node colours identify manufacturers; grey links are inferred backbone connections and red arrows are candidate HT links. Radius measures backbone generation, not calendar time. The 7 undated families are isolated crosses on the outer dotted ring, not identified historical founders. Cross-subcategory links and the optional numerical-VI overlay are omitted.}\label{app:b_network_arm_microcontrollers_mcu}
\alttxt{Radial network for ARM Microcontrollers - MCU, showing 233 design families, 105 inferred backbone links and 17 within-category external-precedent pairs. Node colours distinguish recorded manufacturers. Radius indicates generation, not elapsed time.}
\end{figure}
\clearpage
\begin{figure}[p]\centering
\includegraphics[width=.98\linewidth,height=.54\textheight,keepaspectratio]{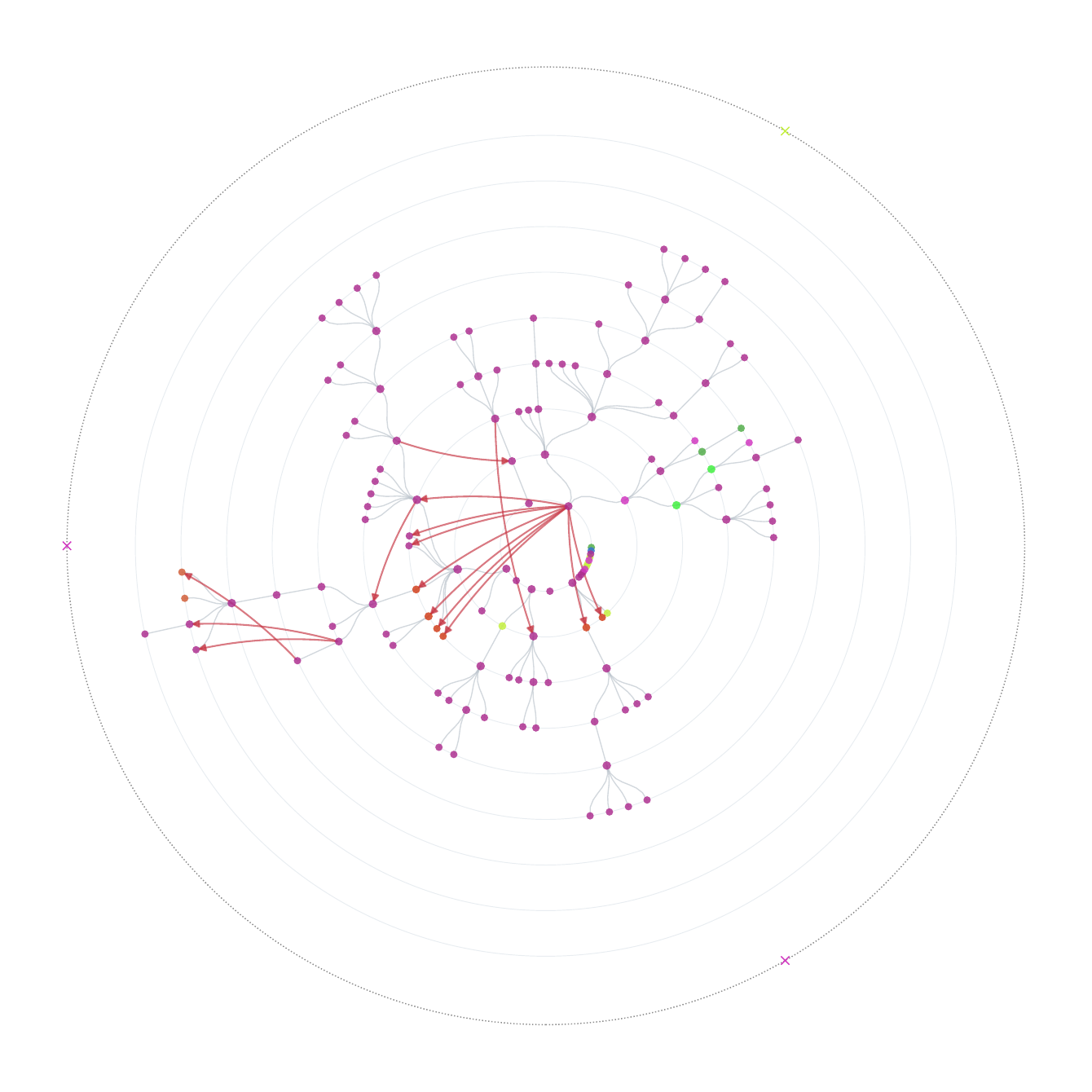}
\caption{\textbf{16-bit Microcontrollers - MCU.} All 137 design families, 116 backbone links, and 15 within-subcategory candidate-HT donor--recipient pairs are shown. Recorded first appearances span 2002--2023. Node colours identify manufacturers; grey links are inferred backbone connections and red arrows are candidate HT links. Radius measures backbone generation, not calendar time. The 3 undated families are isolated crosses on the outer dotted ring, not identified historical founders. Cross-subcategory links and the optional numerical-VI overlay are omitted.}\label{app:b_network_16_bit_microcontrollers_mcu}
\alttxt{Radial network for 16-bit Microcontrollers - MCU, showing 137 design families, 116 inferred backbone links and 15 within-category external-precedent pairs. Node colours distinguish recorded manufacturers. Radius indicates generation, not elapsed time.}
\end{figure}
\clearpage
\begin{figure}[p]\centering
\includegraphics[width=.98\linewidth,height=.54\textheight,keepaspectratio]{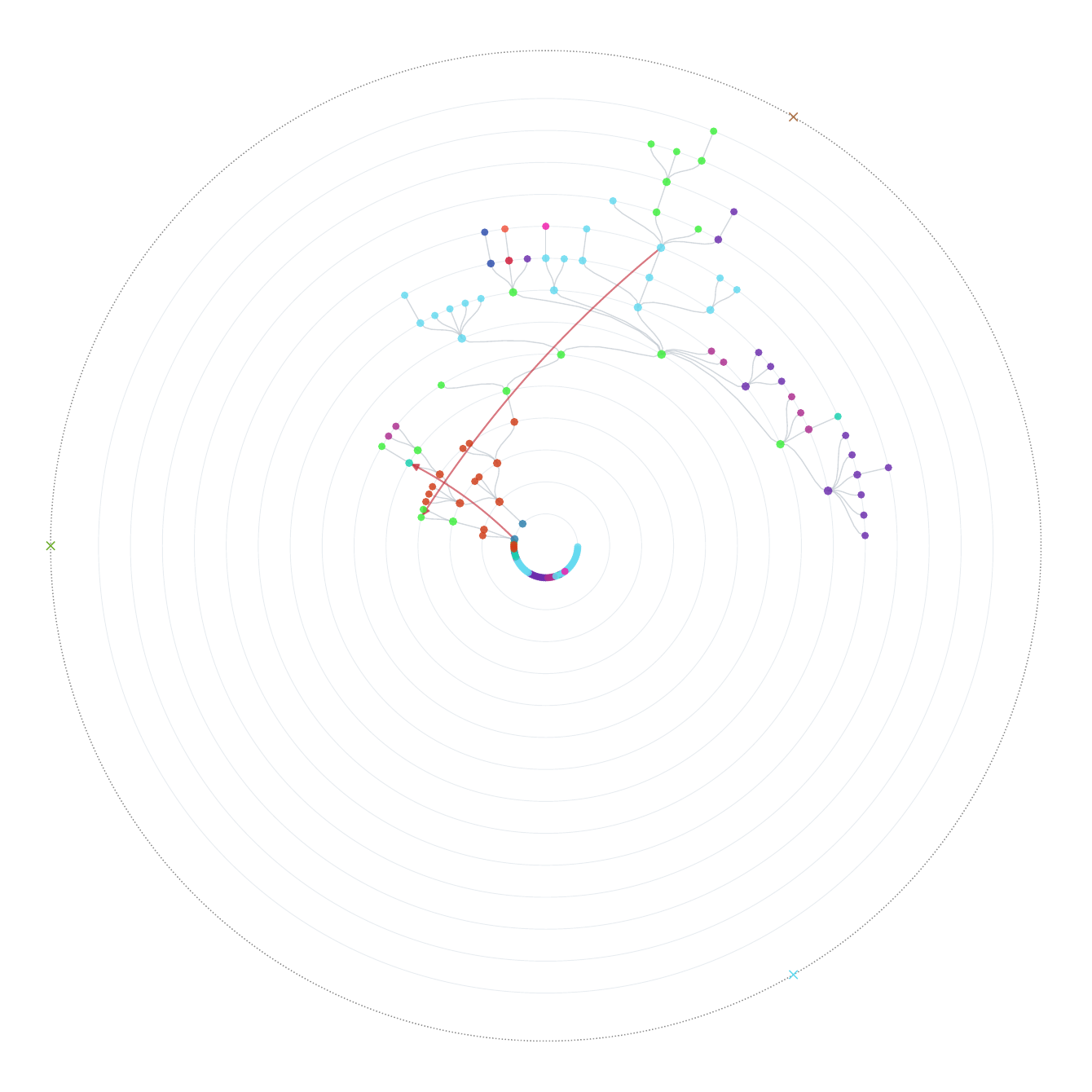}
\caption{\textbf{8-bit Microcontrollers - MCU.} All 136 design families, 80 backbone links, and 2 within-subcategory candidate-HT donor--recipient pairs are shown. Recorded first appearances span 1985--2022. Node colours identify manufacturers; grey links are inferred backbone connections and red arrows are candidate HT links. Radius measures backbone generation, not calendar time. The 3 undated families are isolated crosses on the outer dotted ring, not identified historical founders. Cross-subcategory links and the optional numerical-VI overlay are omitted.}\label{app:b_network_8_bit_microcontrollers_mcu}
\alttxt{Radial network for 8-bit Microcontrollers - MCU, showing 136 design families, 80 inferred backbone links and 2 within-category external-precedent pairs. Node colours distinguish recorded manufacturers. Radius indicates generation, not elapsed time.}
\end{figure}
\clearpage
\begin{figure}[p]\centering
\includegraphics[width=.98\linewidth,height=.54\textheight,keepaspectratio]{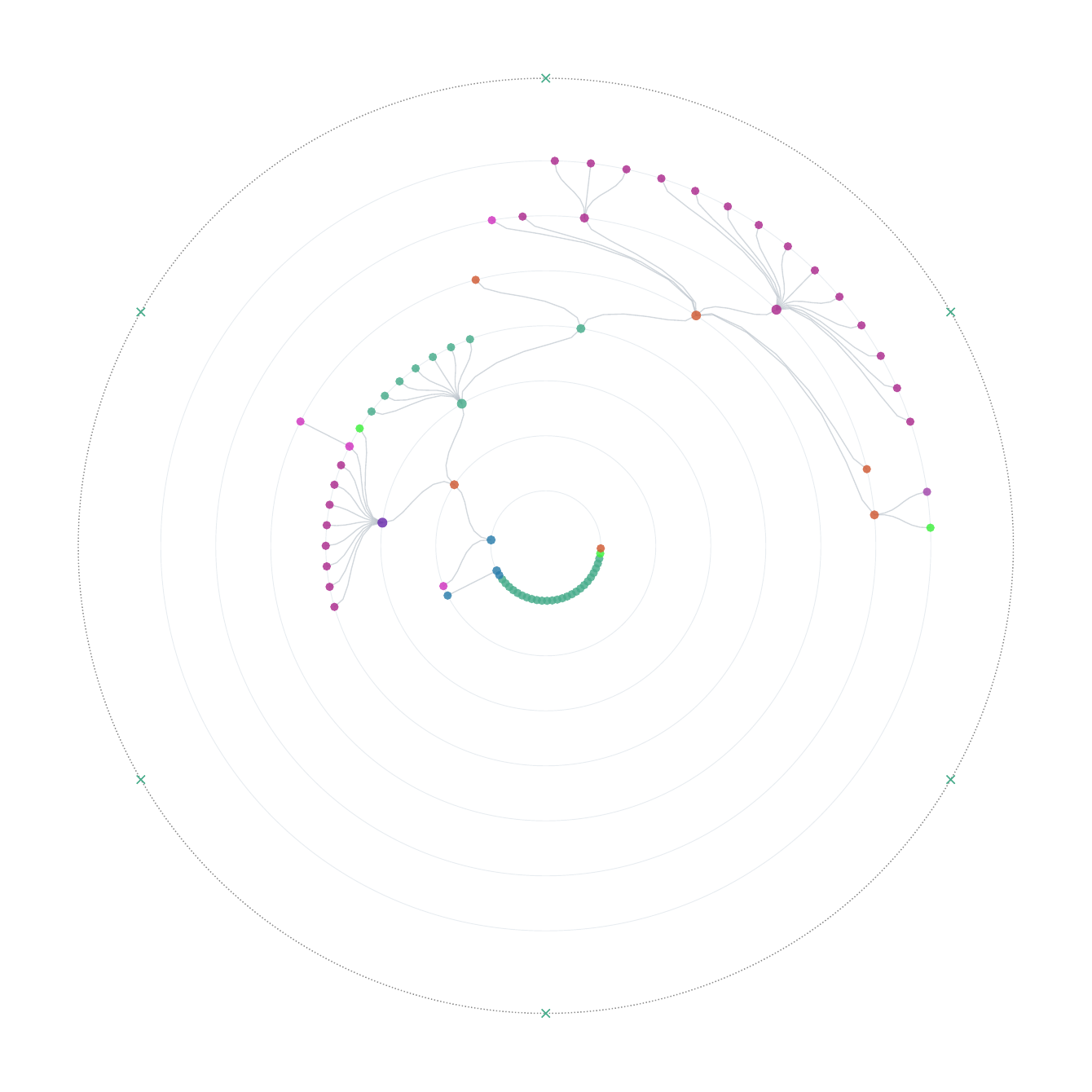}
\caption{\textbf{Microprocessors - MPU.} All 84 design families, 48 backbone links, and 0 within-subcategory candidate-HT donor--recipient pairs are shown. Recorded first appearances span 1984--2023. Node colours identify manufacturers; grey links are inferred backbone connections and red arrows are candidate HT links. Radius measures backbone generation, not calendar time. The 6 undated families are isolated crosses on the outer dotted ring, not identified historical founders. Cross-subcategory links and the optional numerical-VI overlay are omitted.}\label{app:b_network_microprocessors_mpu}
\alttxt{Radial network for Microprocessors - MPU, showing 84 design families, 48 inferred backbone links and 0 within-category external-precedent pairs. Node colours distinguish recorded manufacturers. Radius indicates generation, not elapsed time.}
\end{figure}
\clearpage
\begin{figure}[p]\centering
\includegraphics[width=.98\linewidth,height=.54\textheight,keepaspectratio]{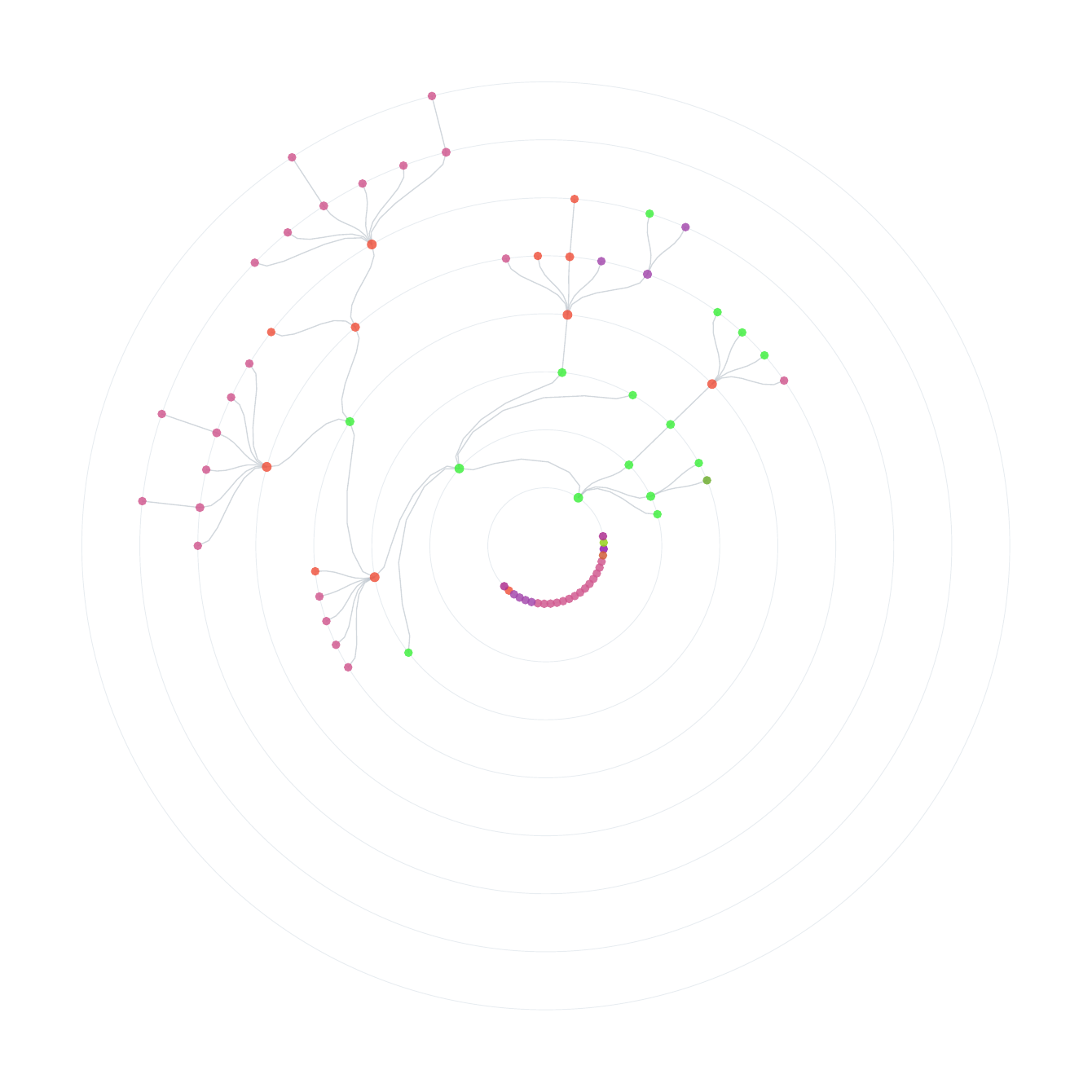}
\caption{\textbf{EEPROM.} All 76 design families, 51 backbone links, and 0 within-subcategory candidate-HT donor--recipient pairs are shown. Recorded first appearances span 1987--2022. Node colours identify manufacturers; grey links are inferred backbone connections and red arrows are candidate HT links. Radius measures backbone generation, not calendar time. Cross-subcategory links and the optional numerical-VI overlay are omitted.}\label{app:b_network_eeprom}
\alttxt{Radial network for EEPROM, showing 76 design families, 51 inferred backbone links and 0 within-category external-precedent pairs. Node colours distinguish recorded manufacturers. Radius indicates generation, not elapsed time.}
\end{figure}
\clearpage
\begin{figure}[p]\centering
\includegraphics[width=.98\linewidth,height=.54\textheight,keepaspectratio]{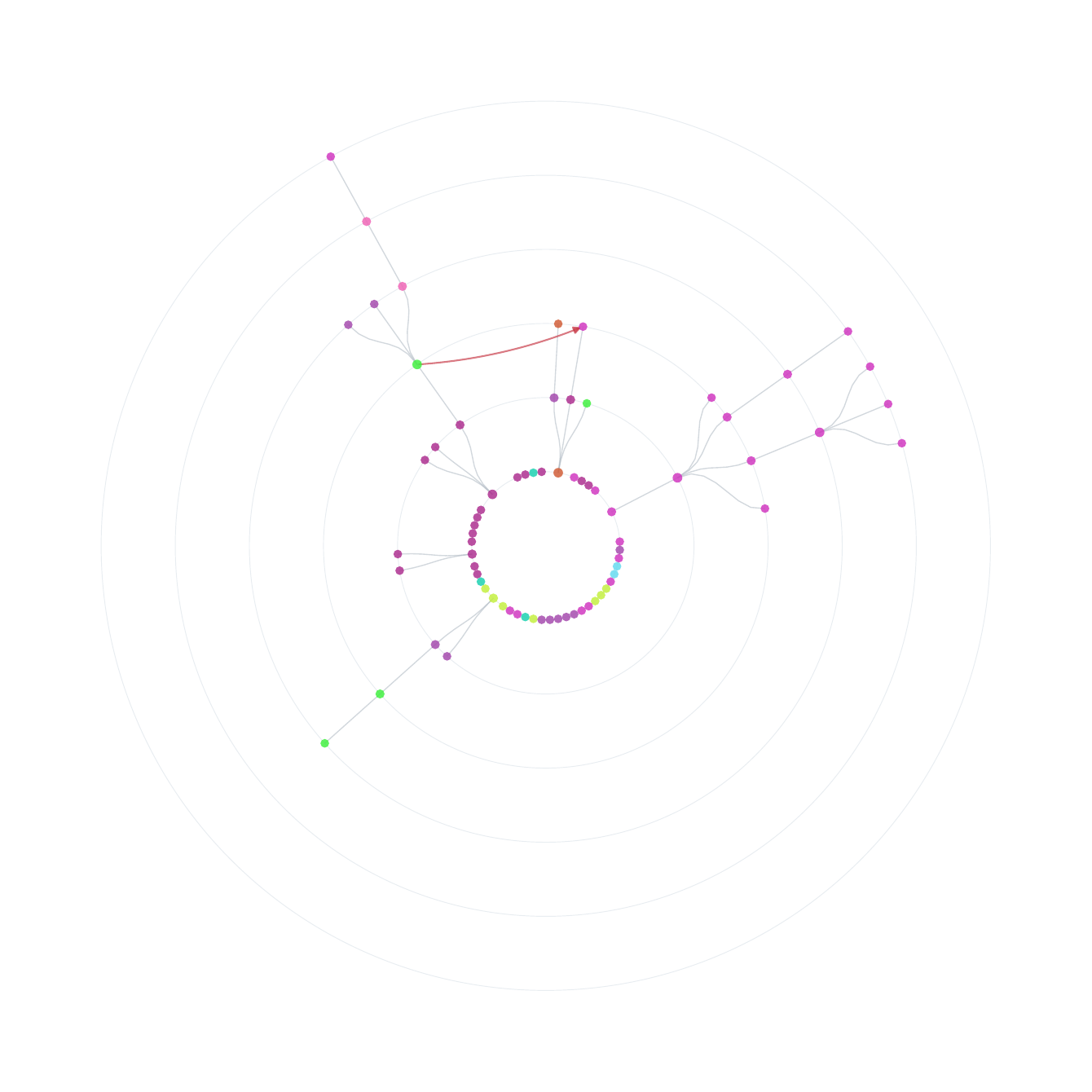}
\caption{\textbf{32-bit Microcontrollers - MCU.} All 74 design families, 31 backbone links, and 1 within-subcategory candidate-HT donor--recipient pair are shown. Recorded first appearances span 2001--2023. Node colours identify manufacturers; grey links are inferred backbone connections and red arrows are candidate HT links. Radius measures backbone generation, not calendar time. Cross-subcategory links and the optional numerical-VI overlay are omitted.}\label{app:b_network_32_bit_microcontrollers_mcu}
\alttxt{Radial network for 32-bit Microcontrollers - MCU, showing 74 design families, 31 inferred backbone links and 1 within-category external-precedent pairs. Node colours distinguish recorded manufacturers. Radius indicates generation, not elapsed time.}
\end{figure}
\clearpage
\begin{figure}[p]\centering
\includegraphics[width=.98\linewidth,height=.54\textheight,keepaspectratio]{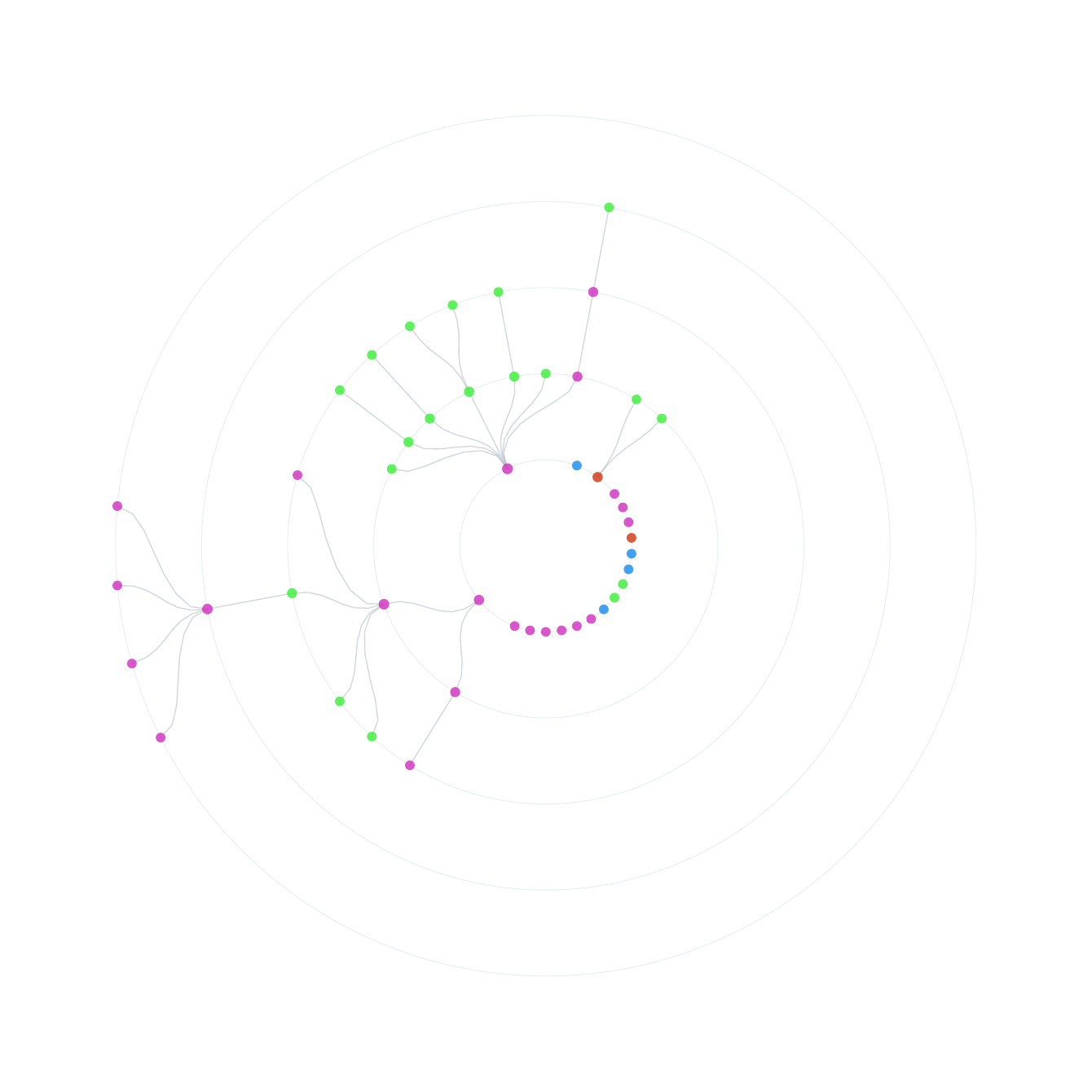}
\caption{\textbf{Digital Signal Processors \& Controllers - DSP, DSC.} All 47 design families, 28 backbone links, and 0 within-subcategory candidate-HT donor--recipient pairs are shown. Recorded first appearances span 1989--2023. Node colours identify manufacturers; grey links are inferred backbone connections and red arrows are candidate HT links. Radius measures backbone generation, not calendar time. Cross-subcategory links and the optional numerical-VI overlay are omitted.}\label{app:b_network_digital_signal_processors_controllers_dsp_dsc}
\alttxt{Radial network for Digital Signal Processors \& Controllers - DSP, DSC, showing 47 design families, 28 inferred backbone links and 0 within-category external-precedent pairs. Node colours distinguish recorded manufacturers. Radius indicates generation, not elapsed time.}
\end{figure}
\clearpage
\begin{figure}[p]\centering
\includegraphics[width=.98\linewidth,height=.54\textheight,keepaspectratio]{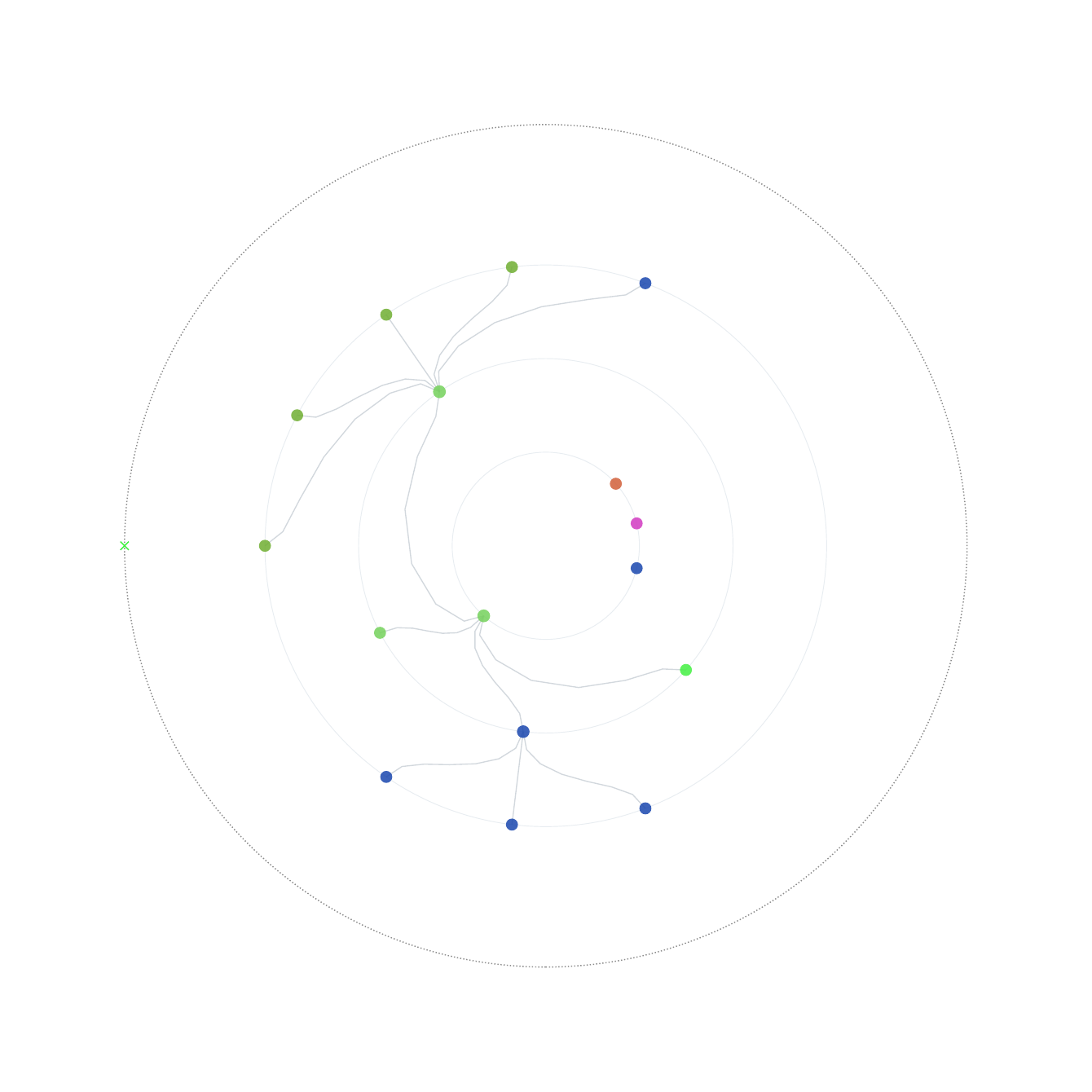}
\caption{\textbf{Processors - Application Specialized.} All 17 design families, 12 backbone links, and 0 within-subcategory candidate-HT donor--recipient pairs are shown. Recorded first appearances span 2015--2023. Node colours identify manufacturers; grey links are inferred backbone connections and red arrows are candidate HT links. Radius measures backbone generation, not calendar time. The single undated family is an isolated cross on the outer dotted ring, not an identified historical founder. Cross-subcategory links and the optional numerical-VI overlay are omitted.}\label{app:b_network_processors_application_specialized}
\alttxt{Radial network for Processors - Application Specialized, showing 17 design families, 12 inferred backbone links and 0 within-category external-precedent pairs. Node colours distinguish recorded manufacturers. Radius indicates generation, not elapsed time.}
\end{figure}
\clearpage
\begin{figure}[p]\centering
\includegraphics[width=.98\linewidth,height=.54\textheight,keepaspectratio]{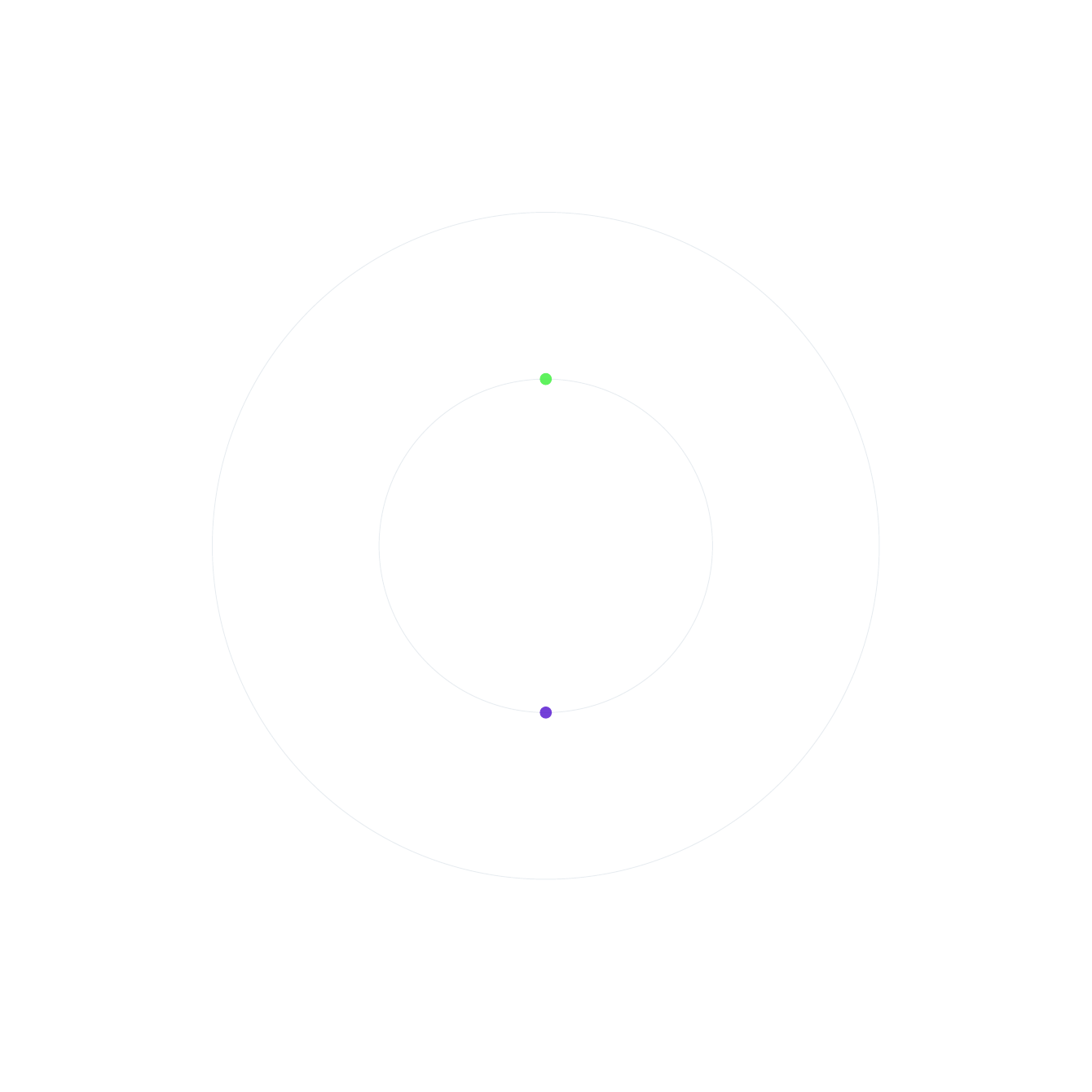}
\caption{\textbf{CPLD - Complex Programmable Logic Devices.} All 2 design families, 0 backbone links, and 0 within-subcategory candidate-HT donor--recipient pairs are shown. Recorded first appearances span 2002--2004. Node colours identify manufacturers; grey links are inferred backbone connections and red arrows are candidate HT links. Radius measures backbone generation, not calendar time. No admissible backbone link joins the observed families. Cross-subcategory links and the optional numerical-VI overlay are omitted.}\label{app:b_network_cpld_complex_programmable_logic_devices}
\alttxt{Radial network for CPLD - Complex Programmable Logic Devices, showing 2 design families, 0 inferred backbone links and 0 within-category external-precedent pairs. Node colours distinguish recorded manufacturers. Radius indicates generation, not elapsed time.}
\end{figure}
\clearpage

\FloatBarrier\clearpage

\section{Supplementary combinations of first-appearance and carriage predictions}\label{app:subcategory_scenarios}\label{app:revised_ranges}
This appendix combines the first-appearance and subsequent-carriage models on specified target profiles. The calculations illustrate the assumptions required to move from conditional associations to probability-weighted carriage. The scenarios combine pooled regression relationships on explicitly defined product profiles.

\subsection{Construction and interpretation of the scenarios}\label{app:scenario_methods}
The calculation combines a first-appearance probability with a conditional expected carriage count. This asks a different question from either conditional model alone. On the same retained target profiles $\mathcal R_c$ in subcategory $c$, we calculate two separate scenario curves,
\begin{equation}\label{eq:joint-within}
Q_{s,c}(a)=\frac{1}{|\mathcal R_c|}\sum_{r\in\mathcal R_c}\widehat p(a,X_{T,r})\widehat\mu_s(a,X_{s,r}),\qquad s\in\{W,D\}.
\end{equation}
Predictions are multiplied within target profiles before averaging. Within-family and descendant outcomes are separate consequences, not additional factors in a three-term product. Applying the follow-up models to a common opportunity population requires transport between the outcome-specific samples.

Targets must have observed capability--manufacturer--subcategory combinations in all participating fits and estimable full-design predictions. Their raw Anchor grid intersects each stage's subcategory-specific 5th--95th percentile range. It also respects $0\leq a/\log(1+G_i^-)\leq1$ for every retained target. Controls and target membership remain fixed as Anchor varies. Because generations remain fixed, the scenario varies implied continuity rather than simultaneously intervening on continuity and lineage length. Continuous-control extrapolation is flagged; these restrictions do not prove full multivariate overlap.

A descriptive near-maximum set retains values for which $Q_{s,c}$ is at least 95\% of its grid maximum. The common range additionally requires both outcomes to meet this criterion and the mean detection slope to be nonpositive. The 95\% threshold is a numerical tolerance, not a confidence level. Boundary maxima and unsupported subcategories are reported explicitly. The models pool Anchor coefficients across categories; the scenarios reflect category intercepts, covariate profiles and support, rather than separately estimated subcategory-specific Anchor effects.

\subsection{Point estimates and their support}
Among the 19 subcategories, only RF SoC permits a three-model scenario under the stated support restrictions. Pairwise detection--descendant scenarios are additionally available for 16-bit and RF microcontrollers. ARM microcontrollers have inadequate Anchor variation in a required stage, while the remaining cases lack the necessary stage support. An unavailable interval indicates that the required fitted-stage support is missing for this calculation.

\begin{figure}[tbp]\centering
\includegraphics[width=\linewidth]{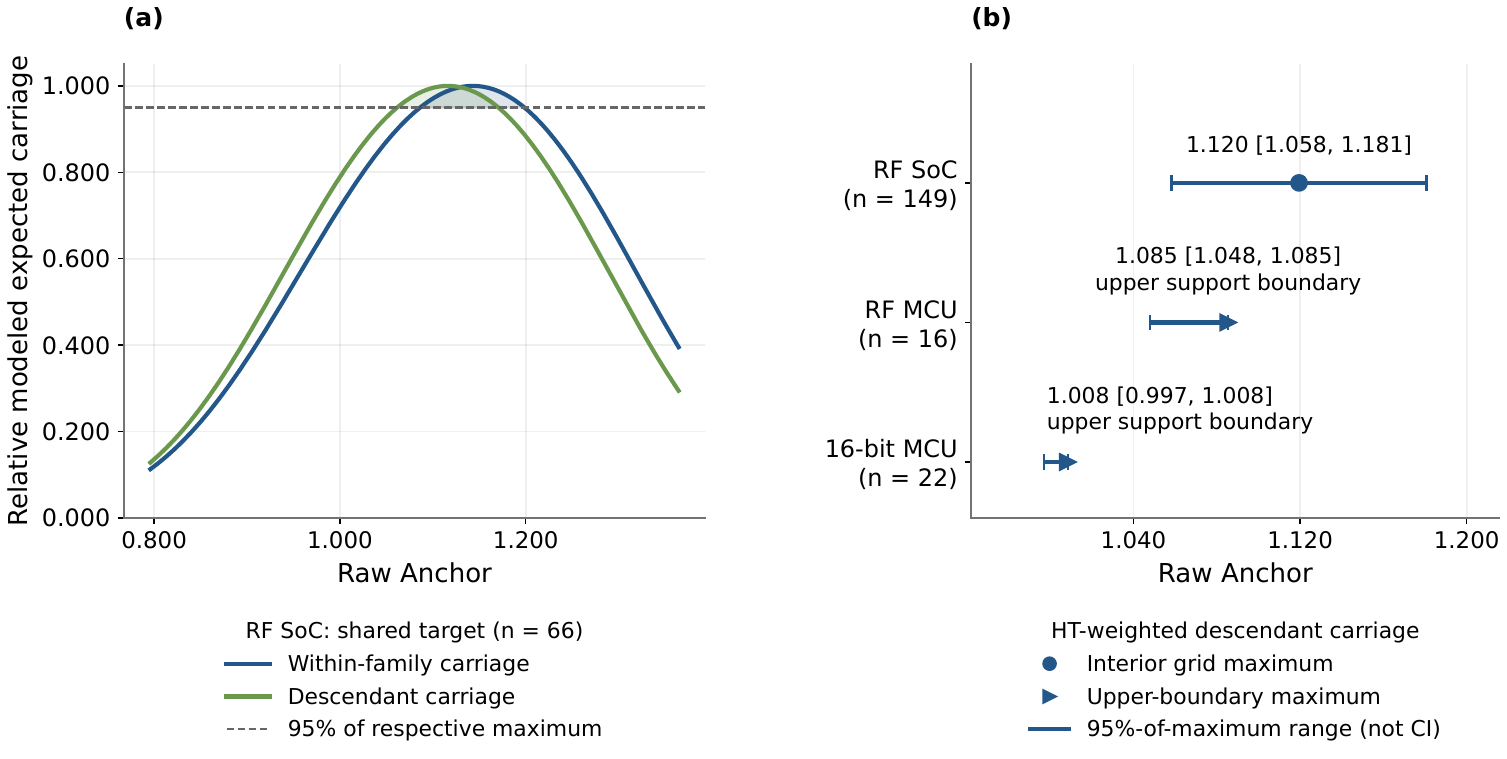}
\caption{\textbf{Acquisition-weighted carriage and supported Raw Anchor peaks.} (a) RF SoC curves on 66 shared three-stage targets, normalised by their permitted-grid maxima. The dashed line marks 95\% of each maximum; Eq.~\eqref{eq:rf-range} additionally requires nonincreasing HT. (b) Descendant-carriage peaks on pairwise targets, ordered from highest to lowest. The circle marks an interior grid maximum; right-pointing triangles mark upper-boundary maxima. Horizontal bars show 95\%-of-maximum near-maximum ranges, not confidence intervals. Panel (b) uses 149 RF SoC targets and supports only three subcategories; it does not impose the declining-HT restriction. Both panels use pooled stage models.}\label{fig:joint-performance}
\alttxt{The left panel shows two normalised RF SoC probability-weighted carriage curves on 66 shared target profiles. Both have interior grid maxima. The right panel compares descendant-carriage grid maxima for RF SoC and two MCU categories; both MCU maxima occur at the upper support boundary. Horizontal ranges are not confidence intervals.}
\end{figure}

For the common RF SoC target of 66 opportunity profiles, the HT-weighted within-family curve peaks at $C=1.143$, and the HT-weighted descendant curve at $C=1.116$. Their respective 95\%-of-maximum intervals are 1.087--1.198 and 1.062--1.170. Requiring both conditions and nonincreasing detection gives the common interval
\begin{equation}\label{eq:rf-range}
1.104\leq C\leq1.170.
\end{equation}
Figure~\ref{fig:joint-performance}(a) illustrates this model-based composition. Over this region, lower detection can offset higher conditional follow-up; neither every conditional stage nor every subcategory needs to exhibit an inverse U for such a product to have an interior peak.

For comparison across subcategories, Figure~\ref{fig:joint-performance}(b) uses one common objective: the permitted-grid maximum of acquisition-weighted descendant carriage, $Q_{D,c}(C)$, on each category's pairwise detection--descendant target. It orders the supported peaks from highest to lowest: RF SoC ($C=1.120$), RF MCU ($C=1.085$), and 16-bit MCU ($C=1.008$). Horizontal ranges retain at least 95\% of that category's fitted maximum; they are numerical tolerances rather than confidence intervals. The RF SoC pairwise target differs from the three-stage target in panel (a), giving slightly different peaks and ranges. The other 16 subcategories lack supported pairwise scenarios and have no plotted scenario maximum.

\begin{table}[tbp]\centering\begin{threeparttable}
\caption{Subcategory scenarios and descriptive near-maximum Anchor ranges}\label{tab:selected_optimal_ranges}
\small\doublespacing\setlength{\tabcolsep}{3pt}
\begin{tabularx}{\linewidth}{@{}p{.20\linewidth}Xrr@{}}\toprule
Subcategory & Outcome and target & Peak $C$ & 95\%-of-peak range\\\midrule
RF SoC & Within-family; joint & 1.143 & 1.087--1.198\\
RF SoC & Descendants; joint & 1.116 & 1.062--1.170\\
RF SoC & Descendants; pairwise & 1.120 & 1.058--1.181\\
16-bit MCU & Descendants; pairwise & 1.008 & 0.997--1.008\\
RF MCU & Descendants; pairwise & 1.085 & 1.048--1.085\\\midrule
RF SoC & Both outcomes and declining HT & --- & 1.104--1.170\\\bottomrule
\end{tabularx}
\begin{tablenotes}\footnotesize\doublespacing
\item All values use raw $C=A^-\log(1+G^-)$ and a 401-point permitted grid. Displayed values are rounded to three decimals; range membership uses unrounded values. RF SoC joint curves share 66 targets; its pairwise descendant curve uses 149. Pairwise MCU curves use 22 and 16 targets respectively, peak at the upper permitted boundary, and contain no declining-HT region. The 95\% threshold is a numerical tolerance, not a confidence interval. Scenario construction uses pooled coefficients; the support diagnostics are reported below.
\end{tablenotes}\end{threeparttable}\end{table}

The RF SoC target contains 66 profiles from one root; 36 lie outside at least one stage's subcategory-specific range for a continuous control, and one positive RF SoC event appears in all three fitted samples. In the MCU pairwise scenarios, the maxima occur at the upper boundaries and HT increases throughout permitted support. These are model-based compositions on the stated targets; their near-maximum intervals summarise fitted curves rather than product-design optima.
The following support register covers all 19 categories. Subsequent figures present the category-specific scenarios.
\FloatBarrier

\subsection{Complete fitted-stage coverage and category figures}
\begingroup\small\doublespacing\setlength{\tabcolsep}{4pt}
\begin{longtable}{@{}>{\raggedright\arraybackslash}p{.48\linewidth}rrrrrr@{}}
\caption{Fitted-stage coverage and scenario availability for all subcategories}\label{app:c_support}\\
\toprule
 & \multicolumn{3}{c}{Fitted observations} & \multicolumn{3}{c}{Scenario scope}\\
Subcategory & HT & Within & Desc. & W & D & J\\\midrule
\endfirsthead
\caption[]{Fitted-stage coverage and scenario availability (continued)}\\
\toprule
 & \multicolumn{3}{c}{Fitted observations} & \multicolumn{3}{c}{Scenario scope}\\
Subcategory & HT & Within & Desc. & W & D & J\\\midrule
\endhead
\midrule\multicolumn{7}{r}{\emph{Continued on next page}}\\\endfoot
\bottomrule\endlastfoot

16-bit MCU & 43 & 0 & 3 & M & A & M\\
32-bit MCU & 0 & 0 & 0 & M & M & M\\
8-bit MCU & 0 & 4 & 8 & M & M & M\\
ARM MCU & 75 & 1 & 1 & V & V & V\\
Application-specific MCU & 0 & 141 & 145 & M & M & M\\
CPLD & 0 & 0 & 0 & M & M & M\\
CPU & 0 & 0 & 0 & M & M & M\\
DRAM & 0 & 0 & 0 & M & M & M\\
DSP / DSC & 0 & 0 & 0 & M & M & M\\
EEPROM & 0 & 0 & 0 & M & M & M\\
FPGA & 0 & 0 & 0 & M & M & M\\
Microcontrollers & 0 & 0 & 0 & M & M & M\\
MPU & 0 & 0 & 0 & M & M & M\\
NAND Flash & 0 & 0 & 1 & M & M & M\\
NOR Flash & 0 & 0 & 1 & M & M & M\\
Application-specialized processors & 0 & 1 & 0 & M & M & M\\
RF MCU & 48 & 0 & 6 & M & A & M\\
RF SoC & 223 & 5 & 14 & A & A & A\\
SRAM & 0 & 0 & 0 & M & M & M\\

\end{longtable}
\noindent{\footnotesize\doublespacing HT, Within and Desc. count fitted observations, not families or target profiles. W pairs HT with Within; D pairs HT with Desc.; J requires all three stages. A: a descriptive scenario is available. M: a required fitted stage is missing. V: a required stage contains only one Anchor value. Only categories meeting the specified stage-availability rule receive scenario curves. Covariate overlap and cluster support are assessed separately.\par}
\endgroup

\FloatBarrier

\begin{figure}[tbp]\centering

\includegraphics[width=\linewidth]{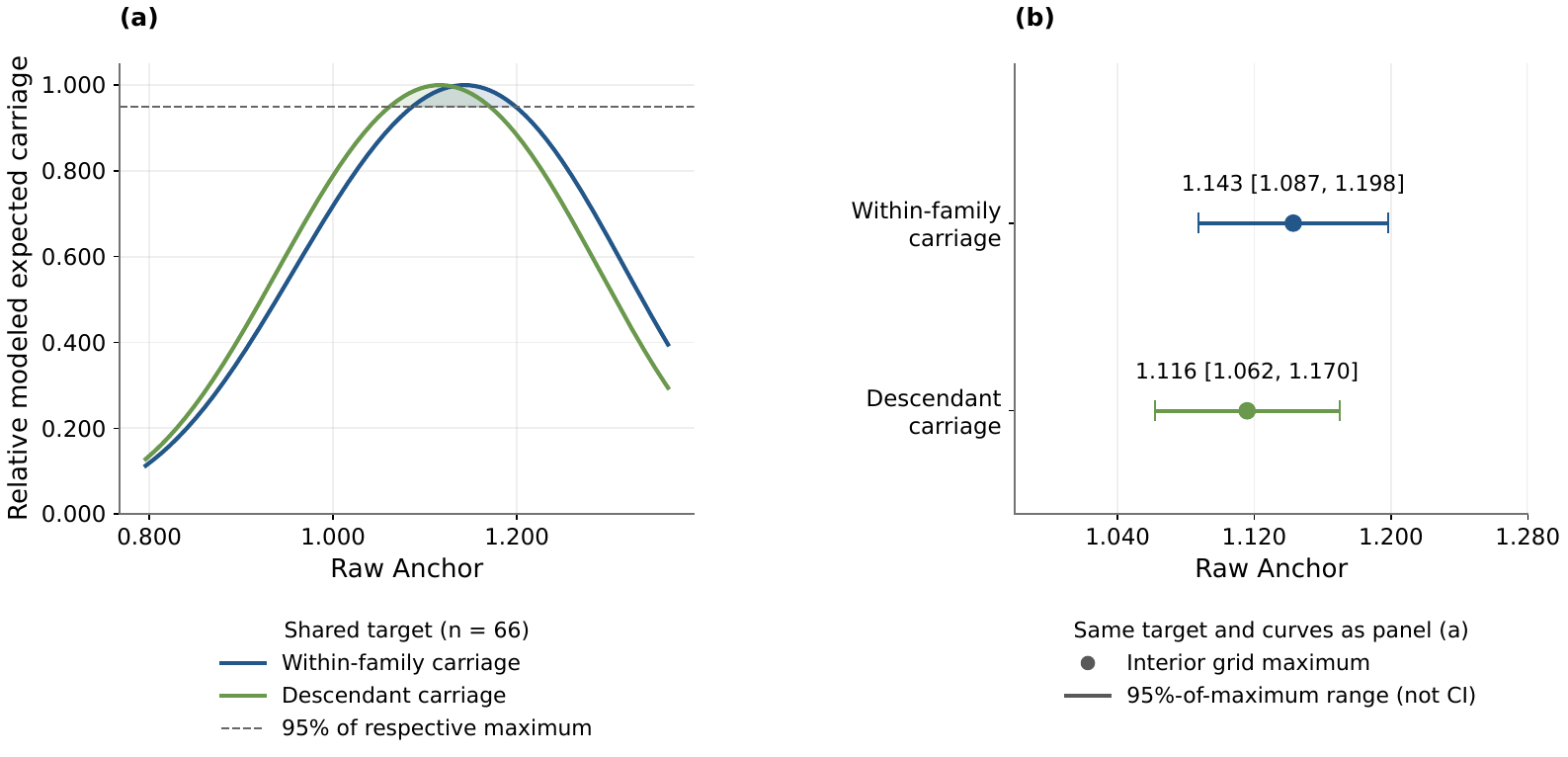}

\caption{\textbf{RF SoC: shared-target Anchor scenarios.} Both panels use the same 66 target rows, rather than the 149-row pairwise descendant target used in Figure~\ref{fig:joint-performance}(b). Within-family and descendant carriage peak at raw Anchor 1.143 and 1.116, respectively. Their joint 95\%-of-maximum set with nonpositive HT slope is [1.104, 1.170]. The target has one root; joint covariate overlap is unestablished. Horizontal ranges are descriptive near-maximum sets, not confidence intervals.}\label{app:c_rf_soc}

\alttxt{Normalised within-family and descendant scenario curves for the same 66 RF SoC targets peak at raw Anchor 1.143 and 1.116. The second panel marks these maxima and their descriptive near-maximum ranges. The targets belong to one root; the displayed ranges are not confidence intervals.}
\end{figure}

\FloatBarrier

\begin{figure}[tbp]\centering

\includegraphics[width=\linewidth]{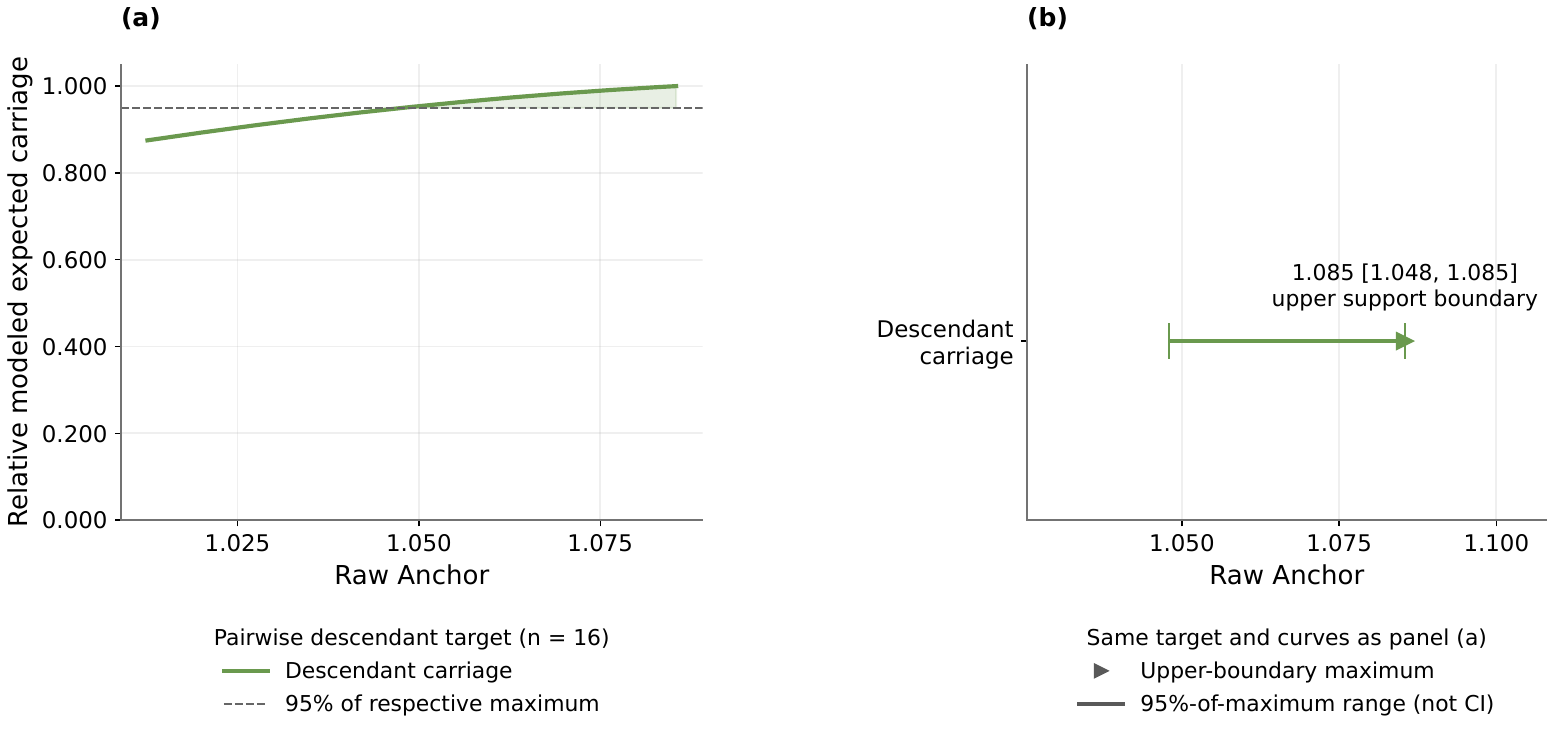}

\caption{\textbf{RF MCU: acquisition--descendant scenarios.} Both panels use the same 16 target rows. Descendant carriage peaks at raw Anchor 1.085, with a 95\%-of-maximum range of [1.048, 1.085]. The triangle marks the upper support boundary, so the grid does not identify an interior optimum. Recorded HT probability increases throughout this supported interval. Within-family carriage is not estimated because this category is absent from that fitted stage. The ranges are not confidence intervals.}\label{app:c_rf_mcu}

\alttxt{The RF MCU descendant scenario increases across the supported Anchor range and reaches its maximum at raw Anchor 1.085, the upper support boundary. The second panel marks that boundary maximum and the range 1.048--1.085. Only 16 target profiles are represented; no within-family curve is available.}
\end{figure}

\FloatBarrier

\begin{figure}[tbp]\centering

\includegraphics[width=\linewidth]{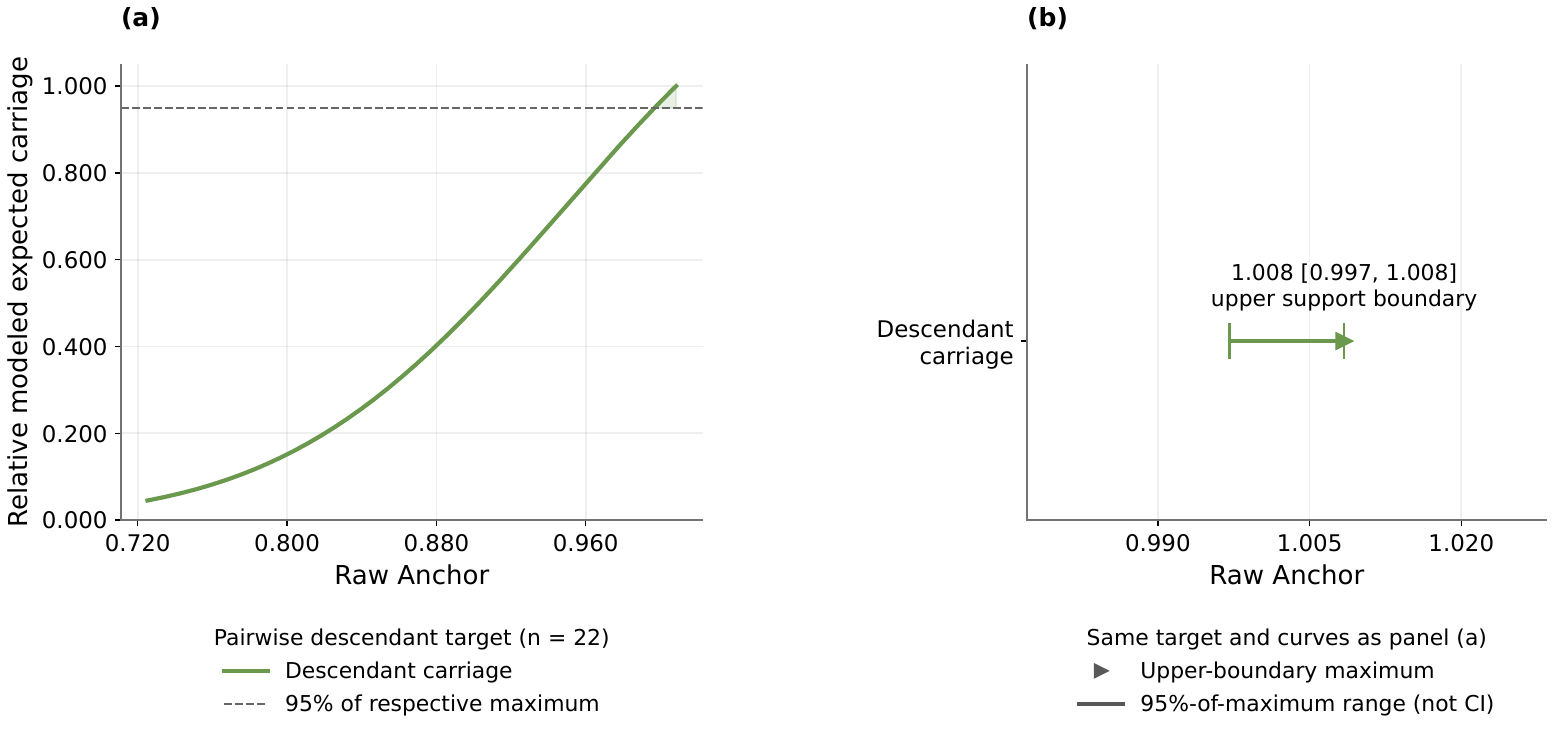}

\caption{\textbf{16-bit MCU: acquisition--descendant scenarios.} Both panels use the same 22 target rows. Descendant carriage peaks at raw Anchor 1.008, with a 95\%-of-maximum range of [0.997, 1.008]. The triangle marks the upper support boundary, so the grid does not identify an interior optimum. Recorded HT probability increases throughout this supported interval. Within-family carriage is not estimated because this category is absent from that fitted stage. The ranges are not confidence intervals.}\label{app:c_16bit_mcu}

\alttxt{The 16-bit MCU descendant scenario increases over its supported range, reaching raw Anchor 1.008 at the upper boundary. The second panel shows the near-maximum range 0.997--1.008. There are 22 target profiles and no within-family scenario.}
\end{figure}

\FloatBarrier

The scenario register records both numerical availability and target support. All target rows lie in a single root for RF SoC and RF MCU and in two roots for 16-bit MCU. Thirty of 66 RF-SoC shared-target rows, five of 16 RF-MCU rows, and one of 22 16-bit-MCU rows lie within every stage-specific marginal control range. These are marginal rather than joint overlap checks. RF MCU and 16-bit MCU have increasing HT curves and upper-boundary carriage maxima within their permitted intervals.

\source{Notebook 09; saved \code{curves.csv}, \code{near_optimal_ranges.csv}, \code{subcategory_support.csv}, and \code{joint_tradeoff_ranges.csv}; presentation and point-estimate checks in \code{build_appendix_anchor_scenarios.py}.}

\FloatBarrier\clearpage

\section{Reconstruction and statistical robustness}\label{app:robustness}
\subsection{Reconstruction assumptions}
The baseline's mean selected similarity is 0.983. Scores are approximately one for 85.22\% of linked families, and 58.26\% have tied best-score candidates. The least-supported transported match uses exactly two comparable fields for 42.53\% of linked families; 44.48\% compare less than one quarter of available historical entropy weight. Parent selection follows the deterministic support, date and identifier tie-breaking rule described in the main article.

\begin{figure}[p]\centering
\includegraphics[width=\linewidth,height=.54\textheight,keepaspectratio]{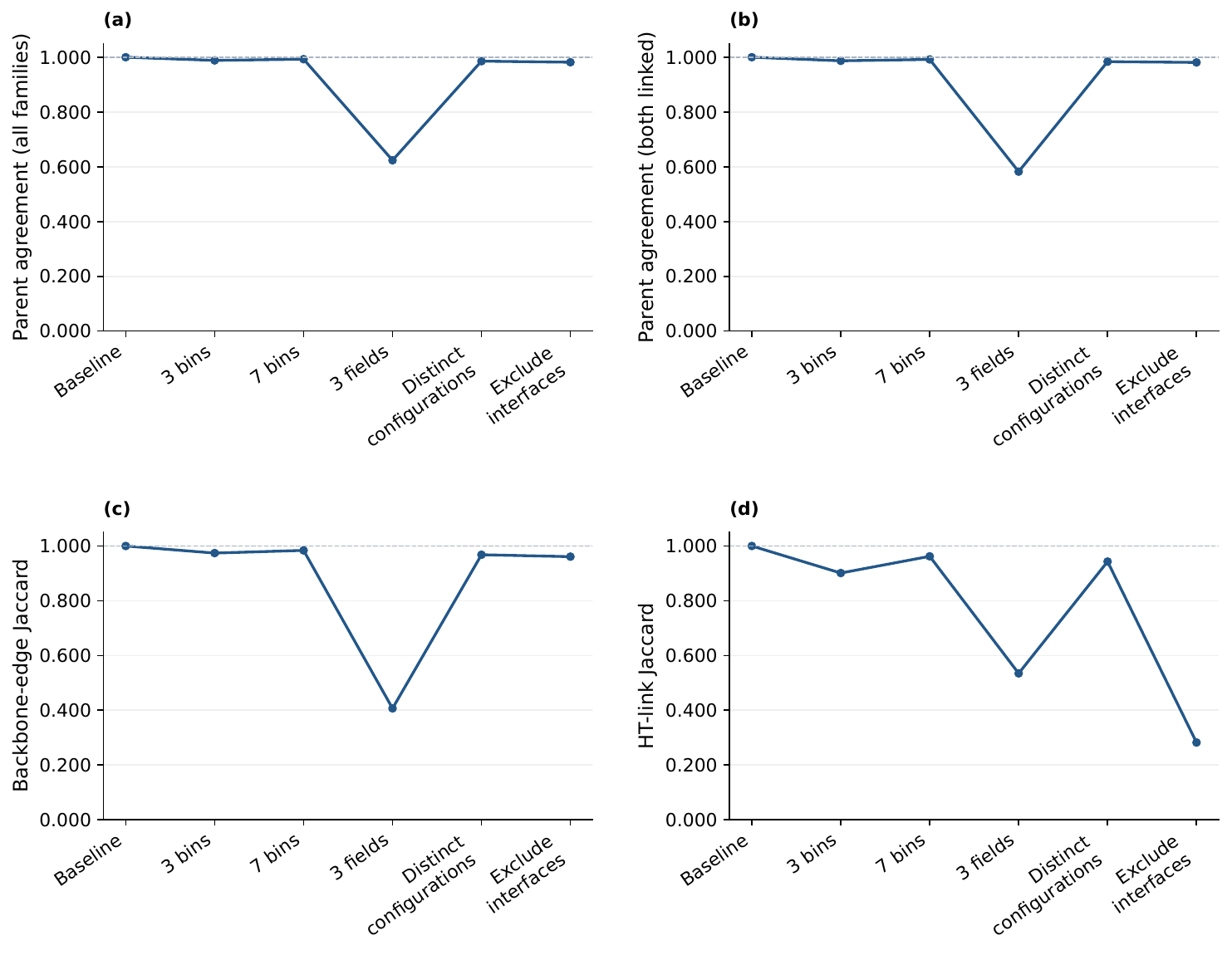}
\caption{\textbf{Reconstruction sensitivity.} The four panels report: (a) parent agreement over all families, (b) agreement among families linked in both reconstructions, (c) backbone-edge Jaccard, and (d) distinct HT-link Jaccard. Values compare each setting with the baseline. Parent agreement includes unchanged roots in (a), whereas (b) conditions on retained linkage.}\label{app:d_reconstruction}
\alttxt{Four panels compare reconstruction settings. Parent agreement remains high under alternative entropy bins and distinct-configuration weighting but declines when three comparable fields are required. Excluding interface fields leaves most parents unchanged while sharply reducing external-precedent link overlap. The panels measure different aspects of network stability.}
\end{figure}

Figure~\ref{app:d_reconstruction} reports reconstruction diagnostics. Each setting changes one declared part of the baseline reconstruction while retaining the same 45,996 family identities and launch-year SKU evidence. The baseline uses five numerical entropy bins, at least two prior observed families and two comparable positive-entropy fields, equal mass for observed launch SKUs, and the full retained feature vocabulary. Alternatives use three or seven bins, require three comparable fields, give equal mass to distinct SKU configurations, or exclude the two interface fields from similarity ranking. Candidate parents and donors are recomputed; the named-capability parser is held fixed.

\FloatBarrier

\begingroup\small\doublespacing\setlength{\tabcolsep}{3pt}
\begin{longtable}{@{}>{\raggedright\arraybackslash}p{0.260\linewidth}>{\raggedright\arraybackslash}p{0.130\linewidth}>{\raggedright\arraybackslash}p{0.130\linewidth}>{\raggedright\arraybackslash}p{0.130\linewidth}>{\raggedright\arraybackslash}p{0.130\linewidth}>{\raggedright\arraybackslash}p{0.120\linewidth}@{}}
\caption{Reconstruction sensitivity under the six declared settings}\label{tab:appendix_d_network_metrics}\\
\toprule Setting & Backbone links & All-parent agreement & Both-linked agreement & Backbone Jaccard & HT Jaccard \\ \midrule\endfirsthead
\toprule Setting & Backbone links & All-parent agreement & Both-linked agreement & Backbone Jaccard & HT Jaccard \\ \midrule\endhead
Baseline & 40,713 & 1.000 & 1.000 & 1.000 & 1.000 \\
Three entropy bins & 40,713 & 0.988 & 0.987 & 0.974 & 0.901 \\
Seven entropy bins & 40,713 & 0.993 & 0.992 & 0.984 & 0.962 \\
Three comparable fields & 40,191 & 0.624 & 0.582 & 0.407 & 0.535 \\
Distinct-configuration mass & 40,713 & 0.986 & 0.984 & 0.968 & 0.943 \\
Interface fields excluded & 40,644 & 0.982 & 0.981 & 0.961 & 0.282 \\
\bottomrule\end{longtable}\endgroup
\noindent{\footnotesize Agreement and Jaccard values are proportions. All-parent agreement includes unchanged roots; both-linked agreement conditions on a parent in both reconstructions. HT metrics compare distinct directed donor-recipient pairs.}

Changing bin count preserves 0.988--0.993 of parent assignments; distinct-configuration weighting preserves 0.986. Requiring three fields lowers all-family agreement to 0.624 and agreement among commonly linked families to 0.582, leaving 40,191 backbone links rather than 40,713. Excluding interface fields preserves 0.982 of parent assignments but yields HT-link Jaccard 0.282. Thus aggregate continuity and external-precedent assignment respond differently to the information available for comparison. These checks compare the network reconstructions; downstream regressions are not recomputed under every alternative graph.

\subsection{Global main-path sensitivity}
\begingroup\small\doublespacing\setlength{\tabcolsep}{3pt}
\begin{longtable}{@{}>{\raggedright\arraybackslash}p{0.260\linewidth}>{\raggedright\arraybackslash}p{0.130\linewidth}>{\raggedright\arraybackslash}p{0.120\linewidth}>{\raggedright\arraybackslash}p{0.120\linewidth}>{\raggedright\arraybackslash}p{0.140\linewidth}>{\raggedright\arraybackslash}p{0.130\linewidth}@{}}
\caption{Whole-network global main-path sensitivity}\label{tab:appendix_d_path_metrics}\\
\toprule Setting & Graph & Path edges & Shared edges & Edge Jaccard & Node Jaccard \\ \midrule\endfirsthead
\toprule Setting & Graph & Path edges & Shared edges & Edge Jaccard & Node Jaccard \\ \midrule\endhead
Baseline & backbone & 14 & 14 & 1.000 & 1.000 \\
Baseline & combined & 19 & 19 & 1.000 & 1.000 \\
Three entropy bins & backbone & 14 & 12 & 0.750 & 0.875 \\
Three entropy bins & combined & 20 & 9 & 0.300 & 0.367 \\
Seven entropy bins & backbone & 14 & 12 & 0.750 & 0.875 \\
Seven entropy bins & combined & 16 & 7 & 0.250 & 0.276 \\
Three comparable fields & backbone & 18 & 2 & 0.067 & 0.133 \\
Three comparable fields & combined & 17 & 6 & 0.200 & 0.226 \\
Distinct-configuration mass & backbone & 14 & 12 & 0.750 & 0.875 \\
Distinct-configuration mass & combined & 18 & 11 & 0.423 & 0.444 \\
Interface fields excluded & backbone & 14 & 14 & 1.000 & 1.000 \\
Interface fields excluded & combined & 20 & 4 & 0.114 & 0.171 \\
\bottomrule\end{longtable}\endgroup
\noindent{\footnotesize These are whole-network global SPC paths, distinct from the within-subcategory case paths expanded in the main article. Shared edges and Jaccard values compare each setting with the baseline.}

The path comparison uses whole-network global SPC paths, not the two subcategory case histories highlighted in Figures~2 and~3 of the main article. A small fraction of changed parent assignments can alter a structurally prominent route. Under three comparable fields, global backbone edge Jaccard is 0.067 and combined-network edge Jaccard is 0.200. Removing interface fields leaves the baseline backbone global path unchanged (1.000), while the combined-network global path has edge Jaccard 0.114. High overall parent agreement therefore does not imply stable main-path interpretation.

\subsubsection*{Documentary comparisons and date audit}
Ten purposively chosen cases compare product architecture and broad historical relationships with documentary evidence. The date audit flags 131 catalogue nodes under its criteria; none lies on the displayed global paths. Two displayed backbone-path edges have documented core identities at both endpoints, including one edge spanning a core change. The displayed combined path has no adjacent pair documented at both endpoints under that criterion. These cases provide partial documentary checks of the reconstructed histories.

\subsection{Stage definitions, controls and uncertainty}\label{app:revised_events}
Birth-time HT candidates require a strictly earlier carrier outside the receiving family's prior backbone ancestry. The selected regression opportunities additionally require cross-manufacturer availability. This is evidence of an external precedent rather than evidence of an actual transaction or a uniquely identified donor. Earlier own-family and ancestral records exclude a first-introduction opportunity; missing capability evidence is not treated as a verified nonadoption.

For each focal family, $A_i^-$ is mean similarity over complete backbone links strictly before its incoming link, and $G_i^-$ is the number of those prior links. The selected composite $C_i=A_i^-\log(1+G_i^-)$ is frozen at family birth and requires at least three prior generations. Each stage standardises $C_i$ using its eligible sample before separation exclusions and then squares the standardised value for the quadratic term. Table~\ref{tab:selected_anchor_scaling} retains the stage scales.

\begingroup\small\doublespacing\setlength{\tabcolsep}{3pt}
\begin{longtable}{@{}>{\raggedright\arraybackslash}p{0.360\linewidth}>{\raggedright\arraybackslash}p{0.180\linewidth}>{\raggedright\arraybackslash}p{0.180\linewidth}>{\raggedright\arraybackslash}p{0.180\linewidth}@{}}
\caption{Stage-specific standardization of the same raw log-generation Anchor}\label{tab:selected_anchor_scaling}\\
\toprule Model & Eligible mean & Population SD & Eligible N \\ \midrule\endfirsthead
\toprule Model & Eligible mean & Population SD & Eligible N \\ \midrule\endhead
Recorded HT detection & 1.237 & 0.241 & 889 \\
Within-family carriage & 1.611 & 0.327 & 153 \\
Descendant carriage & 1.577 & 0.393 & 179 \\
\bottomrule\end{longtable}\endgroup
\noindent{\footnotesize The scenario analysis evaluates common raw C = prior mean similarity times log(1+G), then applies the appropriate row of this table to each stage. Coefficients are not directly comparable on a shared z scale.}

All three models control for standardised log-transformed prior generations and prior outside-path carrier counts (using $\log(1+x)$ for each), and standardised linear year, with capability, subcategory and manufacturer fixed effects. Detection additionally controls for standardised log current-release SKU count. Because generations also enter $C_i$, its conditional association varies similarity at a fixed lineage length. Count outcomes retain their natural units and use no exposure offset in the selected specification. Post-event SKU or descendant totals are not added as baseline controls; recording-rate specifications constitute distinct estimands in the saved sensitivity analyses.

\begingroup\small\doublespacing\setlength{\tabcolsep}{3pt}
\begin{longtable}{@{}>{\raggedright\arraybackslash}p{0.240\linewidth}>{\raggedright\arraybackslash}p{0.110\linewidth}>{\raggedright\arraybackslash}p{0.150\linewidth}>{\raggedright\arraybackslash}p{0.090\linewidth}>{\raggedright\arraybackslash}p{0.080\linewidth}>{\raggedright\arraybackslash}p{0.120\linewidth}>{\raggedright\arraybackslash}p{0.130\linewidth}@{}}
\caption{Selected model eligibility, separation, and root concentration}\label{tab:selected_stage_support}\\
\toprule Model & Eligible & Removed by separation & Fitted & Roots & Effective roots & Largest root \\ \midrule\endfirsthead
\toprule Model & Eligible & Removed by separation & Fitted & Roots & Effective roots & Largest root \\ \midrule\endhead
Recorded HT detection & 889 & 500 & 389 & 8 & 2.583 & 57.33\% \\
Within-family carriage & 153 & 1 & 152 & 5 & 1.144 & 93.42\% \\
Descendant carriage & 179 & 0 & 179 & 10 & 1.509 & 81.01\% \\
\bottomrule\end{longtable}\endgroup

Detection retains 389 of 889 eligible opportunities after separation exclusions. The two follow-up models retain 152 and 179 events under their respective observation and horizon rules. Within-family carriage places 142 of 152 events in one root; five roots have an effective concentration count of 1.144, and the root-cluster covariance for the two Anchor terms has rank one. Table~\ref{tab:supp_stage_directions} reports all three directional comparisons under root, family and manufacturer clustering. These are alternative covariance calculations for the same fitted relationships.

\begin{table}[tbp]\centering\begin{threeparttable}
\caption{Directional tests and sensitivity to the dependence unit}\label{tab:supp_stage_directions}
\small\doublespacing\setlength{\tabcolsep}{4pt}\begin{tabularx}{\linewidth}{@{}Xrrr@{}}\toprule
Directional comparison & Root $p$ & Family $p$ & Maker $p$\\\midrule
HT slope negative at fitted P95 & $<0.001$ & $0.002$ & $0.004$\\
Within-family slope positive at P05 & $<0.001$ & $0.132$ & $0.099$\\
Descendant slope positive at P05 & $0.027$ & $0.252$ & $0.186$\\
\bottomrule\end{tabularx}\begin{tablenotes}\footnotesize\doublespacing
\item Entries are nominal, unadjusted \emph{one-sided endpoint} $p$-values under the dependence unit in each column. They differ from the two-sided coefficient tests in Table~3 of the main article.
\item P05 and P95 refer to the corresponding fitted stage's Anchor distribution. The joint directional comparison and its multiplicity adjustment are reported in the next subsection. The within-family root covariance for the two Anchor terms has rank one.
\end{tablenotes}\end{threeparttable}\end{table}

\subsubsection*{Fitted shapes and turning points}
For stage $s$, the slope on the linear-predictor scale is proportional to $\beta_s+2\kappa_s z_s(C)$. A fitted stationary point in raw units is
\begin{equation}
C_s^*=\overline C_s-\sigma_s\frac{\beta_s}{2\kappa_s}.
\end{equation}
Using the saved unrounded estimates, the HT-occurrence and descendant turning points are approximately 1.104 and 1.200. The within-family stationary point is 4.089, beyond its fitted support. Thus, the reported within-family curve is increasing over observed support even though its regression includes a quadratic term. A negative quadratic coefficient, an endpoint-direction comparison and a supported inverse-U relationship are different summaries.

\subsection{Specification search and largest-root exclusions}
\begingroup\small\doublespacing\setlength{\tabcolsep}{3pt}
\begin{longtable}{@{}>{\raggedright\arraybackslash}p{0.250\linewidth}>{\raggedright\arraybackslash}p{0.230\linewidth}>{\raggedright\arraybackslash}p{0.100\linewidth}>{\raggedright\arraybackslash}p{0.100\linewidth}>{\raggedright\arraybackslash}p{0.120\linewidth}>{\raggedright\arraybackslash}p{0.100\linewidth}@{}}
\caption{Specification combinations meeting the three nominal endpoint criteria}\label{tab:appendix_d_nominal_candidates}\\
\toprule Anchor & Risk set & HT p & Within p & Descendant p & Holm p \\ \midrule\endfirsthead
\toprule Anchor & Risk set & HT p & Within p & Descendant p & Holm p \\ \midrule\endhead
Mean similarity times log(1+G) & Prior cross-firm carrier & $<0.001$ & $<0.001$ & 0.027 & 1.000 \\
Mean similarity times G/(G+3) & Prior cross-firm carrier & 0.003 & $<0.001$ & 0.030 & 1.000 \\
Mean similarity times log(1+G) & Any outside-path carrier & 0.022 & $<0.001$ & 0.034 & 1.000 \\
Mean similarity times G/(G+3) & Any outside-path carrier & 0.036 & $<0.001$ & 0.033 & 1.000 \\
\bottomrule\end{longtable}\endgroup
\noindent{\footnotesize All four specifications use quadratic Anchor terms, manufacturer fixed effects and five-year count outcomes. One-sided endpoint p-values are unadjusted; the final column reports the Holm-adjusted joint comparison across the family of 128 combinations.}

The specification-selection record contains 128 three-model combinations and 288 unique full-sample fits. Four combinations meet the three directional endpoint criteria at nominal thresholds; each has a Holm-adjusted joint $p=1.000$. For the main combination, the largest unadjusted directional $p$ is 0.027 and the adjusted value is 1.000. The number of combinations meeting nominal inverse-U criteria in all three outcomes is zero. A separate sign-agnostic association criterion yields one adjusted passing combination with low-Anchor directions opposite to the directional pattern examined in the main article. Joint directional, complete-shape and sign-agnostic comparisons are therefore reported as distinct tests.

Removing each input sample's largest root yields 144 completed fits among 288 exclusion attempts. The within-family variants are unidentified under these exclusions, leaving the three-model comparison unavailable. Unidentified fits are recorded as unavailable rather than as zero effects. The electronic record retains all covariance alternatives, exclusion attempts and observation identities alongside the full selection registry.

\subsection{Support and uncertainty of the Anchor scenarios}
Appendix~\ref{app:subcategory_scenarios} displays predictions only where fitted-stage support permits them. For each target, the permitted raw-Anchor grid intersects stage-specific subcategory fifth-to-ninety-fifth-percentile ranges, requires observed module--manufacturer--subcategory combinations and estimable prediction rows, and imposes $0\leq C/\log(1+G)\leq1$ at its fixed generation count. These checks do not establish joint overlap of all continuous controls.

The shared RF SoC target contains 66 opportunity profiles from one root; 36 fall outside a participating stage's subcategory range for another continuous control. Only one actual positive RF SoC event belongs to all three fitted samples. The pairwise descendant scenario instead uses 149 targets, explaining its different peak. These scenarios transport pooled coefficients across observed populations and do not estimate independent subcategory-specific Anchor effects. The MCU maxima occur at upper support boundaries while fitted HT increases throughout the permitted intervals.

Four hundred common-cluster Rademacher score perturbations under each clustering scheme (root and family) preserve cross-stage dependence by applying shared signs. They perturb fitted coefficients locally rather than resampling observations and refitting models. For RF SoC, the family-based central 95\% spans of perturbed joint-target peak locations are approximately 0.836--1.276 for within-family carriage and 0.847--1.241 for descendant carriage. These are local sensitivity envelopes; calibration, prior specification selection and graph uncertainty are separate inferential issues. Likewise, the 95\%-of-maximum bars in the figures express a performance tolerance, not a 95\% confidence claim.

\subsection{Reproducibility and supplementary evidence}\label{app:reproduction}
The computational documentation identifies the data release, family membership, initial SKU comparisons, inferred paths, dated capability observations and regression samples. Separate records cover the selected fits, all 128 specification combinations, alternative covariance estimates, largest-root exclusions and scenario calculations. The model-search registry includes unavailable fits rather than recording them as zero effects.

Source fingerprints, graph and event manifests, sample identities and scenario-support registers connect the reported results to their inputs. Display rounding does not alter the underlying values, test statistics or membership of the near-maximum sets. Repository organisation, execution order and source-file requirements are specified in the computational documentation. These records establish computational provenance; historical parentage and physical technology transfer remain distinct questions for documentary validation.

\end{document}